\documentclass[12pt,a4paper]{article}
\usepackage{amssymb}

\usepackage{graphicx}
\usepackage{amsmath}

\begin{document}

\title{Notes on a possible phenomenology of internal transport barriers in tokamak}
\author{F. Spineanu$^{1}$, M. Vlad$^{1}$, K. Itoh$^{1}$ and S.-I. Itoh$^{2}$ \\
$^{1}$ National Institute for Fusion Science\\
322-6 Oroshi-cho, Toki-shi, Gifu-ken 509-5292, Japan\\
$^{2}$ Research Institute for Applied Mechanics, Kyushu University, \\
Kasuga 816-8580, Japan}
\date{}
\maketitle

\begin{abstract}
We propose a new phenomenology of the generation of internal transport
barriers, based on the exact periodic solution of the Flierl-Petvishvili
equation. We examine the stability of this solution and compare the late
stages of the flow with the ensemble of vortices.

\end{abstract}

\tableofcontents

\section{Introduction}

The aim of these notes is to contribute to the clarification of the
dynamical processes leading to the formation, confinement charactersitics
and intermittent behavior of Internal Transport Barriers (ITB). We focus
here on the role of the stationary periodic structure of the flow appearing
in the intermediate spatial range between the highly radially elongated
eddies of the Ion Temperature Gradient (ITG) instability and the Larmor -
radius scale vortices described by the Hasegawa-Mima equation. In this \
``mesoscopic'' range the structures are described by the ion dynamics
equation where the scalar nonlinearity (in contrast to the polarisation
drift nonlinearity) is dominating.

The current physical understanding of the formation of the transport
barriers is based on the effect of the sheared poloidal plasma rotation on
the ITG potential structures. The ITG potential structures are in general of
large radial extension. They may persist even in the turbulent regime and
are a major agent of transport since the plasma convection inside the eddies
is an efficient support for temperature advection between high (small $r$)
and low (high $r$) temperature regions. It is well known that this geometry
of the flow generates velocity stress and that the divergence of this tensor
induces a substantial contribution in the momentum balance equation for
ions. When this momentum drive is sufficiently high, the plasma begins to
rotate, and a space distribution of the poloidal velocity is established.
The shear of the poloidal velocity has a strong effect on the turbulent
structures with comparable space extension and the radial correlation length
is reduced. This correspondingly reduces the transport leading to the
formation of a transport barrier. A theory that explains this in
quantitative terms has been proposed by Diamond and coworkers, as a
``predator-pray'' model, in general formalized with Fischer-type equations.
It consists of the description, in terms of the density of the energy
quanta, of the balance in phase space of two distinct populations,
associated with the turbulent waves and respectively with the zonal flows.
The model naturally leads to the self-organised state consisting of mutual
control of these two physical entities, found in a permanent relative
adjustement. This model is essentially a thermodynamical model, which
probably remains true for a wide variety of dynamical details.

A picture of the dynamics is allways useful, even if the scaling laws (of
threshold and transport rate) are in general based on thermodynamical
balance and conservation properties. The later is a more general (and safer)
level of description, leading to a picture where the invariance properties
of the dynamical equations govern the scaling. The dynamical evolution of
the system is a detailed description and its validity is strongly dependent
on a clear separation of space-time scales and a correct choice of initial
and boundary conditions. However, the physical content can only revealed by
the detailed description.

In this work we develop a possible phenomenological model of generation of
the Internal Transport Barriers in tokamak. This model relies on the role of
the stationary periodic exact solution of the Flierl-Petviashvili equation
as an attracting state for the ion turbulence, when the regularity of the
eddies is still pronounced. We also analyse analytically and numerically the
stability of this solution. Some properties are shown to be connected with
the fact that this equation is closely related with other, exactly
integrable models.

\section{Possible phenomenology of the ITB formation}

Numerical simulations show that the structure of the potential perturbation
in the ion temperature gradient (ITG) instability are elongated and rather
thin on the direction that is transversal to their longest dimension. The
Reynolds stress arising from the broken symmetry configuration induces the
rotation of the plasma. It may be useful here to recall the similar
situation arising in the thermal convection associated with the
Rayleigh-Benard (RB) instability. Experiments carried out by Howard and
Krishnamurti in typical RB geometry at increasing Rayleigh number have
revealed that the classical conduction-convection bifurcation is only the
first of a series of transitions generating new system 's behaviors. The RB
geometry of convective rolls is sensitive to perturbations and this leads to
a symmetry breaking consisting of deformations of the rolls in direction
parallel with the two plates, in one direction for the low (high
temperature) plate and opposite direction for the upper (low temperature)
plate. This is accompanied by a nonvanishing statistical average of the
velocity stress tensor, whose \ nonzero divergence (Reynolds stress) is
equivalent to a momentum drive. The lower part of the convective fluid is
entrained in a flow in one direction (parallel to the plates) while the
upper part of the convective field aquires a flow motion in the opposite
direction. These flows are called ``winds'' and they are seen, together with
the strong deformation of the rolls, on the experimental picture of the
fluid section. In addition, intermittent processes consisting of rising of
finite volumes of cold fluid (from the bottom region) toward the top plate
and fall of warm fluid from the top region into the cold region have been
observed. These events are called ``plumes''. One may consider from this
picture only the part relating to the effect of the Reynolds stress as
particularly relevant for the plasma convection in the eddies of the ITG
instability.

The result of the strong Reynolds stress arising in the ITG instability is
the deformation (``tilting'') of the convection eddies and the generation of
the corresponding winds. The convective structures are strongly elongated
and tend to align with the poloidal direction, as imposed by the momentum
from the Reynolds stress. At a certain moment the deformed ITG pattern of
flow evolves to the solution consisting simply of parallel layers of flow, a
structure which is essentially poloidally oriented and periodic along the
radial direction. We have proved that this is an exact and robust solution
of the ion dynamics on mesoscopic space scales, (approx. $\rho
_{s}/\varepsilon $) where it is dominated by the scalar nonlinearity.
Attaining this state, the plasma has practically vanishing transport in the
radial direction. We have to note that it is not necessary to invoke the
tearing appart of the ITG eddies and the destruction of the radial
correlation via sheared flow. The solution intrinsically has a radial scale
and no transport because there is no phase mismatch between particles and
potential.

The flow represented by this solution may last indefinitely except for the
collisional decay which however may be considered low in high temperature
regimes. Furthermore, in the ideal stationary case, the Reynolds stress
vanishes due to the poloidal symmetry. However there is a limitation of the
persistance of this periodically layered flow and this arises from the
stability against perturbations. We can suppose that this may take place in
two different ways. The first of them is essentially a reversed process as
that which led to the periodic layers, and consists of closing the flow
lines at finite poloidal wavelength, with suppression of the zonal flow and
reformation of the tilted ITG pattern. The finite radial projection leads to
a sudden increase of transport. The other way is a strong qualitative change
of the flow.

We have found that the stability of the periodic flow pattern is determined
both by the amplitude and by the geometry of the initial perturbation. Some
perturbations (like, for example, the monopolar vortices embedded inside a
layer) have a very long stability time since they accomodate with the
background flow by reshaping the distribution of local velocities. However,
perturbations that have a high amplitude relative to the background flow
and/or do not conform to the flow geometry lead to the destabilisation and
eventually destruction of the flow pattern. It is important however to note
that the numerical simulations (presently with limitted precision) and
analytical considerations suggest that the destruction of the flow pattern
does not immediately lead to an arbitrary random field: a fundamental
process is the generation of small ($\sim \rho _{s}$) space scale vortices.
We can show that the regular flow structure is replaced by a lattice of
vortices which is reminescent of the exact solution of a closely related
nonlinear equation. This solution evolves, due to the weak interaction
between the vortices, to an ensemble of quasi-independent vortices that
collide inelastically and become of various amplitudes plus a surrounding
drift wave radiation. From this random field the ITG instability may
regenerate the structure of eddies and reinstate the transport. The time
scale for doing this is $\gamma _{ITG}$, followed by saturation, as shown by
many simulations.

The successive steps of this qualitative scenario can repeate themselves
over and over but the senstivity to perturbation of the time scales involved
may lead to an intermittent (bursty) behavior rather than to a limit cicle.
From purely theoretical point of view it is hard to expect that this
evolution can be lumped into a single equation whose phase space could
provide the successive transitions.

We note that the transport is low in the phase where the flow is described
by the stationary periodic solution of the Flierl-Petviashvili equation and
in the state where the ensemble of vortices has replaced it (since the
elementary space scale is very small, of the order of few Larmor radius, as
we will show below. The large eddies have a substantial transport.

\section{Derivation of the equation at the intermediate space scales}

\subsection{The equation for the two-dimensional ion dynamics}

Consider the equations for the ITG model in two-dimensions with adiabatic
electrons: 
\begin{eqnarray*}
\frac{\partial n_{i}}{\partial t}+\mathbf{\nabla \cdot }\left( \mathbf{v}%
_{i}n_{i}\right) &=&0 \\
\frac{\partial \mathbf{v}_{i}}{\partial t}+\left( \mathbf{v}_{i}\cdot 
\mathbf{\nabla }\right) \mathbf{v}_{i} &=&\frac{e}{m_{i}}\left( -\mathbf{%
\nabla }\phi \right) +\frac{e}{m_{i}}\mathbf{v}_{i}\times \mathbf{B}
\end{eqnarray*}
We assume the quasineutrality 
\begin{equation*}
n_{i}\approx n_{e}
\end{equation*}
and the Boltzmann distribution of the electrons along the magnetic field
line 
\begin{equation*}
n_{e}=n_{0}\exp \left( -\frac{\left| e\right| \phi }{T_{e}}\right)
\end{equation*}
In general the electron temperature can be a function of the radial variable 
\begin{equation*}
T_{e}\equiv T_{e}\left( x\right)
\end{equation*}

The velocity of the ion fluid is perpendicular on the magnetic field and is
composed of the diamagnetic, electric and polarization drift terms 
\begin{eqnarray*}
\mathbf{v}_{i} &=&\mathbf{v}_{\perp i} \\
&=&\mathbf{v}_{dia,i}+\mathbf{v}_{E}+\mathbf{v}_{pol,i} \\
&=&\frac{T_{i}}{\left| e\right| B}\frac{1}{n_{i}}\frac{dn_{i}}{dr}\widehat{%
\mathbf{e}}_{y} \\
&&+\frac{-\mathbf{\nabla }\phi \times \widehat{\mathbf{n}}}{B} \\
&&-\frac{1}{B\Omega _{i}}\left( \frac{\partial }{\partial t}+\left( \mathbf{v%
}_{E}\cdot \mathbf{\nabla }_{\perp }\right) \right) \mathbf{\nabla }_{\perp
}\phi
\end{eqnarray*}
Introducing this velocity into the continuity equation, one obtains an
equation for the electrostatic potential $\phi $.

In the \textbf{Appendix} it is presented a derivation of three versions of
the nonlinear differential equations that may govern, in certain conditions,
the two-dimensional ion dynamics at intermediate scales.

\subsection{The stationary equation at intermediate space scales
(Flierl-Petviashvili equation)}

The ion drift wave equation has a distinct dynamical behavior according to
the space-time scales involved. The Hasegawa-Mima-Charney (HMC) equation is
obtained for small scales and the stationary states exhibit dipolar
structures of the order of the Larmor radius. On larger scales (intermediate
between the HMC dipole and the large ITG eddies) the scalar (or KdV-type)
nonlinearity is prevailing and the only known structures are monopolar \cite
{HH}. Both are not solitonic but very robust and long lived. In the latter
case a one-dimensional version of the equation has been derived by
Petviashvili \cite{Petv1} who also used it in the study of the Jupiter's Red
Spot. The equation has been rederived along with a careful analysis of the
scales involved \cite{Spatschek2}, \cite{Su}, resolving a controversy on the
role of the temperature gradient \cite{Lakhin}, \cite{LaedkeSpatschek1}. In
the study of ocean flows, Flierl \cite{Flierl} has independently formulated
an equation with the same structure. The one dimensional equation has been
solved on an infinite domain \cite{Meiss}, \cite{Lakhin}, obtaining as
solution the KdV soliton. The two dimensional equation has vortical
monopolar solutions, well studied numerically \cite{Boyd1}, \cite{Iwasaki}.
Other applications include long waves on thin liquid films and Rossby waves
in rotating atmosphere.

The Flierl-Petviashvili equation (derived several times in the Appendix) is 
\begin{equation*}
\Delta \phi =\alpha \phi -\beta \phi ^{2}
\end{equation*}
where 
\begin{eqnarray}
\alpha &=&\frac{1}{\rho _{s}^{2}}\left( 1-\frac{v_{\ast }}{u}\right)
\label{alphabeta} \\
\beta &=&\frac{T_{e}}{2u^{2}eB_{0}^{2}\rho _{s}^{2}}\frac{\partial }{%
\partial x}\left( \frac{1}{L_{n}}\right) =\frac{e}{2m_{i}u^{2}}\frac{%
\partial }{\partial x}\left( \frac{1}{L_{n}}\right)  \notag
\end{eqnarray}

By defining a functional of the solution expressed in one-dimensional
(radial) geometry and taking the extremum of the functional under the
condition ofasymptotic decay, Petviashvili and Pokhotelov \cite{Petv2} have
found the solution 
\begin{equation}
\phi \left( r\right) =\frac{4.8\alpha }{2\beta }\left[ \mathrm{sech}\left( 
\frac{3}{4}\sqrt{x^{2}+\left( y-ut\right) ^{2}}r\right) \right] ^{4/3}
\label{sech43}
\end{equation}
This is only an approximate solution and the reason to look for such form
resides in the previous indication that the one dimensional version of this
equation, with generalized nonlinearity, presents exact solutions as powers
of the $\mathrm{sech}$ function.

\subsection{The monopolar vortex and the similarity with integrable equations%
}

It is well known that the Flierl-Petviashvili equation has long been studied
in relation with the possibility that it has as solution an isolated vortex
with nontrivial stability properties. This vortex has never been determined
analytically and the explanation seems to be that the equation is not
integrable. However, analytical approximations exists Eq.(\ref{sech43}) and
also a very well documented study of numerically-calculated vortex solution
is available \cite{Iwasaki}, \cite{Boyd1}.

The monopolar vortex has been considered in relation with the Great Red
Spot, a vortical structure with a long time life (more than three hundread
years), which is observed to be embedded in the zonal flow of the atmosphere
of Jupiter. The condition imposed to the Flierl-Petviashvili equation is to
provide an isolated, finitely extended vortical solution, obeying boundary
conditions at infinity (on the $2D$) of smooth decaying.

We consider that the insuccess in determining the analytical vortex solution
has a fundamental motivation.

First of all the equation does not pass the Painlev\'{e} test, so it is very
likely that it is not integrable. This, in the definition that is suggested
by the Inverse Scattering Transform method, means that there is no Lax pair
of operators for this equation. In other words, one cannot find a system of
two linear differential equations, whose condition of compatibility to be
expressed as the nonlinear FP equation.

This equation has close ressemblance with some approximative form of other
nonlinear equations. For example, consider the equation 
\begin{equation*}
\Delta \phi =\exp \left( -\phi \right)
\end{equation*}
This is the Liouville equation, with numerous applications. In particular it
has been invoked in the description of the distribution of current in the
magnetic confinement systems (tokamak). This equation is integrable and
exact solutions are known. In addition, exact analytical solutions can be
constructed on periodic domains. The expansion of the right hand side of
this equation takes a form close to the FP equation.

Even more close is the differential equation 
\begin{equation*}
\Delta \phi =\exp \left( \phi \right) -1
\end{equation*}
sometimes known as Abelian-Higgs equation (AH), governing the vortices of
the superconducting media. This equation has also been derived by
Jacobs-Rebbi. It is clear that an expansion of the right hand side leads to
something close to the FP equation, especially because the FP equation has
actually been derived under the neglect of the third order powers of $\phi $%
. 
\begin{eqnarray*}
\Delta \phi &=&\exp \left( \phi \right) -1 \\
&\approx &\phi -\frac{1}{2}\phi ^{2}
\end{eqnarray*}
Or, this equation is exactly integrable on periodic domains. We have proved 
\cite{flmadi7} that it possesses a Lax pair of operators and we have
constructed exact, analytical solutions, expressed as ratios of Riemann $%
\Theta $ functions.

\begin{figure}[tbp]
\centerline{\includegraphics[height=10cm]{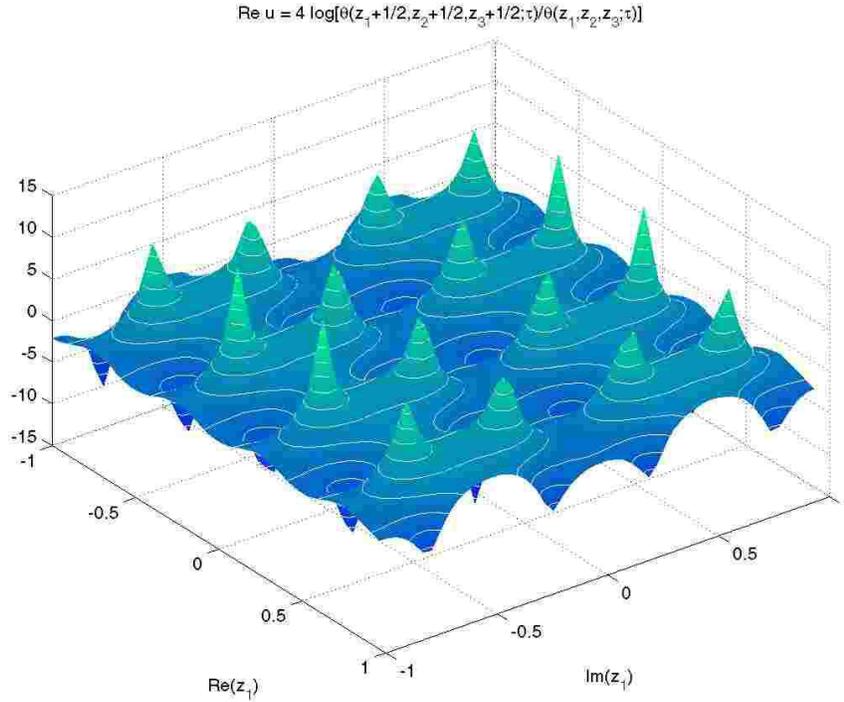}}
\caption{The exact solution of the Jacobs-Rebbi (Abelian-Higgs) equation for
ideal fluids obtained analytically as a ratio of Riemann $\Theta $
functions. }
\label{FigA_2}
\end{figure}

The solutions of the AH equation, as shown in the Fig.\ref{FigA_2} have the
configuration of a lattice of vortices. This solution can be regarded in two
different ways: as a function with a periodic set of maxima and minima; or,
in a more physical way, as an ensemble of vortices in plane that interact
weakly. Then one can easily assume that under the influence of
perturbations, the precise periodic geometry can be lost and the vortices
begin to move rather freely in plane and the interactions take the form of
inelastic collisions.

The transformation of ridges having as section the \emph{sechyperbolic}
function, into an array of vortices has been known previously and is well
documented numerically \cite{Iwasaki}, \cite{Boyd1}. We claim that the
actual reason for this evolution, in the numerical simulation of the FP
equation, is not the attracting nature of the monopolar vortex (which
actually it is not even an exact solution) but the fact that the initial
function evolves to the nearest soliton-like solution of the closely related
equation, the AH equation. Or, this solution is a lattice of monopolar
vortices.

We can also explain why the monopolar vortex of the FP equation, obtained
only numerically and studied under a certain numerical imprecision, has an
amazing stability: it actually represents one only exemplar of vortex from
the lattice, isolated and extended to $2D$ infinity by simply ignoring its
appartenence to the full, periodic solution. This is only numerically
acceptable, as an approximation, if the distance between neighboring
vortices is large enough.

\begin{figure}[!tbp]
\centerline{\includegraphics[height=7cm]{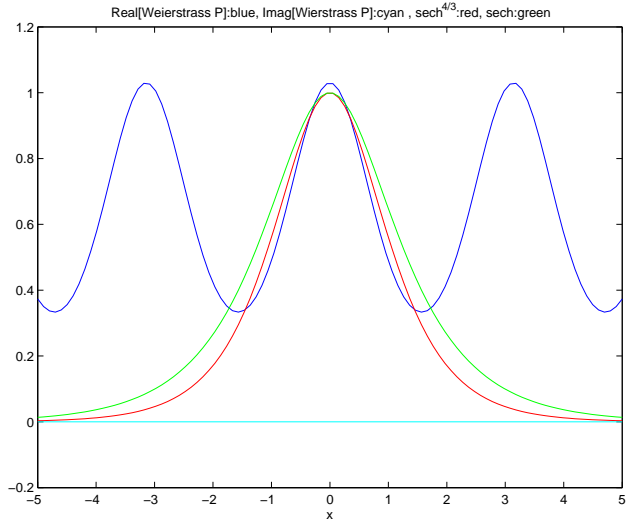}}
\caption{Comparison between the $sech^{4/3}$ with the Weierstrass function,
which is the exact solution of the FP equation}
\label{FigA_3}
\end{figure}

We note, in addition, that the FP equation has also ressemblance with the
nonlinear differential equation 
\begin{equation*}
\left( \Delta +\lambda \right) \phi =\sinh \phi
\end{equation*}
which describes the stationary states of the Hasegawa-Mima equation. This
equation, again, is exactly integrable on periodic domains, and its solution
can be analytically expressed as ratios of Riemann $\Theta $ function. A
comparison between different functions and the exact solution, represented
by $\wp $ is shown in Fig.\ref{FigA_3}. 

\section{Coherent structures and drift wave radiation}

In previous studies \cite{PRLflmadi}, \cite{flmadiPRE} we have developed a
theoretical approach to describe a coherent structure (the monopolar vortex
of the FP equation) immersed in a turbulent environment \cite{Pf25-82-1838}, 
\cite{Pf26-83-990}, \cite{PRep192-90-1}. The theoretical instruments are
inspired by the quantum theory of fields, in particular the calculation of
field correlations from the effective action \cite{Amit}, \cite{qcd}. We
have defined the action functional according to the Martin-Siggia-Rose
description \cite{MSR}, \cite{Krommes}, and calculated explicitely the
generating functional of correlations 
\begin{equation}
\mathit{Z}_{J}=\exp \left( iS_{Js}\right) \frac{1}{2^{n}i^{n}}\left( 2\pi
\right) ^{n/2}\left( \det \left. \frac{\delta ^{2}\widehat{\mathit{O}}}{%
\delta \varphi \delta \chi }\right| _{\varphi _{Js},\chi _{Js}}\right)
^{-1/2}  \label{exp22}
\end{equation}
where the determinant is obtained from the product of the eigenvalues of the
operator 
\begin{equation}
\det \left( \left. \frac{\delta ^{2}\widehat{\mathit{O}}}{\delta \varphi
\delta \chi }\right| _{\varphi _{Js},\chi _{Js}}\right) =\prod_{n}\lambda
_{n}  \label{exp4}
\end{equation}
Then the final expression has been derived in the form 
\begin{eqnarray}
\mathit{Z}_{J} &=&\exp \left( iS_{J}\right) \left( \prod_{n}\frac{\left(
2\pi \right) }{i}\right) \left[ \det \left( \left. \frac{\delta ^{2}S_{J}}{%
\delta \varphi \delta \chi }\right| _{\varphi _{Js},\chi _{Js}}\right) %
\right] ^{-1/2}  \label{oper56} \\
&=&const\;\exp \left( iS_{J}\right) \,\left[ \frac{\beta /2}{\sinh \left(
\beta /2\right) }\right] ^{1/4}\left[ \frac{\sigma /2}{\sin \left( \sigma
/2\right) }\right] ^{1/2}  \notag
\end{eqnarray}
and the correlation results from functional derivatives. 
\begin{equation*}
\left\langle \varphi (y_{2})\varphi (y_{1})\right\rangle =\left. \mathit{Z}%
_{J}^{-1}\frac{\delta ^{2}\mathit{Z}_{J}}{i\delta J(y_{2})\,i\delta J\left(
y_{1}\right) }\right| _{J=0}
\end{equation*}
This theoretical approach may be very useful in the study of the late stages
of the perturbed flow, where the potential consists of both coherent
vortices and drift wave radiation.

\section{The analytical solution of the FP eq.}

In a previous work \cite{Flomadi} we have determined an exact solution of
the FP equation. The method consisted in looking to the trajectories of the
one-dimensional solution singularities, in the complex plane of the spatial
variable. The form of the solution has been assumed as 
\begin{equation}
\phi \left( x,y\right) =-2\phi _{0}\sum_{n=1}^{N}\sum_{l=-\infty }^{\infty }%
\frac{1}{\left[ \gamma \left( y-y_{n}\right) -ilD\right] ^{2}}  \label{phiin}
\end{equation}
and differential equations (plus a constraint) have been derived for the
positions of the complex poles $y_{n}\left( x\right) $. An exact solution to
the Petviashvilli equation with \emph{constant} coefficients $\alpha $ and $%
\beta $ is 
\begin{equation}
\phi \left( x,y\right) =\frac{\alpha }{2\beta }+s\wp \left( iay+ibx+\omega
|g_{2}=\frac{3\alpha ^{2}}{\left( s\beta \right) ^{2}}\right)   \label{sol}
\end{equation}
with the condition 
\begin{equation}
a^{2}+b^{2}=\frac{s\beta }{6}  \label{ab}
\end{equation}
where $\wp $ is the doubly periodic elliptic Weierstrass function. 
\begin{figure}[tbp]
\centerline{\includegraphics[height=7cm]{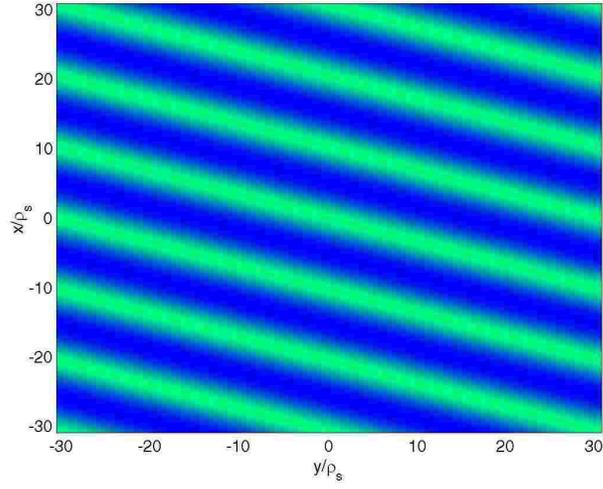}}
\caption{Exact solution of the FP equation.}
\label{FigA_4}
\end{figure}

\begin{figure}[tbp] \centerline{\includegraphics[height=8cm]{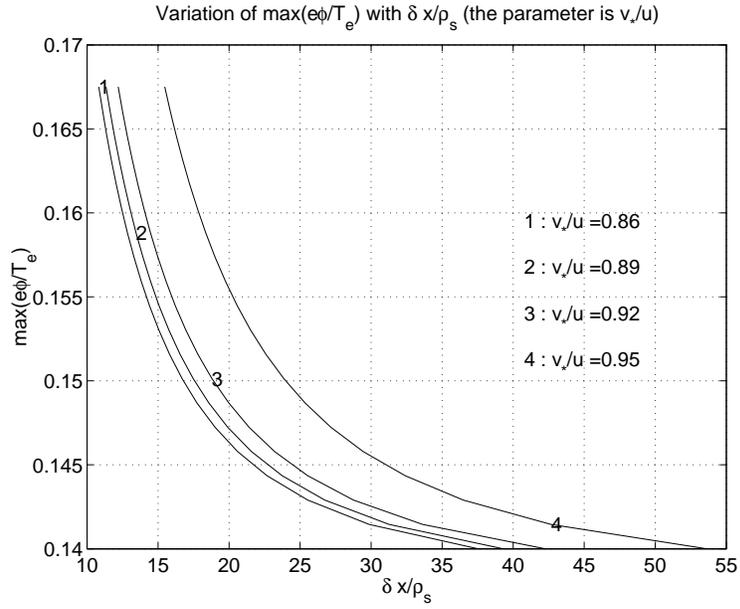}} 
\caption{The realation between the amplitude of the potential of the flow and
the width of the periodicity layer, parametrised by the ratio $v_{*}/u$.} %
\label{Fig1_2} \end{figure} 

\section{Comparison with experiments}

Experimental measurements of the characteristics of the zonal flow have been
performed on Doublet III-D tokamak (\cite{Coda}). In Ohmic and L-mode
plasmas it has been found a perturbed potential $\widetilde{\varphi }%
_{rms}>10\,V$ and a flow shear $\omega _{E\times B}\sim 2\times
10^{5}\,s^{-1}$. The value for the radial wavelength is in the range $%
k_{r}\rho _{s}\in \left[ 0.1,0.6\right] $ which means $\lambda _{r}\in
\left( 15...30\right) \rho _{s}$. Due to the different sensitivity of the
solution $\phi _{s}$ to the parameters, this set is sufficiently restrictive
to determine its form. We have to take $v_{\ast }/u$ very close to unity and 
$g_{3}\sim -1500$ which gives: $\widetilde{\varphi }_{rms}>17\,V$, $\lambda
_{r}\simeq 17.4\rho _{s}$, $\omega _{E\times B}\sim 2.2\times 10^{5}\,s^{-1}$%
. The relative amplitude of the perturbation results $\sim 4\%$. An
important experimental result is the radial spectrum of the perturbation. We
have calculated $S\left( k_{r}\right) $ from the Fourier transform of the
correlation of $\phi _{s}\left( x\right) $. The result is very close to
Fig.3 of Ref.\cite{Coda}, with two symmetric peaks at $k_{0r}\sim 4\,cm^{-1}$%
. The same sharp decay for $\left| k_{r}\right| <k_{0r}$ is observed, as
described in Ref.\cite{Coda}. We also note that the analytic spectrum is
structured with very narrow dips on its range $k_{r}\in \left[ 1,10\right]
\,cm^{-1}$ which, according to Ref.\cite{Coda} cannot be resolved in the
measurements. Again, a small fluctuation of the parameters generate an
average spectrum very close to Fig.3 of the mentioned work. 
\begin{figure}[!tbp]
\centerline{\includegraphics[height=8cm]{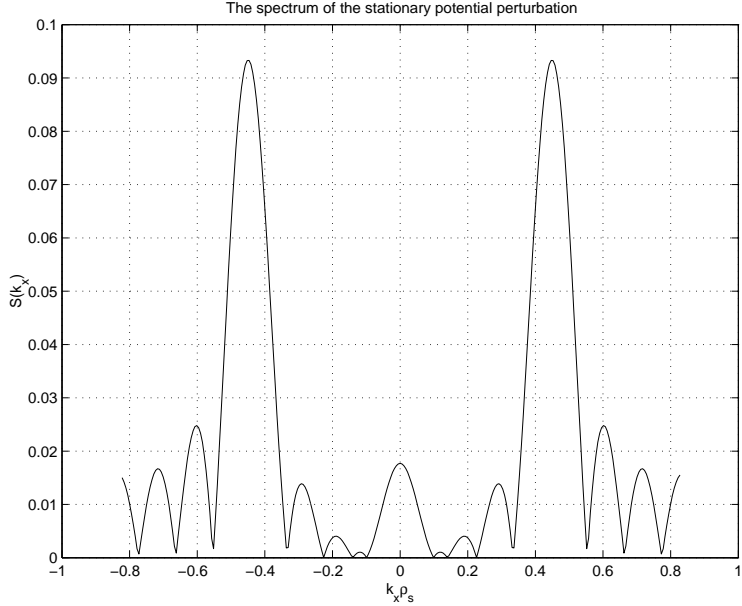}}
\caption{The spectrum obtained from the Fourier transform of the correlation
of the solution of the FP equation.}
\label{Fig2_1}
\end{figure}
\begin{figure}[!tbp]
\centerline{\includegraphics[height=8cm]{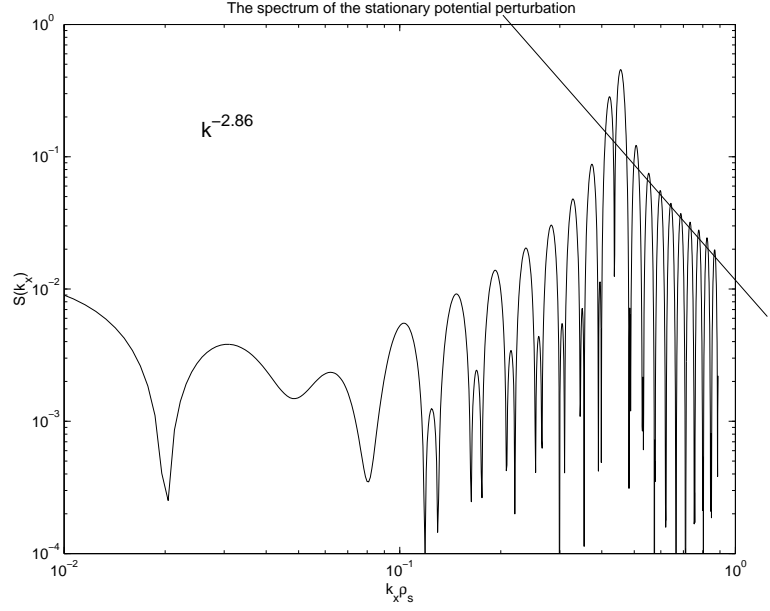}}
\caption{Graph of the spectrum in $log-log$ scale and estimation of the
decay.}
\label{Fig2_2}
\end{figure}
.

\section{Comparisons with results from numerical simulations}

\subsection{Numerical simulations at Lausanne}

The conditions are the following: 
\begin{eqnarray*}
L_{T_{i}} &=&a/3\;\left( m\right) \\
B_{T} &=&2.5\;\left( T\right) \\
\Omega _{i} &=&120\;\left( MHz\right) \\
\rho _{s} &=&1.8\times 10^{-3}\;\left( m\right) \\
T_{e} &=&T_{i}=5\;\left( KeV\right)
\end{eqnarray*}

In these conditions the results obtained were presented in several plots in
Ref.\cite{Lausanne1}.

\subsubsection{Geometry of the zonal flow}

The radial electric field is shown in Figure 4 of this reference. The values
are in the range 
\begin{eqnarray*}
E_{r} &\sim &\left( -0.02\cdots +0.02\right) \times 1.2\times 10^{6}\;\left(
V/m\right) \\
&\sim &\left( -24\cdots +24\right) \times 10^{3}\;\left( V/m\right)
\end{eqnarray*}

Another characteristic of the flow obtained numerically in this reference is
the periodicity width, $\delta x$. This may be estimated on the basis of the
Fig.4. 
\begin{equation*}
\delta x\sim \left( 0.06\cdots 0.1\right) a
\end{equation*}
and $a$ in this calculation has been taken 
\begin{equation*}
a=135\rho _{L}
\end{equation*}
which gives 
\begin{equation*}
\delta x\sim \left( 8.1\cdots 13.5\right) \rho _{L}
\end{equation*}

Using our analytical solution we can compute the radial electric field on
the basis of the analytical solution, $\phi _{s}\left( x\right) $. This
gives 
\begin{eqnarray*}
E_{x}^{phys} &=&-\frac{d\phi _{s}^{phys}}{dx^{phys}} \\
&=&-\frac{T_{e}}{\left| e\right| }\frac{1}{\rho _{s}}\left( \frac{d\phi _{s}%
}{dx}\right) \;\;\left( V/m\right)
\end{eqnarray*}
From this we can calculate the flow velocity 
\begin{equation*}
v_{y}^{phys}=\frac{E_{x}^{phys}}{B}\;\left( m/s\right)
\end{equation*}

\subsubsection{The shearing rate}

In the reference \cite{Lausanne1}, the shearing rate is computed as function
of the radial coordinate $s\equiv r/a$ and represented in Fig.5 and in
Fig.11. 
\begin{equation*}
\omega _{E\times B}\equiv r\frac{d}{dr}\left( \frac{1}{r}\frac{E_{r}}{B}%
\right)
\end{equation*}
The shearing rate is oscillatory as expected for the quasi-periodic geometry
of the flow, and a typical value at maximum is 
\begin{equation*}
\omega _{E\times B}\sim 2\times 10^{-3}\Omega _{i}
\end{equation*}

We can calculate this quantity 
\begin{equation*}
\omega _{E\times B}^{phys}=-\frac{v_{y}^{phys}}{r^{phys}}+\frac{dv_{y}^{phys}%
}{dr^{phys}}
\end{equation*}
In the slab geometry assumed in our model, 
\begin{equation*}
\omega _{E\times B}^{phys}=\frac{dv_{y}^{phys}}{dr^{phys}}\;\left(
s^{-1}\right)
\end{equation*}
which can be normaised to $\Omega _{i}$.

The results and the particular choice of parameters compatible with the
simulation Lausanne1 are in figures.

\texttt{Physical parameters}

\texttt{\ \qquad \qquad rhos (m) \qquad \qquad \qquad \qquad \qquad \qquad
\qquad = 0.0018}

\texttt{\ \qquad \qquad cs (m/sec) \qquad \qquad \qquad \qquad \qquad \qquad
= 0.21600E+06}

\texttt{\ \qquad \qquad Omega\_i (sec**(-1)) \ \qquad = 0.12000E+09}

\texttt{\ \qquad \qquad L\_n (m) \qquad \qquad \qquad \qquad \qquad \qquad\
\ \ = 243.0}

\texttt{\ \qquad \qquad L\_T (m) \qquad \qquad \qquad \qquad \qquad \qquad
\qquad\ = 0.073}

\texttt{\ \qquad \qquad d(Ln)/dx ( ) \qquad \qquad \qquad \qquad \qquad = 6.5%
}

\texttt{\ \qquad \qquad Vstar (m/sec) \qquad \qquad \qquad \qquad\ =
0.160E+01}

\texttt{\ Parameters in the equations (normalised)}

\texttt{\ \qquad \qquad ekn (rhos/Ln) \qquad \qquad \qquad \qquad \qquad
\qquad = 0.0000074}

\texttt{\ \qquad \qquad ekt (rhos/LT) \qquad \qquad \qquad \qquad \qquad
\qquad = 0.0246914}

\texttt{\ \qquad \qquad eknp (rhos**2*d/dx(1/Ln)) = 0.0000000}

\texttt{\ \qquad \qquad v* (v*/Omegai*rhos) \qquad \qquad \qquad = 0.741E-05}

\texttt{\ \qquad \qquad u (u/Omegai*rhos) \qquad \qquad \qquad \qquad =
0.100E-04}

\texttt{\ \qquad \qquad cs (cs/Omegai*rhos) \qquad \qquad \qquad = 0.100E+01}

\texttt{\ Parameters in the Flierl-Petviashvili equation}

\texttt{\ \qquad \qquad alpha \ \ \ \ \ \ \ \ \ \ \ \ \ = 0.259259}

\texttt{\ \qquad \qquad beta \ \ \ \ \ \ \ \ \ \ \ \ \ \ = 1.783265}

\texttt{\ \qquad \qquad s \ \ \ \ \ \ \ \ \ \ \ \ \ \ \ \ \ = 0.005000}

\texttt{\ \qquad \qquad g2 = 2536.4 }

\texttt{\ \qquad \qquad g3 = -2704.2 }

\texttt{\ \qquad \qquad Period omega \ \ \ \ \ \ = 0.265228 }

\texttt{\ \qquad \qquad Period omega\_prim \ = 0.257673}

\texttt{\ \qquad \qquad Orientation of the stationary flow pattern}

\texttt{\ \qquad \qquad \qquad \qquad \qquad acoef = 0.0000}

\texttt{\ \qquad \qquad \qquad \qquad \qquad bcoef = 0.0385}

\texttt{\ \qquad \qquad Width on x of the periodicity layer}

\texttt{\ \qquad \qquad \qquad \qquad \qquad deltax (in units of rhos) =
13.3685}

\texttt{\ Space domain :}

\texttt{\ \qquad \qquad y (poloidal) ymin = 0.0000 \ \ ymax = 100.0000}

\texttt{\ \qquad \qquad x (radial) \ \ xmin = -2.0000 \ xmax = 106.9479}

\texttt{\ The minimum of the solution PHIM = 0.727002E-01}

\texttt{\ the maximum of the solution PHIX = 0.107068}

\texttt{\ \qquad \qquad The difference of min/max amplitude, s1 =
0.343680E-01}

\texttt{\ \qquad \qquad The average amplitude, \ \ \ \ \ \ \ \ \ \ \ \ \ \
s0 = 0.898842E-01}

\texttt{\ \qquad \qquad eddy turn-over time tau (sec) = 0.433340E-04}

\texttt{\ \qquad \qquad Electric field on x, in V/m, minimum : -23237.9 }

\texttt{\ \qquad\ \ \ \ \ \ \ \ \ \ \ \ \ \ \ \ \ \ \ \ \ \ \ \ \ \ \ \ \ \
\ maximum : 23238.0 }

\texttt{\ \qquad \qquad Shear rate normalized at OMCI, min: -0.272078E-01}

\texttt{\ \ \ \ \ \ \ \ \ \ \ \ \ \ \ \ \ \ \ \ \ \ \ \ \ \ \ \ \ \ \ \ \ \
\ \ max: 0.136097E-01}

\begin{figure}[!htb]
\centerline{
\includegraphics[width=7cm]{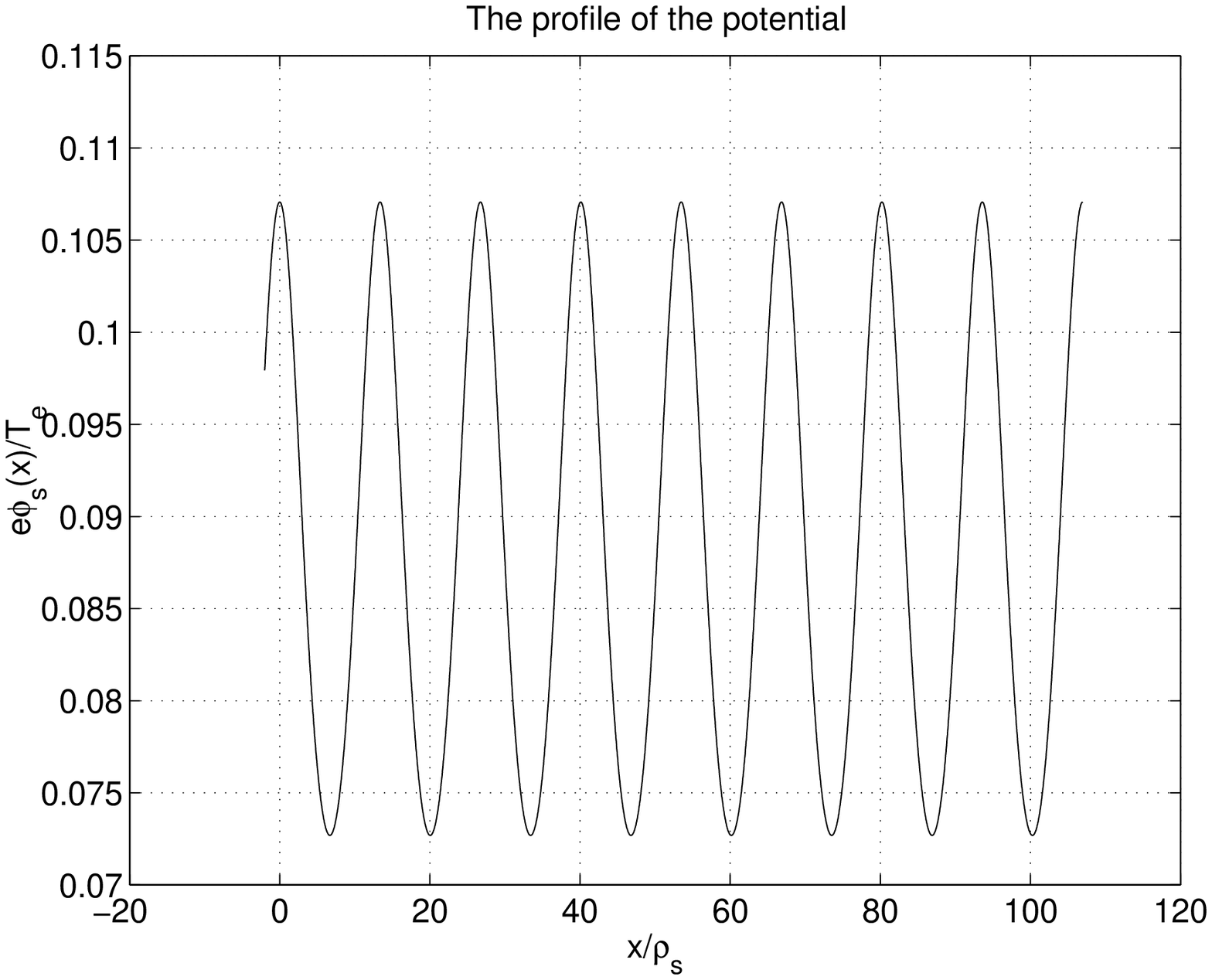}\hfill\includegraphics[width=7cm]{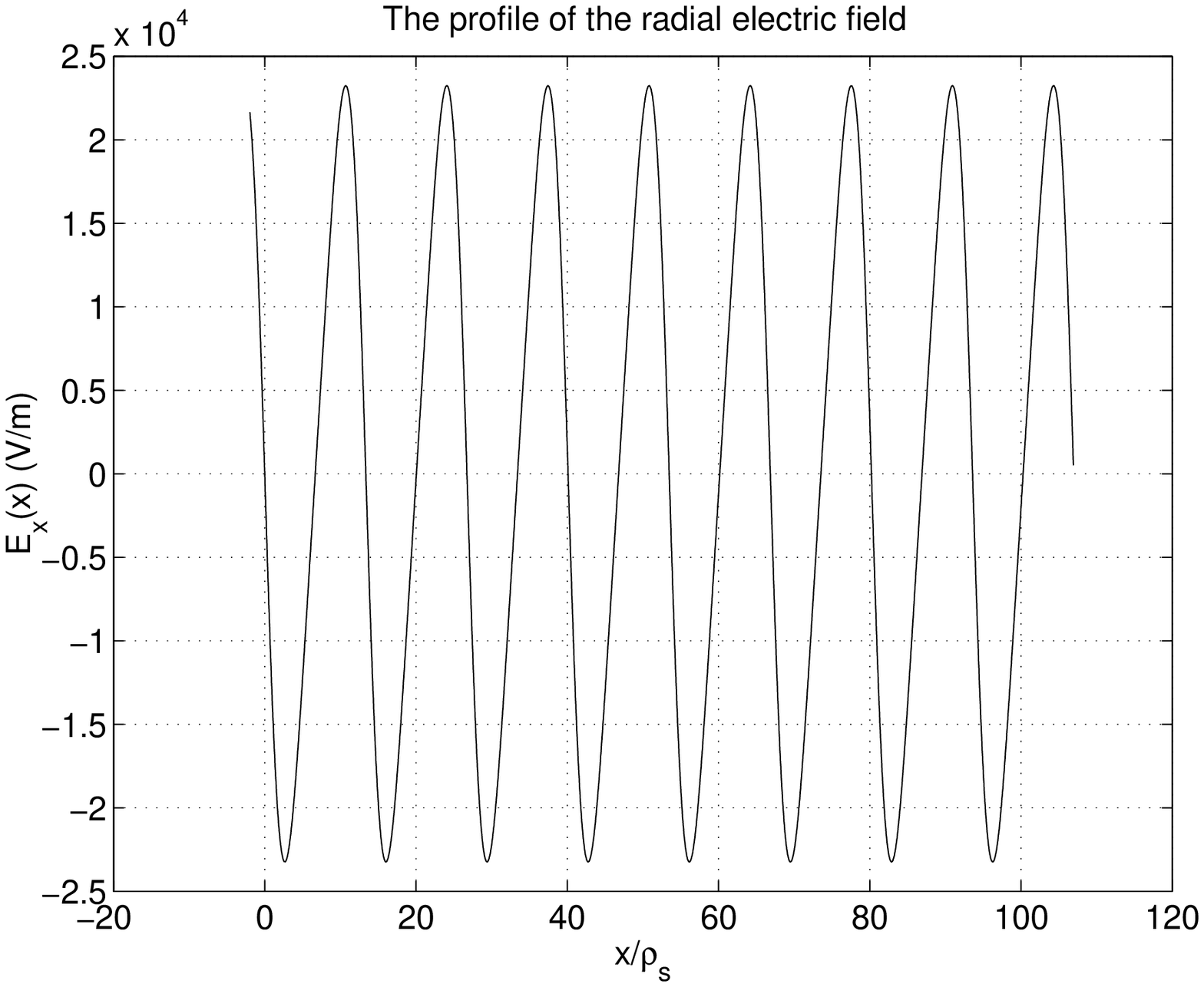}
}
\caption{The profile of the potential and of the radiale electric field}
\end{figure}

\smallskip 
\begin{figure}[!htb]
\centerline{
\includegraphics[width=7cm]{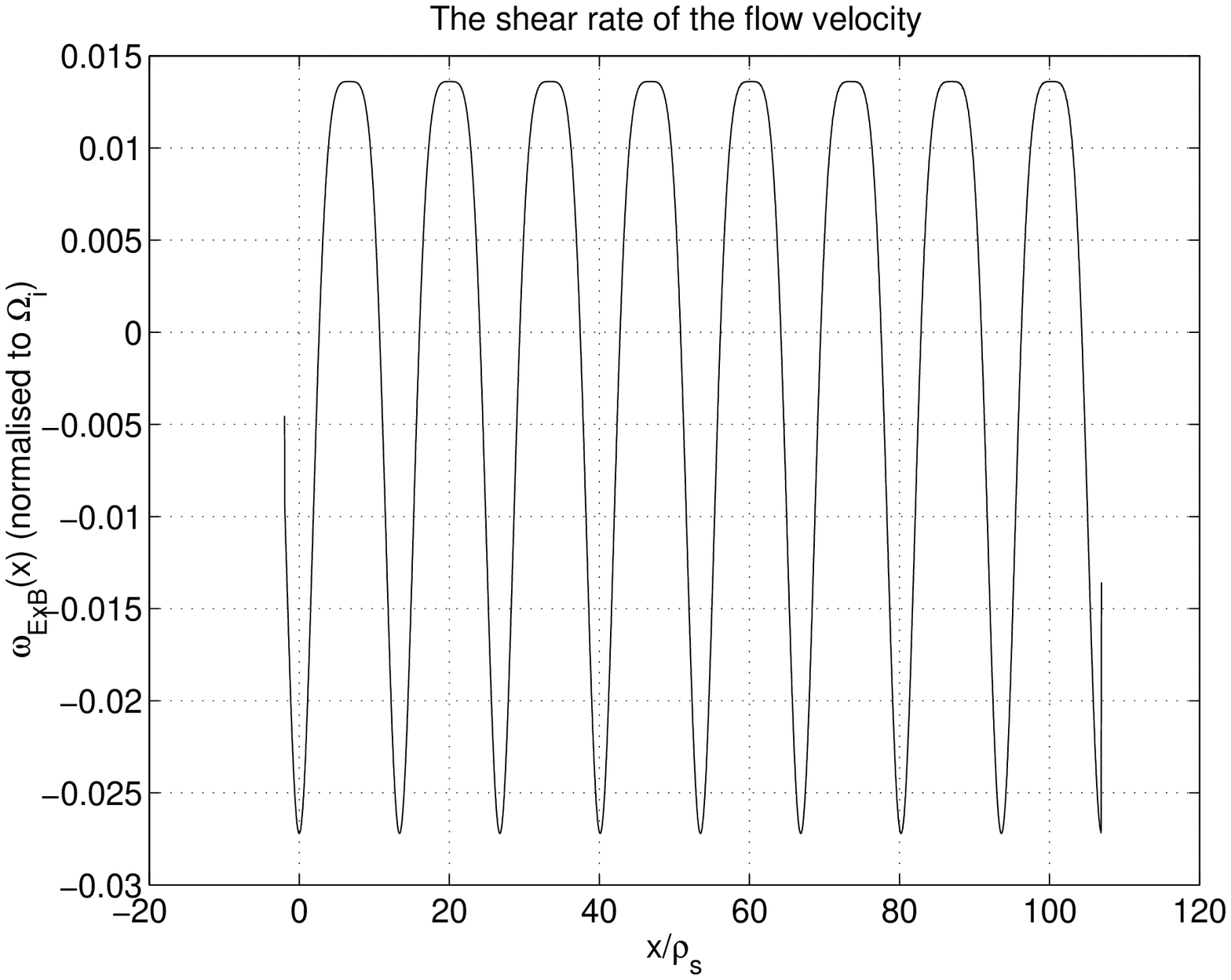}\hfill\includegraphics[width=7cm]{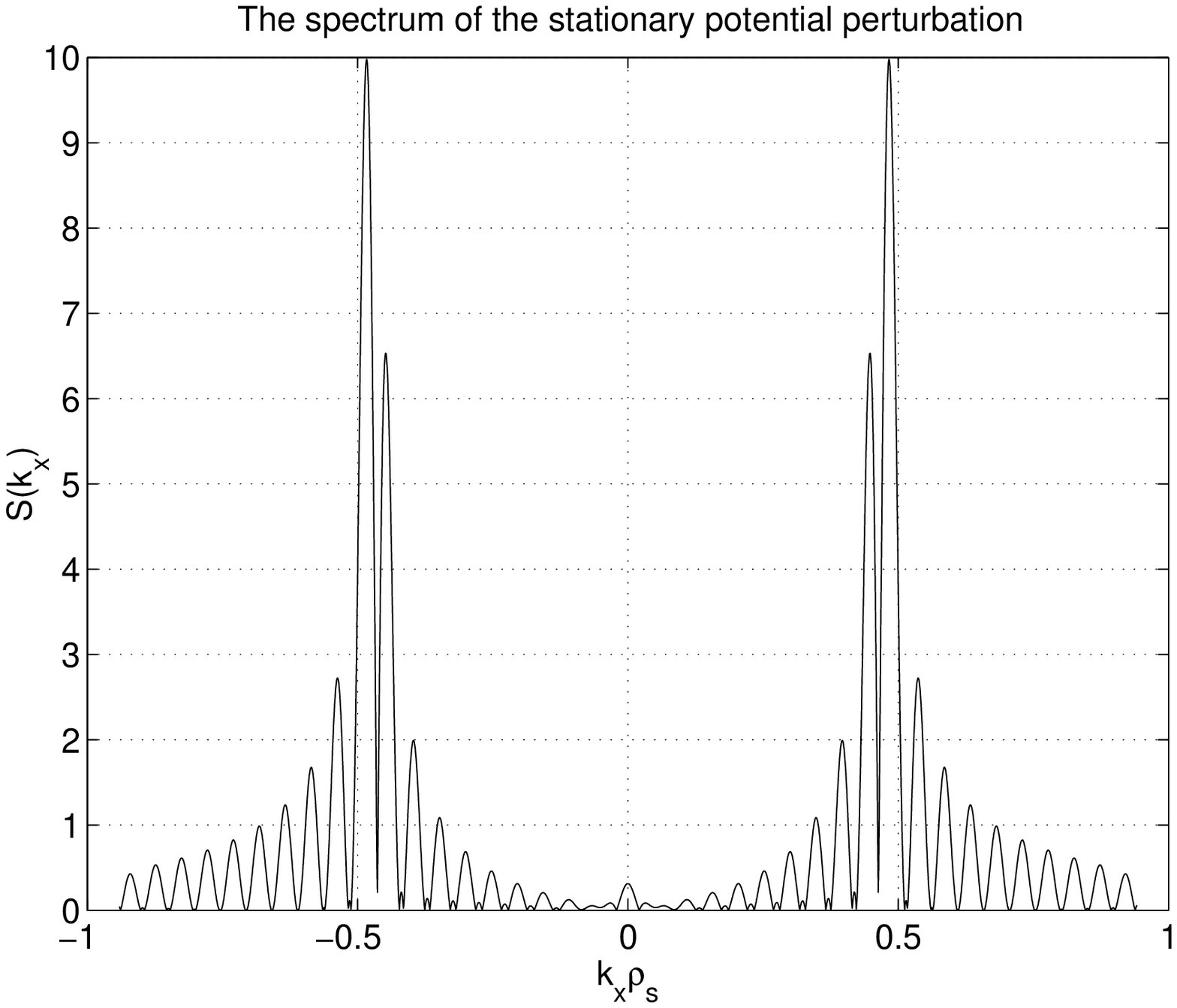}
}
\caption{The flow shear rate and the spectrum $S(k_x)$}
\end{figure}

\section{Linear stability of the stationary periodic solution}

\subsection{Linear dispersion relation}

We start from the equation derived before, with the presence of a
temperature gradient 
\begin{equation*}
\frac{\partial }{\partial t}\left( 1-\mathbf{\nabla }_{\perp }^{2}\right)
\phi =\frac{\partial }{\partial y}\left[ \left( -\mathbf{\nabla }_{\perp
}^{2}+4\eta ^{2}\right) \phi \right] -\phi \frac{\partial \phi }{\partial y}
\end{equation*}

To study the stability we will perturb the stationary solution by a small
function $\varepsilon \left( x,y,t\right) $%
\begin{equation*}
\phi \rightarrow \phi _{s}\left( x,y\right) +\varepsilon \left( x,y,t\right)
\end{equation*}
where $\phi _{s}\left( x,y\right) $ is the periodic flow solution (expressed
in terms of the Weierstrass function). We obtain 
\begin{equation*}
\frac{\partial }{\partial t}\left( 1-\mathbf{\Delta }\right) \varepsilon =%
\frac{\partial }{\partial y}\left( -\Delta \varepsilon +4\eta
^{2}\varepsilon -\phi _{s}\varepsilon -\frac{1}{2}\varepsilon ^{2}\right)
\end{equation*}
after taking into account the fact that $\phi _{s}$ is a solution of the
stationary equation.

This equation must be solved starting from an initial condition $\varepsilon
\left( x,y,t=0\right) $.

If we neglect the quadratic term we have the equation 
\begin{equation}
\frac{\partial }{\partial t}\left( 1-\mathbf{\Delta }\right) \varepsilon +%
\frac{\partial }{\partial y}\Delta \varepsilon -\left( 4\eta ^{2}-\phi
_{s}\right) \frac{\partial \varepsilon }{\partial y}+\varepsilon \frac{%
\partial \phi _{s}}{\partial y}=0  \label{eqz}
\end{equation}
If we consider the stationary periodic solution as being aligned along the
poloidal direction, $y$, then the function $\phi _{s}$ has no $y$ dependence
and the last term can be discarded 
\begin{equation}
\frac{\partial }{\partial t}\left( 1-\mathbf{\Delta }\right) \varepsilon +%
\frac{\partial }{\partial y}\Delta \varepsilon -4\eta ^{2}\frac{\partial
\varepsilon }{\partial y}+\phi _{s}\frac{\partial \varepsilon }{\partial y}=0
\label{eqz1}
\end{equation}

In $\left( k_{y},\omega \right) $ space 
\begin{equation*}
\varepsilon \left( x,y,t\right) =\frac{1}{\left( 2\pi \right) ^{2}}%
\int_{-\infty }^{\infty }dk_{y}d\omega \exp \left( -i\omega t+ik_{y}y\right) 
\widetilde{\varepsilon }\left( x,k_{y},\omega \right)
\end{equation*}
the equation is 
\begin{equation}
\left( 1+k_{y}^{2}-\frac{\partial ^{2}}{\partial x^{2}}\right) \left(
-i\omega \right) \widetilde{\varepsilon }+ik_{y}\left( -k_{y}^{2}+\frac{%
\partial ^{2}}{\partial x^{2}}\right) \widetilde{\varepsilon }-ik_{y}\left(
4\eta ^{2}-\phi _{s}\right) \widetilde{\varepsilon }=0  \label{omky}
\end{equation}
Some rearrangement yields the equation 
\begin{equation*}
\frac{\partial ^{2}\widetilde{\varepsilon }}{\partial x^{2}}+\frac{ik_{y}}{%
i\omega +ik_{y}}\left( -\frac{\omega }{k_{y}}-\omega k_{y}-k_{y}^{2}-4\eta
^{2}+\phi _{s}\right) \widetilde{\varepsilon }=0
\end{equation*}
Let us introduce new notations 
\begin{equation*}
q_{1}\equiv \frac{1}{1+\frac{\omega }{k_{y}}}
\end{equation*}
\begin{equation*}
a_{1}\equiv -\frac{\omega }{k_{y}}-\omega k_{y}-k_{y}^{2}-4\eta ^{2}
\end{equation*}
and the form of the equation is 
\begin{equation*}
\frac{\partial ^{2}\widetilde{\varepsilon }}{\partial x^{2}}+\left(
q_{1}a_{1}+q_{1}\phi _{s}\right) \widetilde{\varepsilon }=0
\end{equation*}

The stationary periodic solution is expressed as 
\begin{equation*}
\phi _{s}\left( x,y\right) =\phi _{0}+s\wp \left( x\right)
\end{equation*}
where $\phi _{0}$ is 
\begin{equation*}
\phi _{0}=\frac{3\alpha }{2\beta }
\end{equation*}
and $s$ is a parameter related to the amplitude of the potential
perturbation. The equation becomes 
\begin{equation}
\frac{\partial ^{2}\widetilde{\varepsilon }}{\partial x^{2}}+\left[
q_{1}a_{1}+q_{1}\phi _{0}+q_{1}s\wp \left( x\right) \right] \widetilde{%
\varepsilon }=0  \label{schrp}
\end{equation}
This is a Calogero-Moser problem but for a single particle and the analysis
of this Schrodinger-type problem is, to our knowledge, not available (see 
\emph{arxiv.org/hep-th/9903002}). In order to obtain an estimation for
physical cases we will retain the periodic character of the Weierstrass
function by the replacement 
\begin{equation*}
\wp \left( x\right) \rightarrow s_{0}+s_{1}\cos \left( px\right)
\end{equation*}
where 
\begin{equation*}
p\equiv \left( \delta x\right) ^{-1}
\end{equation*}
\begin{equation*}
s_{0}\equiv \frac{1}{2}\left[ \wp \left( x\right) |_{\max }+\wp \left(
x\right) |_{\min }\right]
\end{equation*}
\begin{equation*}
s_{1}\equiv \wp \left( x\right) |_{\max }-\wp \left( x\right) |_{\min }
\end{equation*}
The parameter $p$ is the inverse of the periodicity length accross the layer
in the $x$ direction. The latter is obtained as $\delta x$ from the
imaginary-axis semi-period of the Weierstrass function and the physical
parameters. The parameters $s_{0}$ and $s_{1}$ represent respectively the
average amplitude of the function $\wp \left( x\right) $ and respectively
the variation of its amplitude on a period.

This approximation transform the equation (\ref{schrp}) in the Mathieu
equation 
\begin{equation*}
\frac{\partial ^{2}\widetilde{\varepsilon }}{\partial x^{2}}+\left[
a_{2}+b_{2}\cos \left( px\right) \right] \widetilde{\varepsilon }=0
\end{equation*}
with the notations 
\begin{equation*}
a_{2}\equiv q_{1}a_{1}+q_{1}\phi _{0}+q_{1}ss_{0}
\end{equation*}
\begin{equation*}
b_{2}\equiv q_{1}ss_{1}
\end{equation*}
We then change the space variable 
\begin{equation*}
x\rightarrow x^{\prime }\equiv \frac{px}{2}=\frac{x}{2\left( \delta x\right) 
}
\end{equation*}
and arrive at the final form of the equation 
\begin{equation}
\frac{\partial ^{2}\widetilde{\varepsilon }}{\partial x^{2}}+\left[ a-2q\cos
\left( 2x\right) \right] \widetilde{\varepsilon }=0  \label{eqmathieu}
\end{equation}
with the new notations 
\begin{equation*}
a\equiv 4\left( \delta x\right) ^{2}q_{1}\left( a_{1}+\phi _{0}+ss_{0}\right)
\end{equation*}
\begin{equation*}
q\equiv -2\left( \delta x\right) ^{2}q_{1}ss_{1}
\end{equation*}
In detailed form 
\begin{equation}
a=4\left( \delta x\right) ^{2}\frac{1}{1+\omega /k_{y}}\left( -\frac{\omega 
}{k_{y}}-\omega k_{y}-k_{y}^{2}-4\eta ^{2}+\phi _{0}+ss_{0}\right)
\label{eqa}
\end{equation}
\begin{equation}
q=-2\left( \delta x\right) ^{2}ss_{1}\frac{1}{1+\omega /k_{y}}  \label{eqq}
\end{equation}

In order $\widetilde{\varepsilon }\left( x\right) $ to be a periodic
function on $x$ it must be fulfilled one of a discrete set of dispersion
relations. We take the simplest approximative form 
\begin{equation}
a_{0}\left( q\right) \approx -\frac{q^{2}}{2}  \label{rdisM}
\end{equation}
This leads to the following relation between the parameters 
\begin{equation}
-\frac{\omega }{k_{y}}-\omega k_{y}-k_{y}^{2}+P=-Q\frac{1}{1+\omega /k_{y}}
\label{rd1}
\end{equation}
where 
\begin{eqnarray}
P &\equiv &-4\eta ^{2}+\phi _{0}+ss_{0}  \label{PQ} \\
Q &\equiv &\frac{1}{2}\left( \delta x\right) ^{2}s^{2}s_{1}^{2}  \notag
\end{eqnarray}
Eliminating the numitor and grouping the terms 
\begin{equation*}
-\left( k_{y}+\omega \right) ^{2}+\frac{\omega }{k_{y}}\left( P-1\right) -%
\frac{\omega ^{2}}{k_{y}^{2}}+P+Q=0
\end{equation*}
We note 
\begin{equation*}
w\equiv \frac{\omega }{k_{y}}
\end{equation*}
and have the equation 
\begin{equation*}
-k_{y}^{2}\left( 1+w\right) ^{2}+w\left( P-1\right) -w^{2}+P+Q=0
\end{equation*}
or 
\begin{equation*}
w^{2}+2t_{1}w+t_{2}=0
\end{equation*}
\begin{eqnarray*}
2t_{1} &\equiv &\frac{-2k_{y}^{2}+P-1}{-k_{y}^{2}-1} \\
t_{2} &\equiv &\frac{-k_{y}^{2}+P+Q}{-k_{y}^{2}-1}
\end{eqnarray*}

\subsection{Approximations based on evaluations of terms}

Since the spatial variables are measured in $\rho _{s}$ we can already say
that the wavelength of the $y$-perturbations will be of the order of several
units 
\begin{equation*}
\lambda _{\perp }\geq \left( \text{few units}\right) \rho _{s}
\end{equation*}
We will consider that this is a sufficient indication to take 
\begin{equation*}
\frac{1}{k_{y}^{2}}\ll 1
\end{equation*}
Then we can simplify the expressions 
\begin{eqnarray*}
2t_{1} &=&\left( 2-\frac{P-1}{k_{y}^{2}}\right) \left( 1-\frac{1}{k_{y}^{2}}%
\right) \\
&\approx &2-\frac{P-1}{k_{y}^{2}}
\end{eqnarray*}
\begin{eqnarray*}
t_{2} &=&\left( 1-\frac{P+Q}{k_{y}^{2}}\right) \left( 1-\frac{1}{k_{y}^{2}}%
\right) \\
&\approx &1-\frac{P+Q+1}{k_{y}^{2}}
\end{eqnarray*}
and the discriminant of the second order equation is 
\begin{eqnarray*}
\Delta &\equiv &\left( 1-\frac{P-1}{2k_{y}^{2}}\right) ^{2}-\left( 1-\frac{%
P+Q+1}{k_{y}^{2}}\right) \\
&\approx &\frac{Q}{k_{y}^{2}}
\end{eqnarray*}
The solutions are 
\begin{eqnarray*}
w &=&-\left( 1-\frac{P-1}{2k_{y}^{2}}\right) \pm \sqrt{\frac{Q}{k_{y}^{2}}}
\\
&=&-1+\frac{P+1\pm 2k_{y}Q^{1/2}}{2k_{y}^{2}}
\end{eqnarray*}
The dispersion relation is at this moment 
\begin{equation}
\frac{\omega }{k_{y}}=-1+\frac{P+1}{k_{y}^{2}}\pm \frac{Q^{1/2}}{k_{y}}
\label{rd2}
\end{equation}
Let 
\begin{equation*}
u\equiv \frac{1}{k_{y}}
\end{equation*}
The equation is 
\begin{equation*}
\omega u=-1+\frac{P+1}{2}u^{2}\pm Q^{1/2}u
\end{equation*}
or 
\begin{equation*}
u^{2}+2u\left[ \frac{-\omega \pm Q^{1/2}}{\left( P+1\right) /2}\right] -%
\frac{2}{P+1}=0
\end{equation*}
This leads to 
\begin{equation}
\frac{1}{k_{y}}=\frac{\omega \pm Q^{1/2}}{P+1}\left\{ 1\mp \left[ 1+\frac{%
2\left( P+1\right) }{\left( -\omega +Q^{1/2}\right) ^{2}}\right]
^{1/2}\right\}  \label{rd3}
\end{equation}
We can evaluate 
\begin{eqnarray*}
P &=&-4\eta ^{2}+\phi _{0}+ss_{0} \\
&=&-4\eta ^{2}+\frac{1}{2}\phi _{0}+\frac{1}{2}s\wp \left( x\right) |_{\max
}+\frac{1}{2}\phi _{0}+\frac{1}{2}s\wp \left( x\right) |_{\min } \\
&=&-4\eta ^{2}+\frac{1}{2}\phi _{s}|_{\min }+\frac{1}{2}\phi _{s}|_{\max } \\
&=&-4\eta ^{2}+\overline{\phi }
\end{eqnarray*}
where we can have an estimation for the average level of fluctuation 
\begin{equation*}
\overline{\phi }\sim \frac{\widetilde{n}}{n_{0}}
\end{equation*}
In the numerical experiments the field that clearly evolved to breaking into
discrete monopolar vortices had an amplitude of the order of 
\begin{equation*}
\overline{\phi }\sim 0.05...0.1
\end{equation*}
We chose a case with 
\begin{equation*}
4\eta ^{2}\equiv 1-\frac{v_{\ast }}{u}\sim 1-0.85=0.15
\end{equation*}
The two terms are of similar magnitude so we can expect $P$ to be positive
or negative. In this case 
\begin{equation*}
P=-0.15+0.1=-0.05
\end{equation*}
\begin{eqnarray*}
Q &=&\left( \delta x\right) ^{2}s^{2}s_{1}^{2} \\
&=&\left( \delta x\right) ^{2}\left( \phi _{s}|_{\max }-\phi _{s}|_{\min
}\right) ^{2} \\
&\equiv &\left( \delta x\right) ^{2}\left( \delta \phi _{s}\right) ^{2}
\end{eqnarray*}
We take data from one of the characteristic runs 
\begin{eqnarray*}
\delta \phi _{s} &=&\phi _{s}|_{\max }-\phi _{s}|_{\min } \\
&\sim &0.088-0.019 \\
&=&0.0\,\allowbreak 69
\end{eqnarray*}
and 
\begin{equation*}
\delta x\sim 14.17
\end{equation*}
then 
\begin{eqnarray*}
Q &\sim &\left( 14.17\right) ^{2}\times \left( 0.069\right) ^{2} \\
&=&0.956
\end{eqnarray*}
Since 
\begin{eqnarray*}
P+1 &\sim &1 \\
Q &\lesssim &1
\end{eqnarray*}
We note that $Q$ can be smaller for narrower strips of the stationary
solution.

We can do the following approximations 
\begin{eqnarray*}
\left[ 1+\frac{2\left( P+1\right) }{Q}\right] ^{1/2} &\sim &\left( 1+2\frac{1%
}{Q}\right) ^{1/2} \\
&\sim &\frac{\sqrt{2}}{\sqrt{Q}}
\end{eqnarray*}
The dispersion relation is 
\begin{equation*}
\frac{1}{k_{y}}\simeq \frac{\omega \mp \sqrt{Q}}{P+1}\left( 1\mp \frac{\sqrt{%
2}}{\sqrt{Q}}\right)
\end{equation*}
The frequency $\omega $ is in the range given by the inverse of the typical
time of the change of the stationary solution, after being perturbed. The
units which connect an interval of time in physical and respectively
adimensional form are 
\begin{equation*}
\Delta t|_{phys}=\frac{\rho _{s}}{u|_{phys}}\Delta t
\end{equation*}
with 
\begin{eqnarray*}
\rho _{s} &\sim &0.001\;\left( m\right) \\
u|_{phys} &\sim &v_{\ast }\sim 0.3\times 10^{3}\;\left( m/s\right)
\end{eqnarray*}
\begin{eqnarray*}
\Delta t|_{phys} &\sim &\frac{0.001}{300}\Delta t \\
&\sim &3\times 10^{-6}\Delta t
\end{eqnarray*}
In the numerical simulations the first changes took place after about 
\begin{equation*}
\Delta t|_{phys}\sim 4000\Omega _{i}^{-1}\left( s\right) \sim \frac{4000}{%
0.335\times 10^{9}}=11940\times 10^{-9}=11.94\times 10^{-6}
\end{equation*}
which means 
\begin{equation*}
\Delta t\sim \frac{\Delta t|_{phys}}{3\times 10^{-6}}=\frac{11.94\times
10^{-6}}{3\times 10^{-6}}\sim 4
\end{equation*}
Then the frequency associated with the change is 
\begin{equation*}
\omega \sim 1/4=0.25
\end{equation*}
If we can neglect $\omega $ compared with $\sqrt{Q}$ we have 
\begin{eqnarray*}
\frac{1}{k_{y}} &\simeq &\frac{\omega \mp \sqrt{Q}}{P+1}\left( 1\mp \frac{%
\sqrt{2}}{\sqrt{Q}}\right) \\
&\simeq &\mp \sqrt{Q}\left( 1\mp \frac{\sqrt{2}}{\sqrt{Q}}\right) \\
&=&\sqrt{2}\mp \sqrt{Q}
\end{eqnarray*}
or, chosing the $-$ sign 
\begin{equation*}
\frac{1}{k_{y}}\simeq \sqrt{2}-\left( \delta x\right) \left( \delta \phi
\right)
\end{equation*}
which has the character of an \emph{eigenvalue} selection, relating the
spatial extension of the vortices into which the periodic solution decays to
the amplitude and radial width of periodicity.

Taking from above 
\begin{equation*}
\left( \delta x\right) \left( \delta \phi \right) \sim 1
\end{equation*}
we obtain 
\begin{eqnarray*}
\lambda _{\perp } &\sim &2\pi \times 0.41 \\
&\sim &2.6
\end{eqnarray*}

Our numerical simulations only allow to identify the process of generation
of a chain of vortices, but we cannot pursue too far the simulations due to
accumulation of numerical errors. It apparently supports a dimension of
about $6$.

The stability depends also on the type of perturbation, not only simply of
the amplitude. A monopolar solution is very stable. The HM term renders the
equation fragile to decay into lattice of vortices.

\section{Numerical studies}

\subsection{Equation with the scalar nonlinearity (time dependent
Flierl-Petviashvili equation)}

\subsubsection{Introduction}

The equations (\ref{phitrans}), (\ref{ytrans}), (\ref{ttrans}), (\ref{utrans}%
), help us to define the space and time domain of physical parameters for
the numerical simulation.

For the numerical treatment we start form the time-dependent equation 
\begin{equation*}
\frac{\partial }{\partial t}\left( 1-\mathbf{\nabla }_{\perp }^{2}\right)
\phi =\frac{\partial }{\partial y}\left( -\mathbf{\nabla }_{\perp
}^{2}+4\eta ^{2}\right) \phi -\phi \frac{\partial \phi }{\partial y}
\end{equation*}
and notice that it may be preferable to use two fonctions 
\begin{equation*}
\psi \equiv \left( 1-\mathbf{\nabla }_{\perp }^{2}\right) \phi
\end{equation*}
and $\phi $. The equation is transformed into a system of two differential
equations for two unknown functions $\psi $ and $\phi $ 
\begin{eqnarray*}
-\mathbf{\nabla }_{\perp }^{2}\phi &=&-\phi +\psi \\
\frac{\partial \psi }{\partial t}-\frac{\partial \psi }{\partial y}
&=&\left( -1+4\eta ^{2}\right) \frac{\partial \phi }{\partial y}-\phi \frac{%
\partial \phi }{\partial y}
\end{eqnarray*}
or 
\begin{eqnarray*}
-\mathbf{\nabla }_{\perp }^{2}\phi +\phi &=&\psi \\
\frac{\partial \psi }{\partial t} &=&\frac{\partial \psi }{\partial y}%
+\left( -1+4\eta ^{2}\right) \frac{\partial \phi }{\partial y}-\phi \frac{%
\partial \phi }{\partial y}
\end{eqnarray*}

This version is one of the form taht will be used in the numerical
integration of the equation.

\subsubsection{Initial monopolar perturbation}

We start by initializing the potential with a monopolar vortex localised in
one of the layers of periodicity. The amplitude of the potential at this run
was $\phi _{s}|_{\max }=0.03$ and of the perturbation is $\phi
_{pert}|_{\max }=5\times 10^{-3}$ (normalised as $\left| e\right| \phi
/T_{e} $). The evolution is done with a time step of $10\Omega _{i}$ .

We note a very stable evolution of the total flow, consisting of a slow ($%
v_{\ast }/u\lesssim 1$) displacement of the monopole along the layer. In the
first part of the evolution, the profile of the monopolar vortex is adapted
to the local speed of the flow, but afterwards the evolution is amazingly
stable. The duration of this period of stability is greater than $12\times
10^{3}\Omega _{i}$ the limit where the run was stopped.

This conclusion must be useful in relation with the long stability of the
Red Spot of the Jupiter atmosphere. There, the configuration is precisely
the same as in this run: a localised vortex in a structure of zonal flow, as
shown by the observations of Voyager 1 and 2 (see Busse \cite{Busse}). 
\begin{figure}[tbh]
\centerline{
\includegraphics[width=7cm]{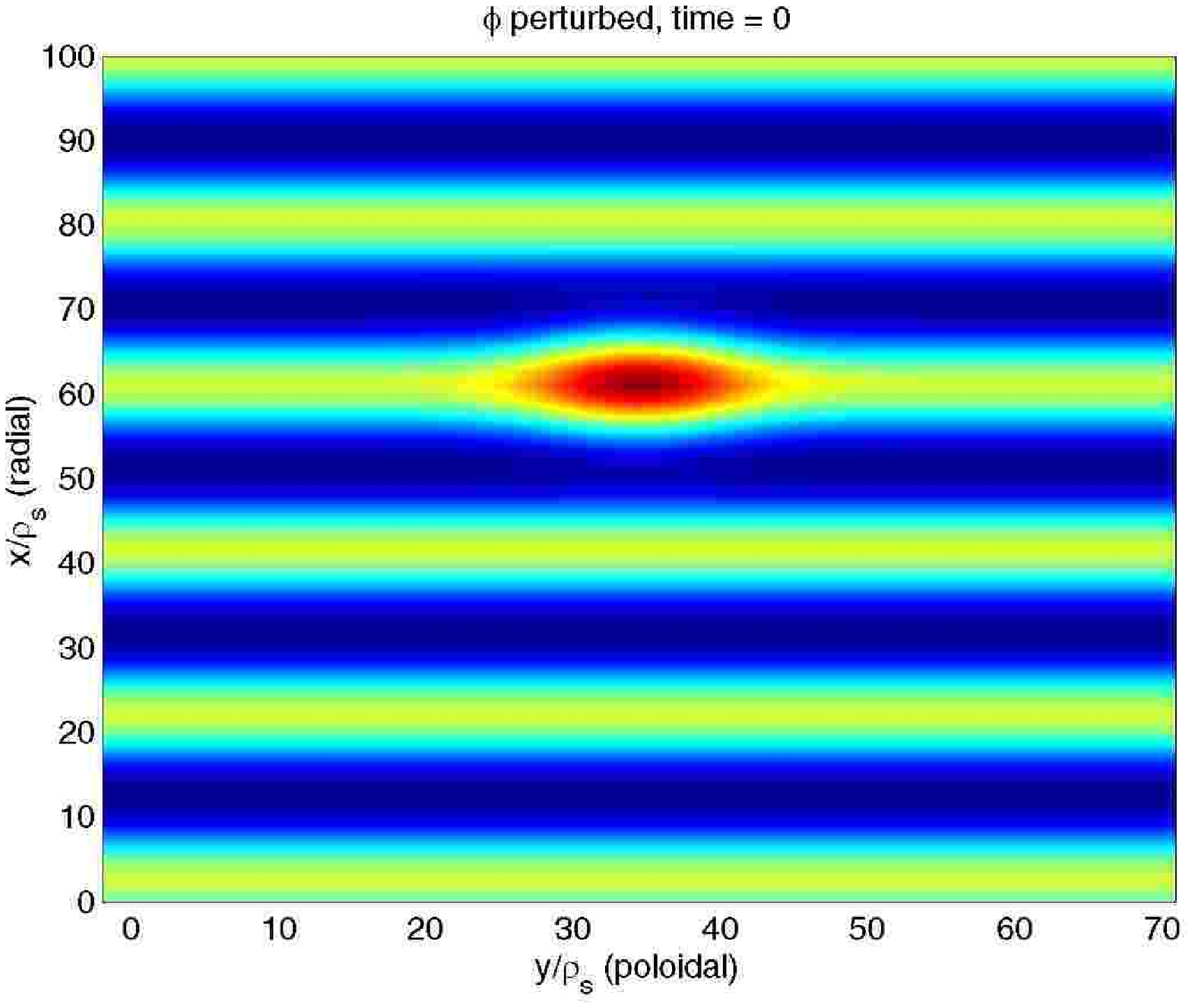}\hfill\includegraphics[width=7cm]{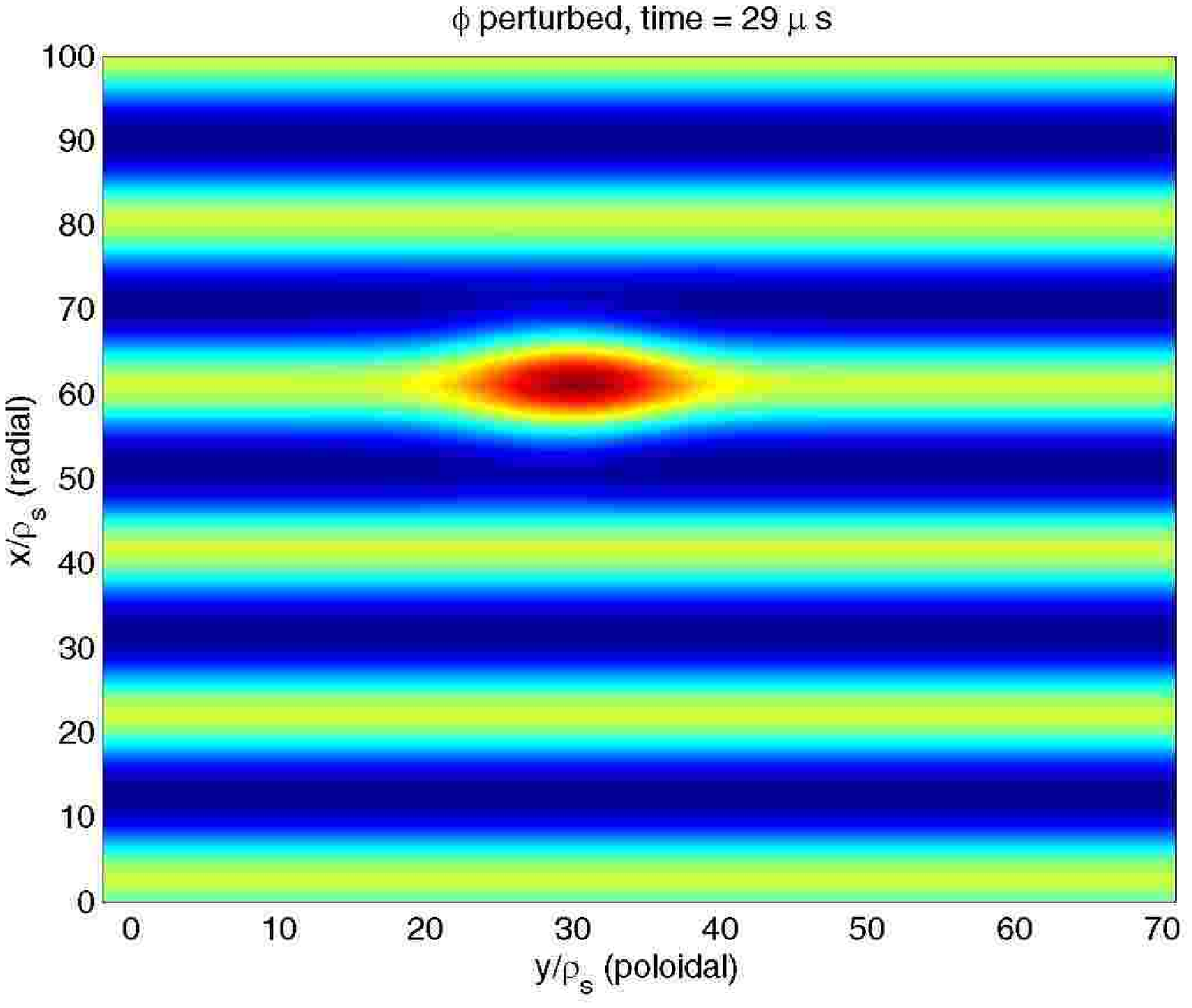}
}
\caption{The initial perturbation is monopolar and located symmetrically in
a flow layer. The initial ($t=0$) and final ($t=29\protect\mu s$) potentials
are shown.}
\end{figure}

\subsubsection{Initial dipolar perturbation}

The initial field is $\phi _{s}$ perturbed by a dipolar vortex placed with
maxima approximately on a line of maximum. The amplitude of $\phi _{s}$ is $%
0.03$ and that of the dipole is $3\times 10^{-3}$. The evolution in this
case is very stable for all the interval of the run ($29\,\mu s$,
corresponding to $12\times 10^{3}\Omega _{i}$). We notice the weak change of
the shape of the dipole due to the interaction with the background flow, in
the first part of the run.

\begin{figure}[!htb]
\centerline{
\includegraphics[width=7cm]{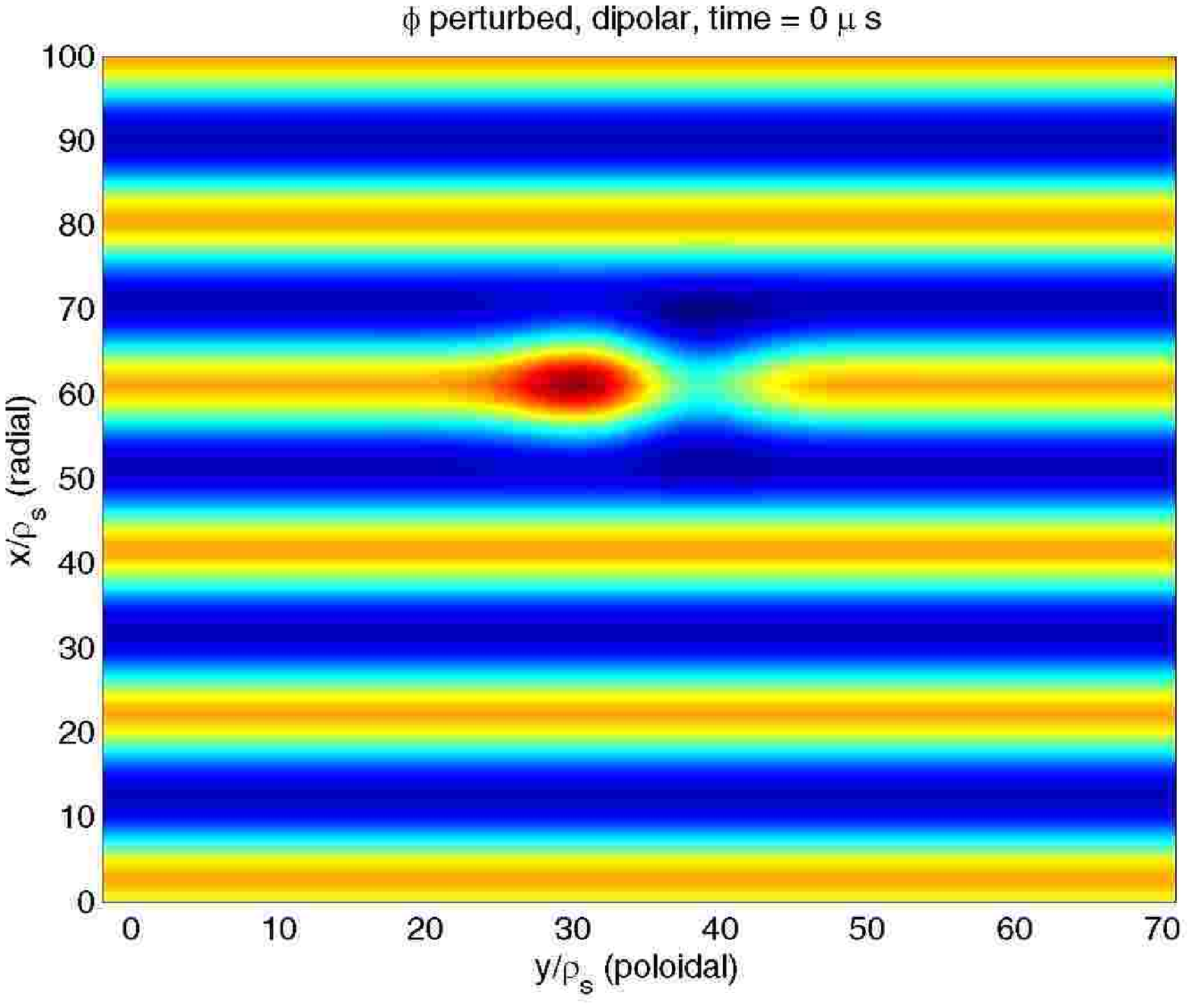}\hfill\includegraphics[width=7cm]{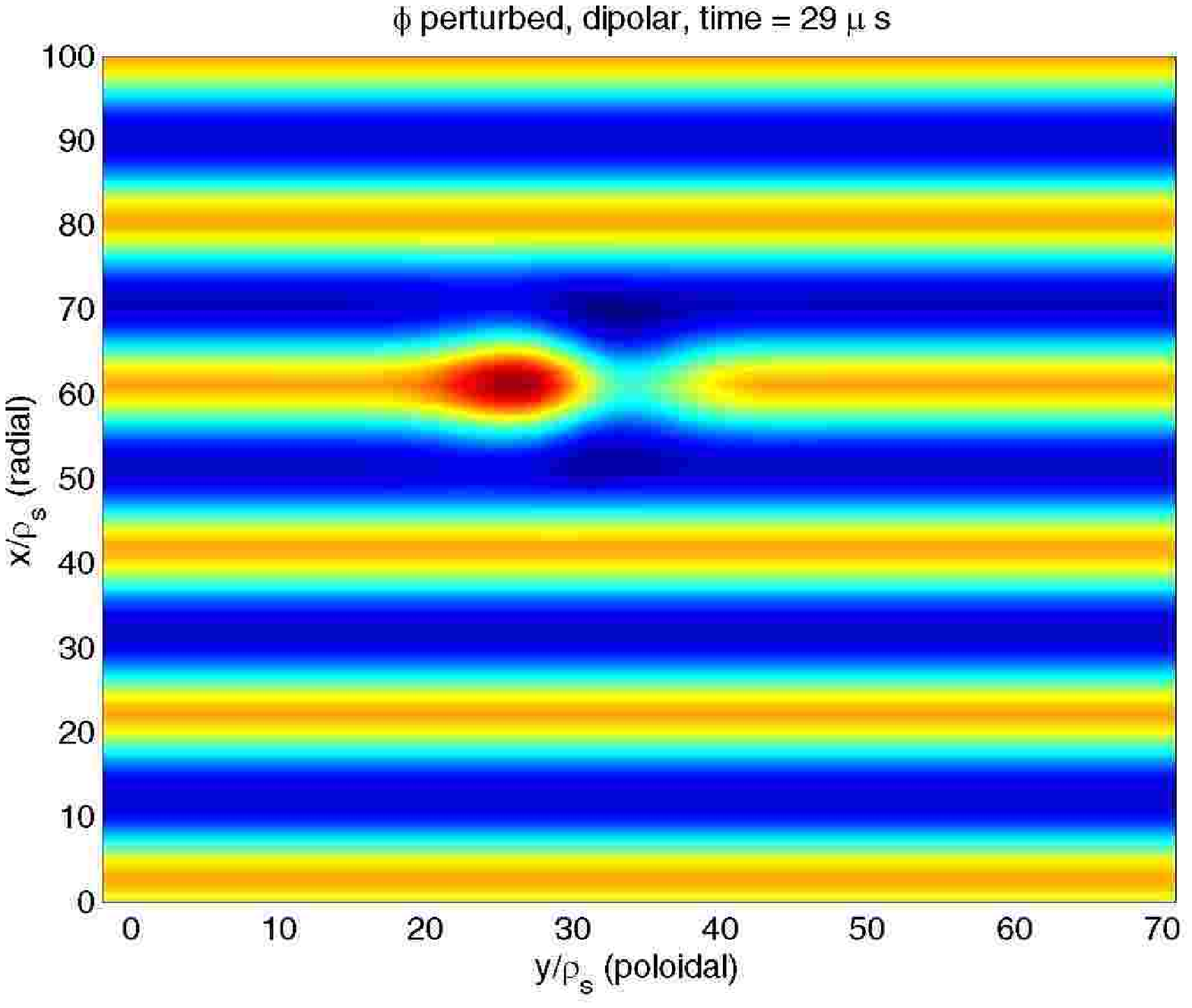}
}
\caption{A dipolar perturbation is initialised. The initial ($t=0$) and
final ($t=29 \protect\mu s$) potentials are shown.}
\end{figure}

In another run, the initial field $\phi _{s}$ of amplitude $0.03$ has been
perturbed by a dipolar vortex with the same localisation as before but with
amplitude comparable to the background field, $0.03$. The run evolves
without destruction of the configuration.

\begin{figure}[!htb]
\centerline{
\includegraphics[width=7cm]{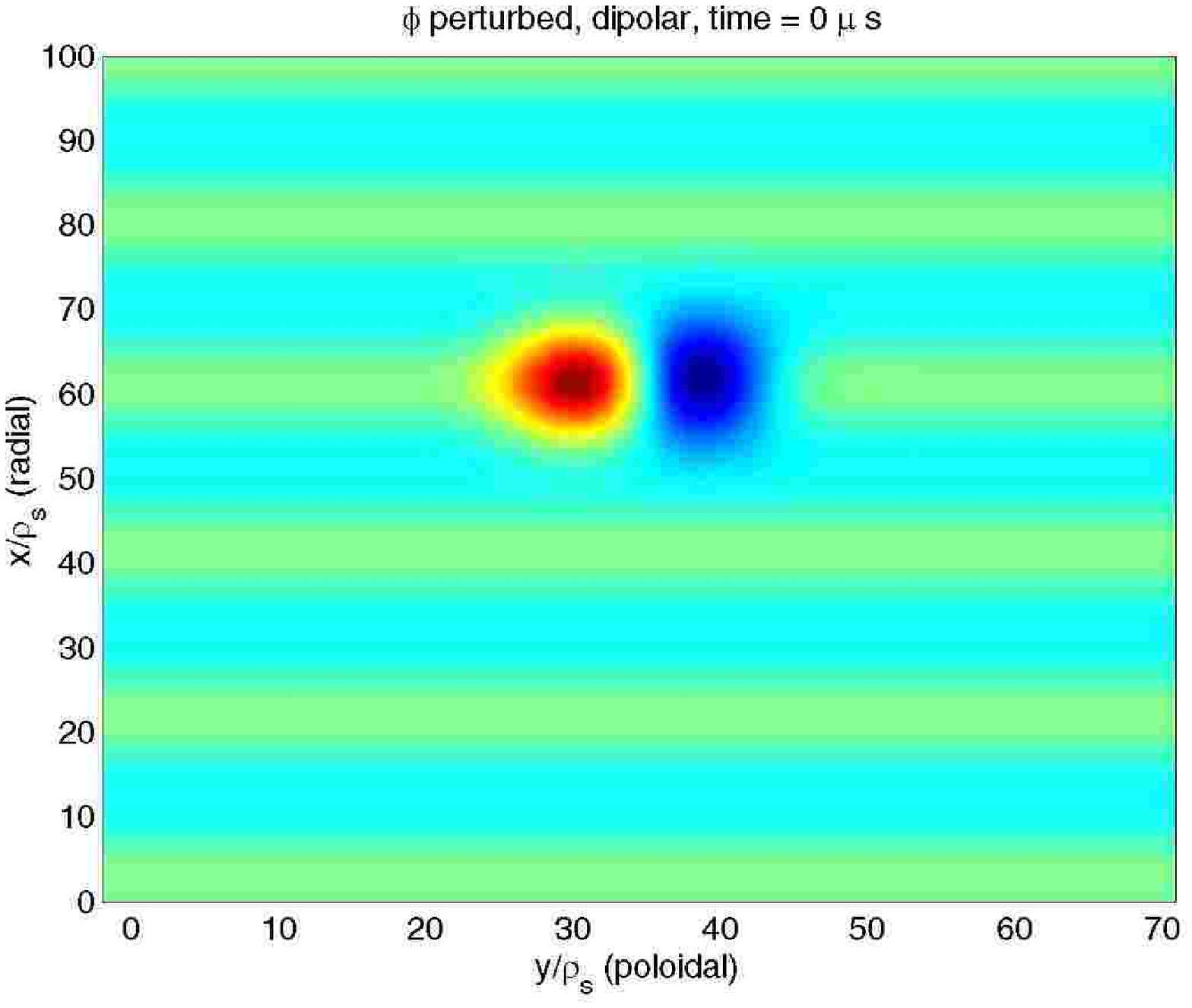}\hfill\includegraphics[width=7cm]{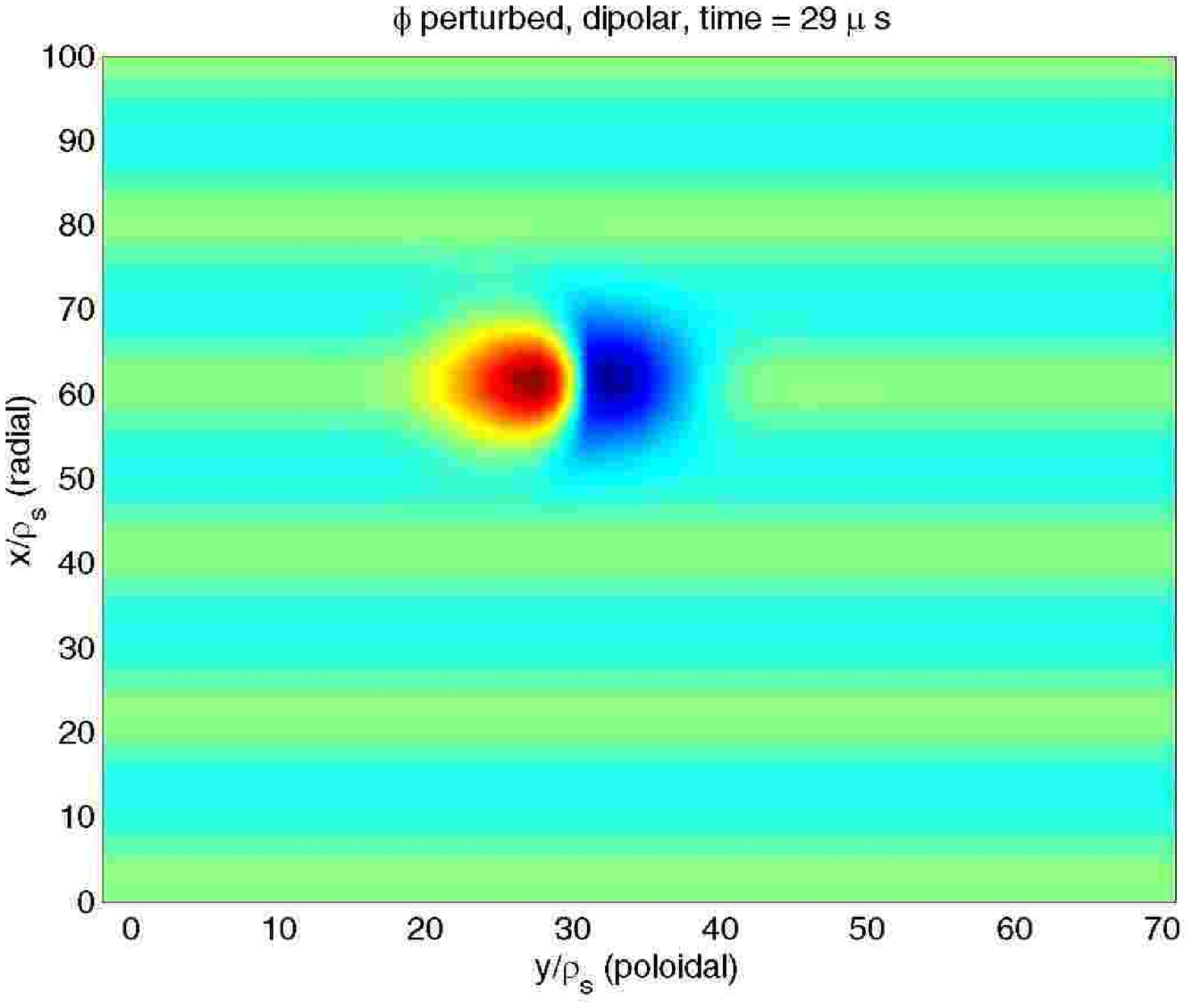}
}
\caption{An initial stationary flow of higher amplitude is perturbed with a
dipolar vortex of comparable amplitude. Only the scalar nonlinearity is
retained.}
\end{figure}

From the numerical experiments with the dipolar perturbations it seems that
low amplitude potential $\phi _{s}$ has weak interaction with the
perturbation and major changes of the flow may probably arise later than
several thousends $\Omega _{i}$.

\subsection{Equation with scalar and polarisation drift nonlinearities (time
dependent Flierl-Petviashvili equation with Hasegawa-Mima term)}

\subsubsection{Initial monopolar perturbation}

In a series of run we take a higher initial background flow $\phi _{s}$,
with amplitude of $0.06$ and perturb it with a monopolar structure strongly
elongated in the radial direction. The amplitude of the parturbation is very
high, $0.12$. We note that the evolution leads first to strong changes of
the perturbation structure which is rotated and deformed. But in the same
time the flow itself tends to be broken into isolated vortices. These runs
require higher numerical precision so that only short period of time can be
afforded at this momant. The evolution was stopped at $6\,\mu s$.

\begin{figure}[!htb]
\centerline{
\includegraphics[width=7cm]{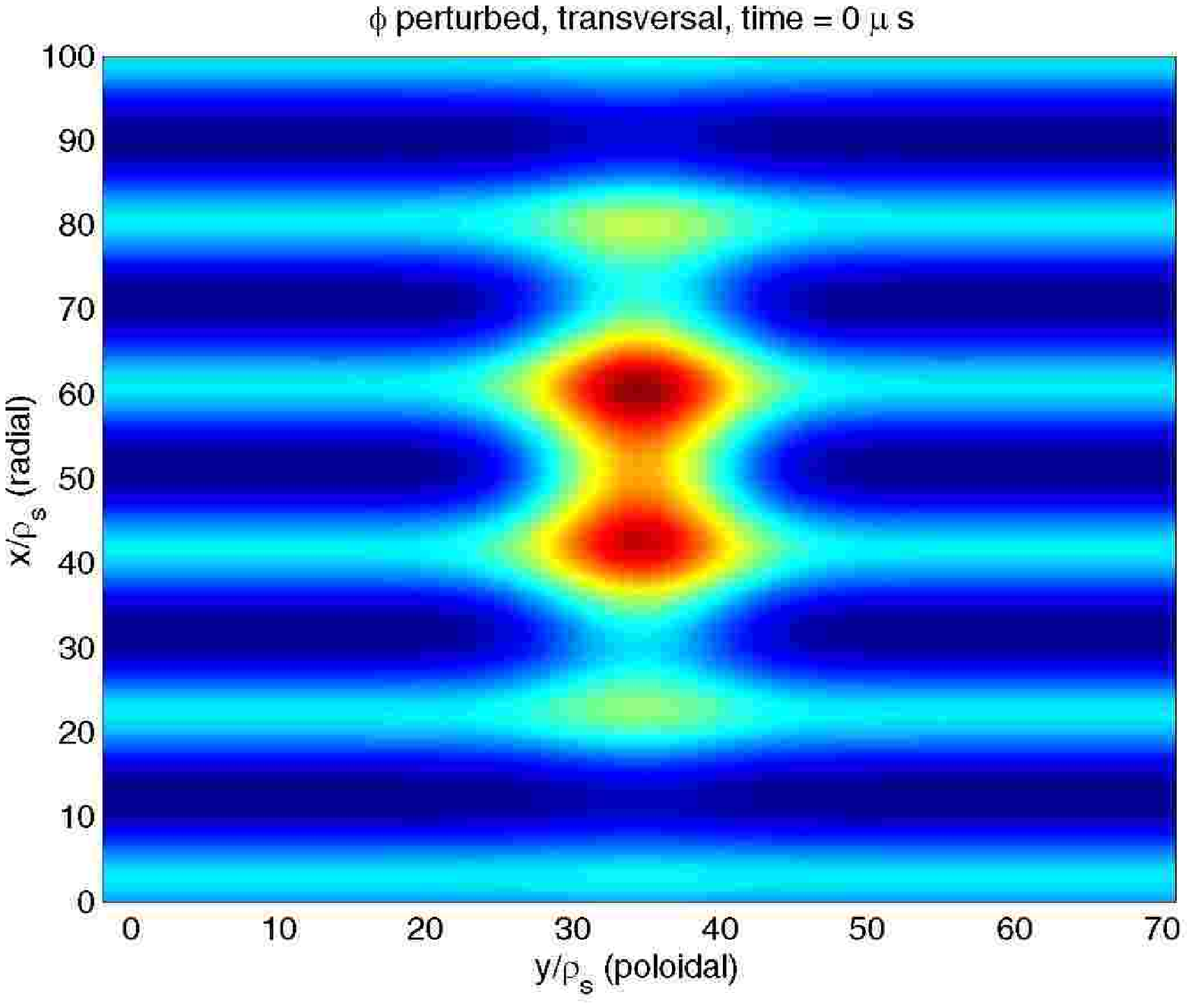}\hfill\includegraphics[width=7cm]{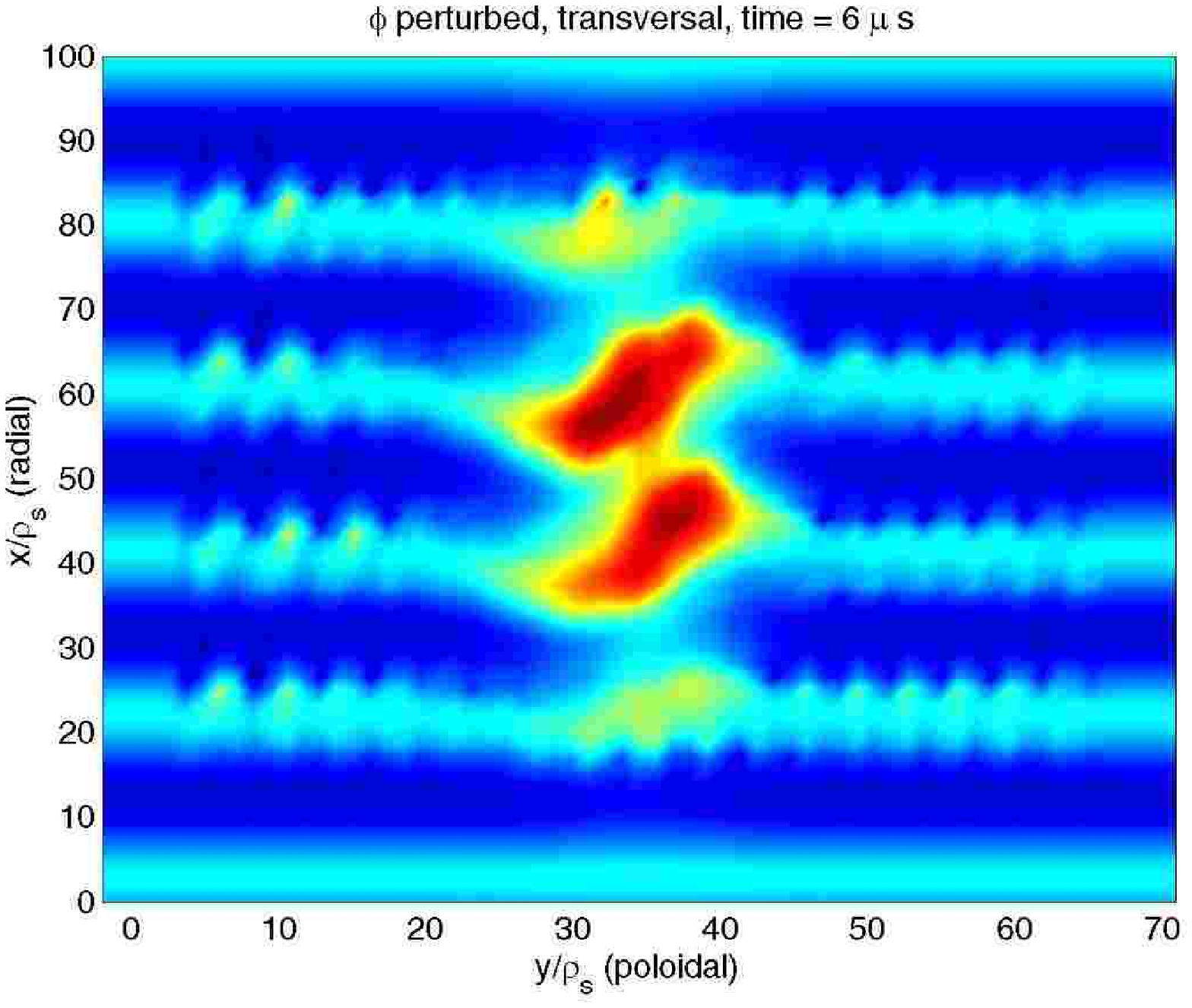}
}
\caption{A run with both scalar and polarisation nonlinearities retained.
The flow is initialised with a monopolar form perturbing the staionary
periodic solution.}
\end{figure}

The late stages of these runs are useful for the ilustration of the breaking
of the background flow into a lattice of vortices.

\begin{figure}[!htb]
\centerline{
\includegraphics[width=7cm]{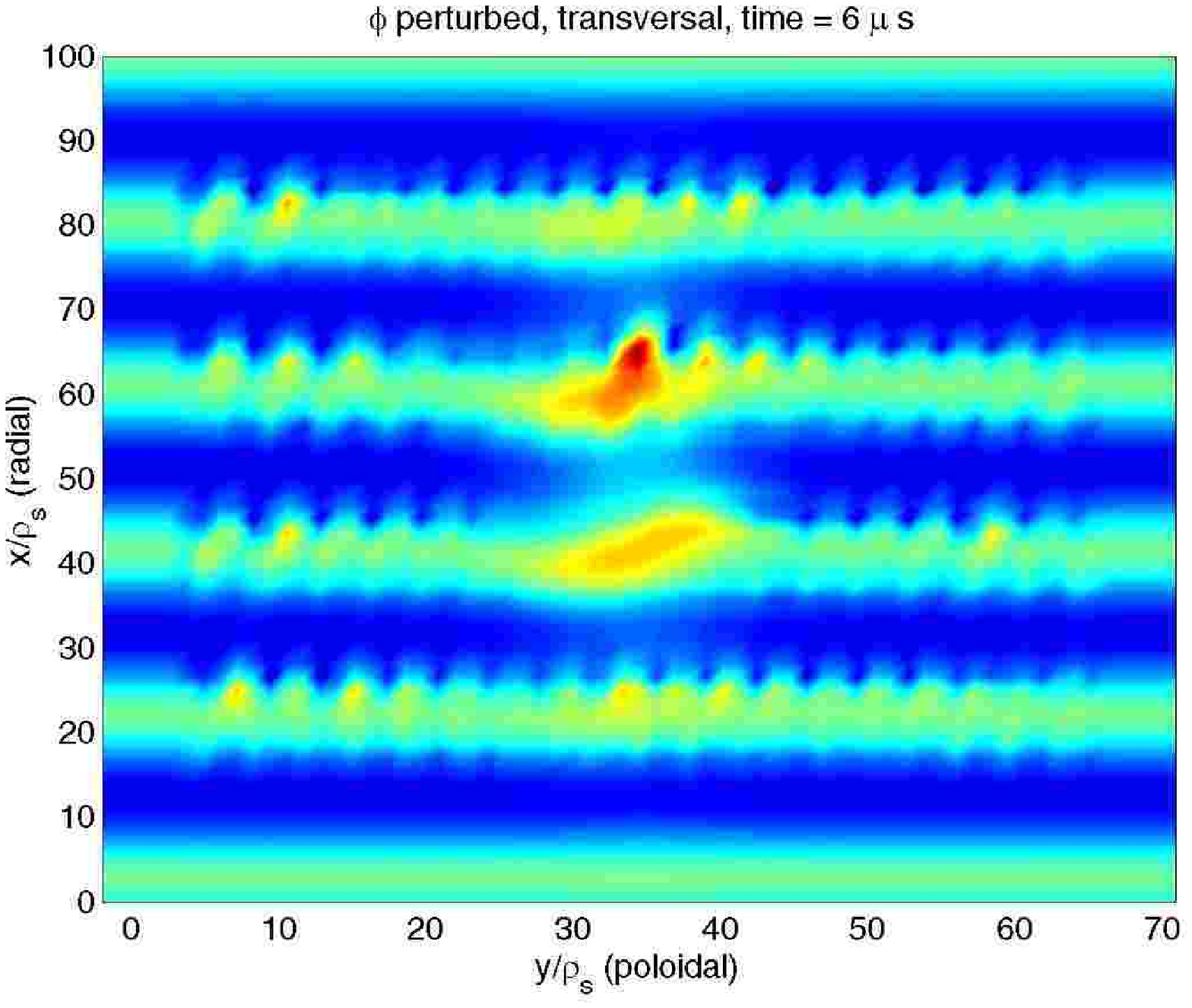}\hfill\includegraphics[width=7cm]{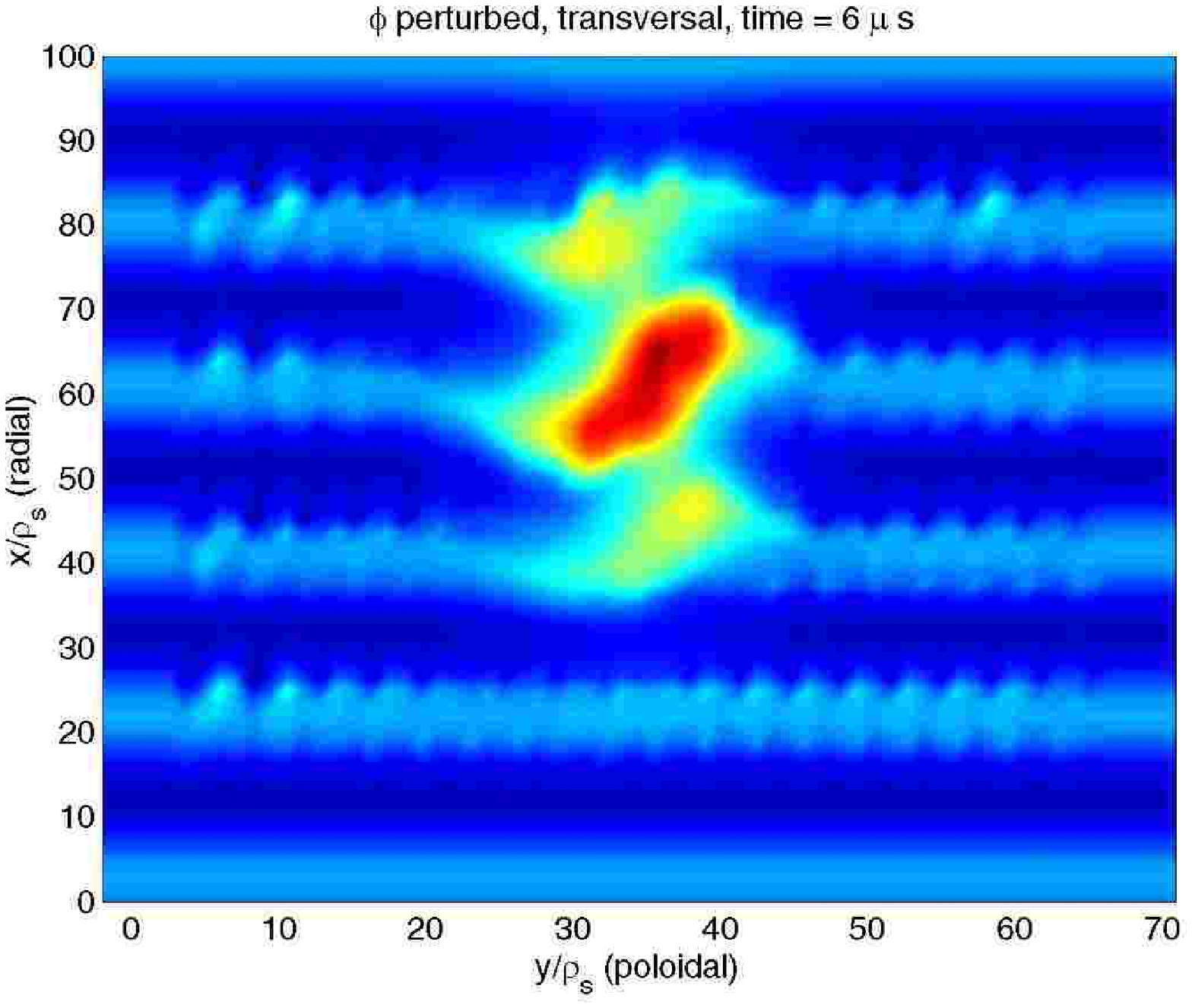}
}
\caption{The late stage of a run with monopolar pertubation.The
breaking of the flow into vortices is apparent.}
\end{figure}

A particular importance takes the examination of the evolution of the flow
when the initial perturbation is monopolar and placed approximately in a
layer of periodicity. This should be directly compared with the previous
runs with the pure scalar nonlinearity.

We notice that the stability is much stranger for this type of geometrically
``compatible'' perturbation. Neither the background flow nor the vortex is
disturbed for more than $25\,\mu s$. The monopole is only advected by the
flow. At late stage, the monopole is deformed and destroyed, while the flow
does not show the previously observed tendency to break into a lattice of
vortices. However, the region of the very high amplitude where the monopole
existed tends to break into vortices.

\begin{figure}[!htb]
\centerline{
\includegraphics[width=7cm]{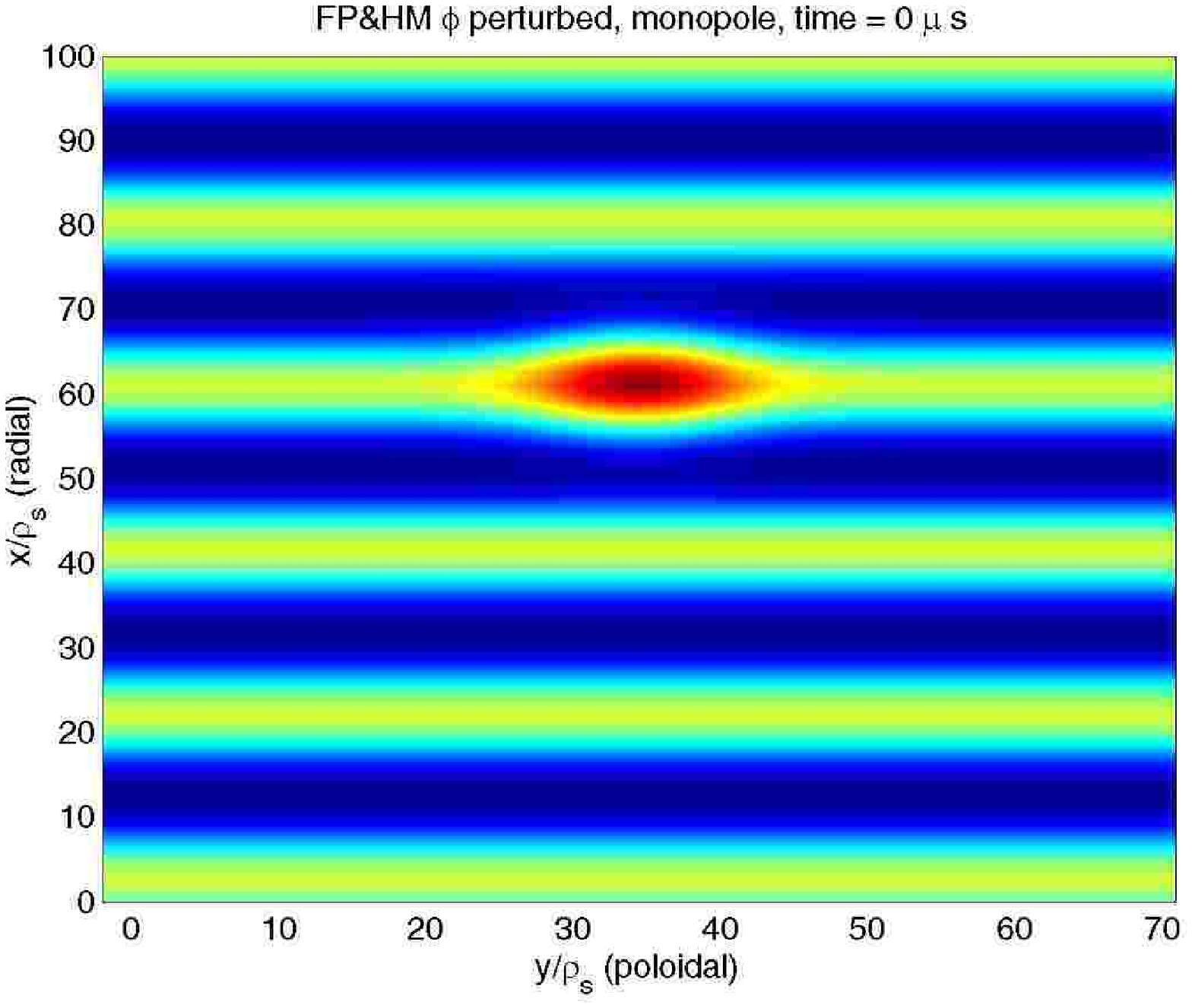}\hfill\includegraphics[width=7cm]{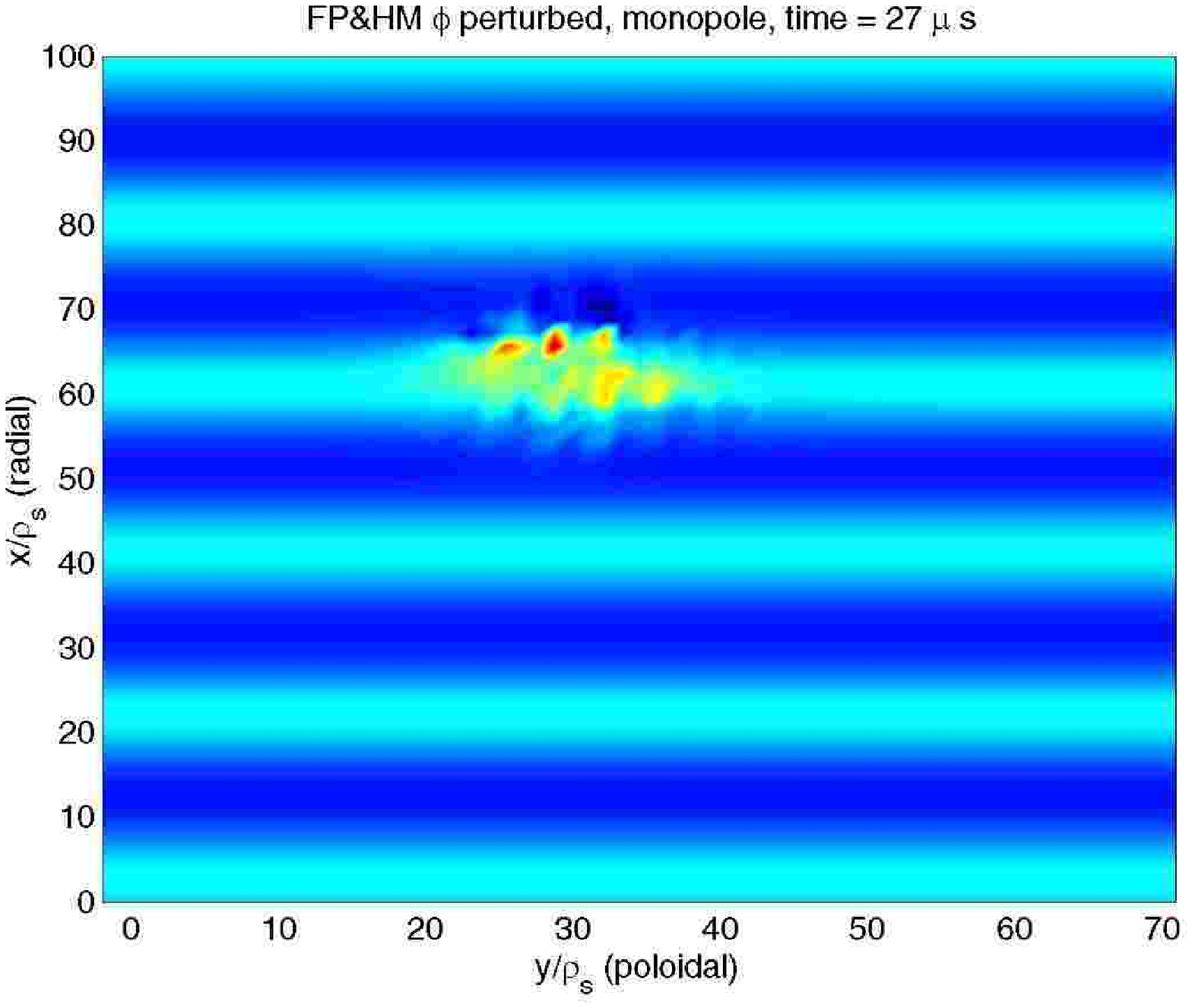}
}
\caption{The late stage of a run with monopolar pertubation
of high amplitude.The
breaking into vortices affects first the strongest perturbation.}
\end{figure}

This shows that the Hasegawa-Mima nonlinearity has the property of inducing
a tandency of the flow to break up, under strong perturbation, into a
lattice of monopolar vortices.

\subsubsection{Initial dipolar perturbation}

A dipolar perturbation shows essentially the same evolution as for monopolar
case. The profile is not changed for rather long period, $20\mu s$ but the
later stages consist of a clear decay into a set of distinct formations with
monopolar structure. We cannot follow too far the simulation due to
accumulation of errors, but it is clear that space scale of the new vortical
structures is close to those found in preceeding cases. This strengthens the
idea that the polarisation drift nonlinearity is the cause of this evolution
toward lattices of vortices.

\begin{figure}[!htb]
\centerline{
\includegraphics[width=7cm]{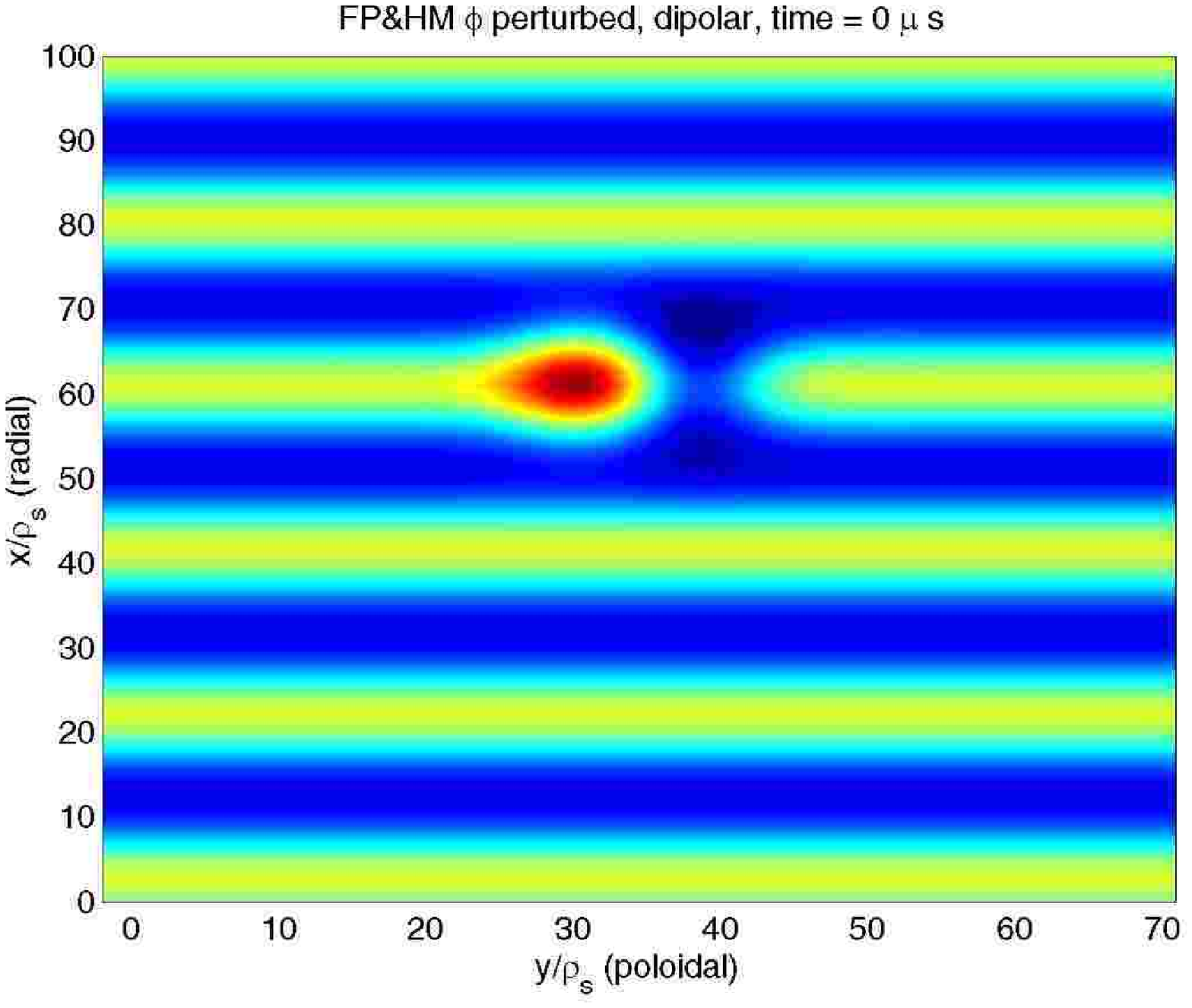}\hfill\includegraphics[width=7cm]{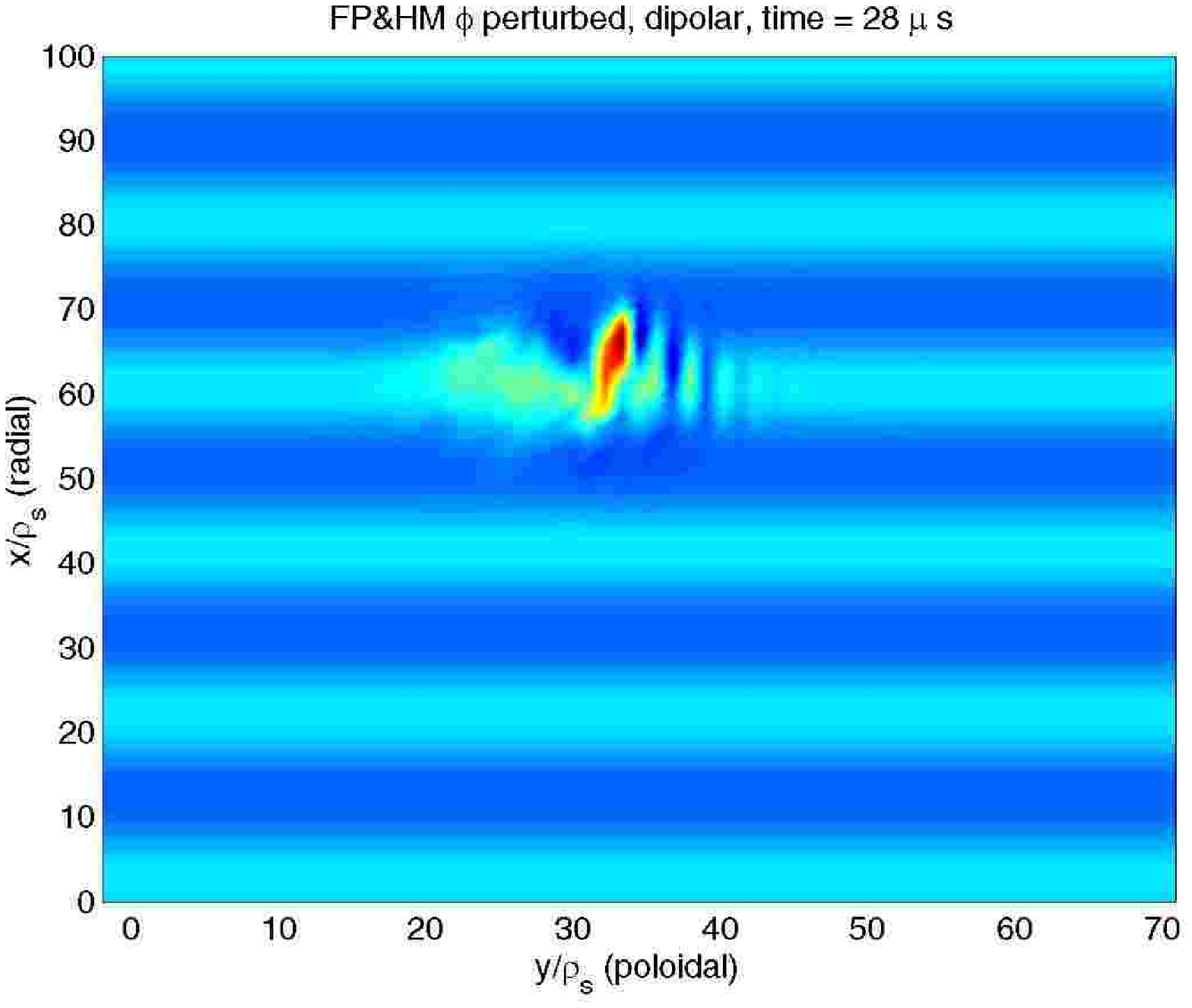}
}
\caption{Initial dipolar perturbation}
\end{figure}

\subsubsection{Initial oscillatory perturbation}

In order to examine the processes of destruction of the flow and decay into
isolated vortices, we initialize with an oscillatory perturbation with a
period much larger than that which seems to be chosen by the system as
extension of the final monopolar vortices. The flow clearly evolves to the
lattice of vortices.

\begin{figure}[!htb]
\centerline{
\includegraphics[width=7cm]{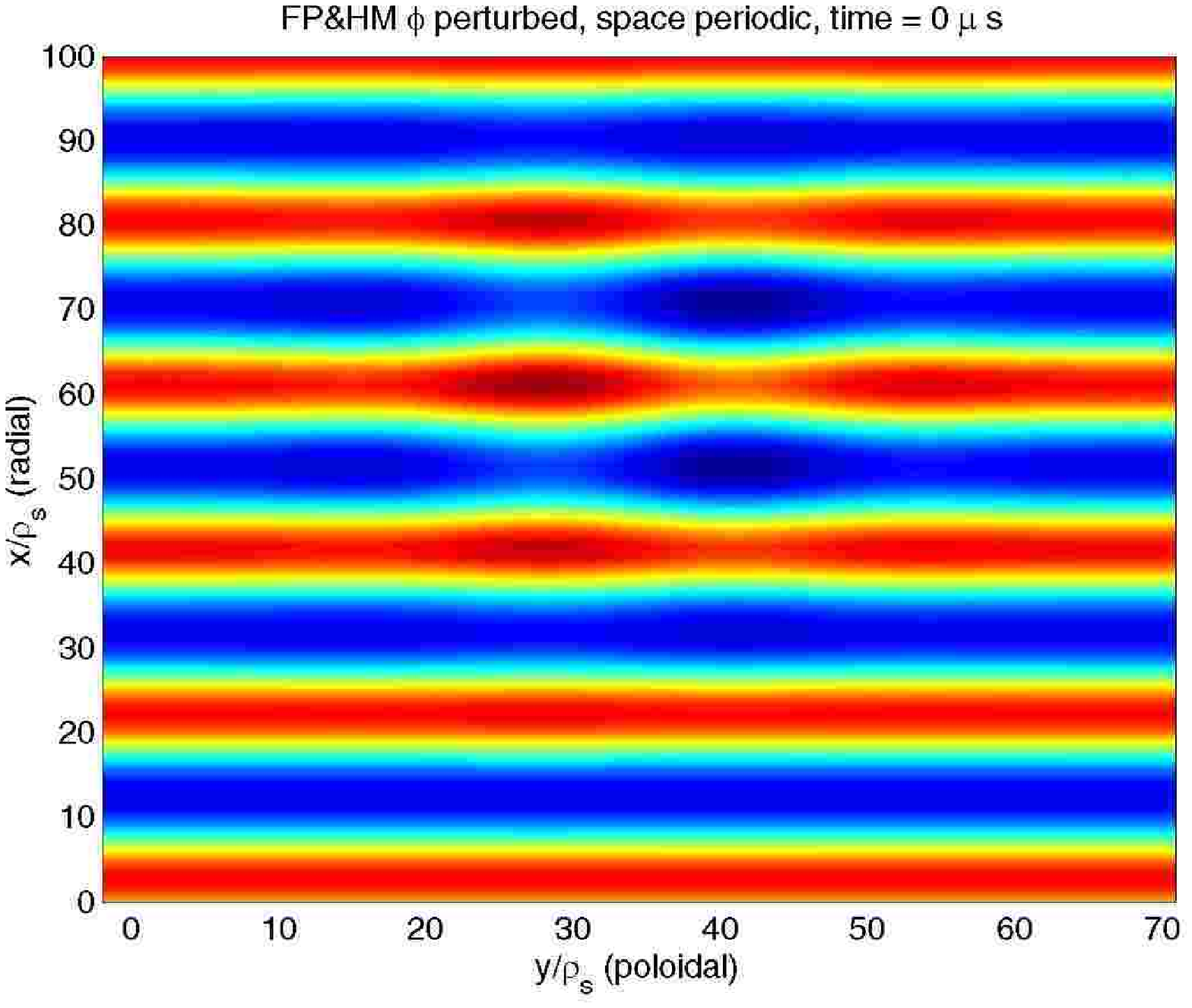}\hfill\includegraphics[width=7cm]{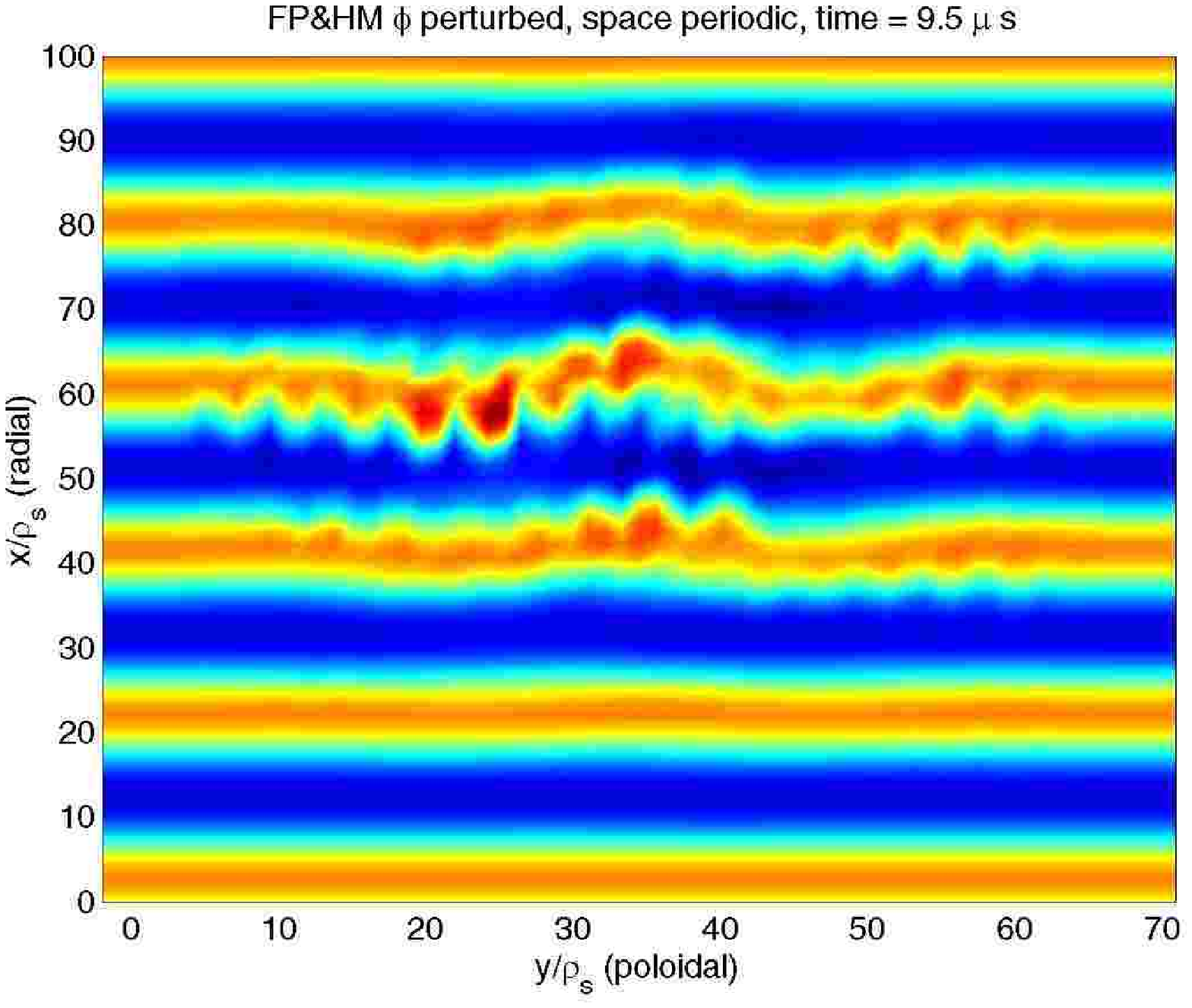}
}
\caption{Perturbation initialised as a spacial oscillation of large
wavelength.}
\end{figure}

For a higher amplitude of the initial perturbation the evolution is
qualitatively similar.

\begin{figure}[!htb]
\centerline{
\includegraphics[width=7cm]{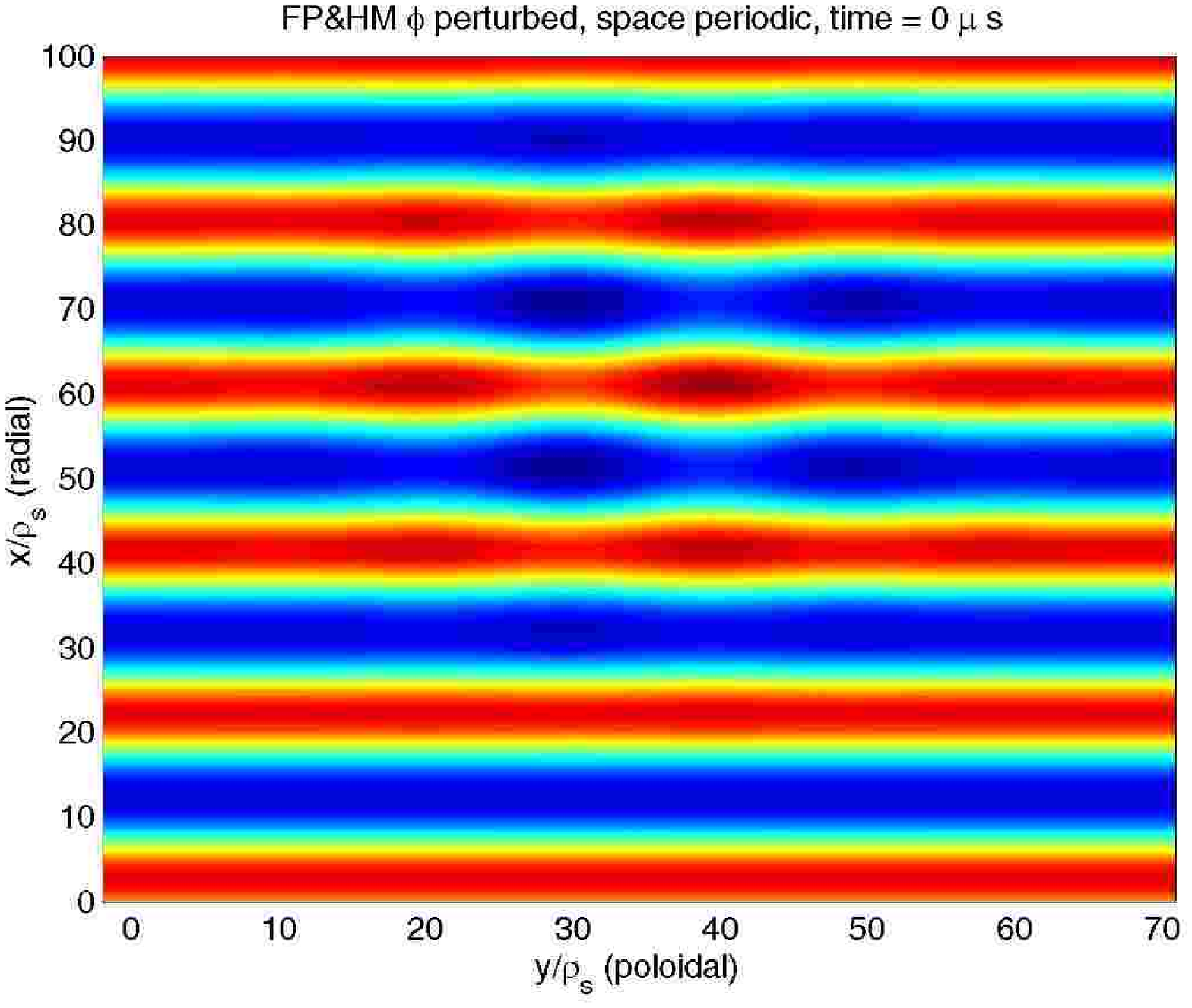}\hfill\includegraphics[width=7cm]{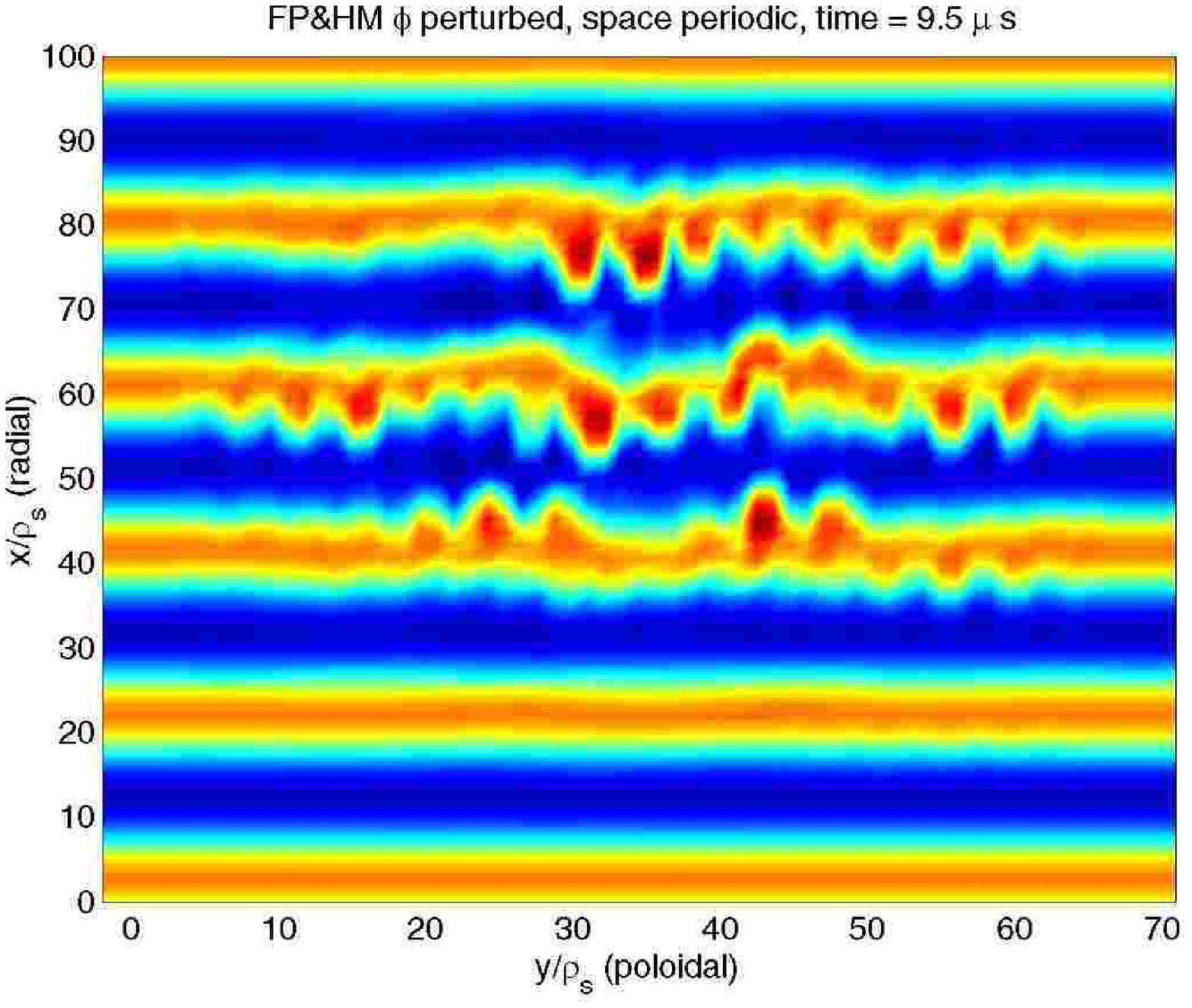}
}
\caption{Initial oscillatory perturbation at higher amplitude.}
\end{figure}

In a series of runs, we initialise with a perturbation consisting of
oscillations with higher spatial wavenumber, \emph{i.e.} slightly closer to
the final vortices. The result shows clearly that the initial conditions do
not influence significantly the space extension of the final monopolar
vortices, which means that the system chooses this dimension according to a
dispersion relation which has the character of an space-eigenvale problem.
This is also suggested by the fact thet the final stages appears to be
similar for rather different time duration which is needed by the system to
reach the stage of breaking into vortices.

\begin{figure}[tbh]
\centerline{
\includegraphics[width=7cm]{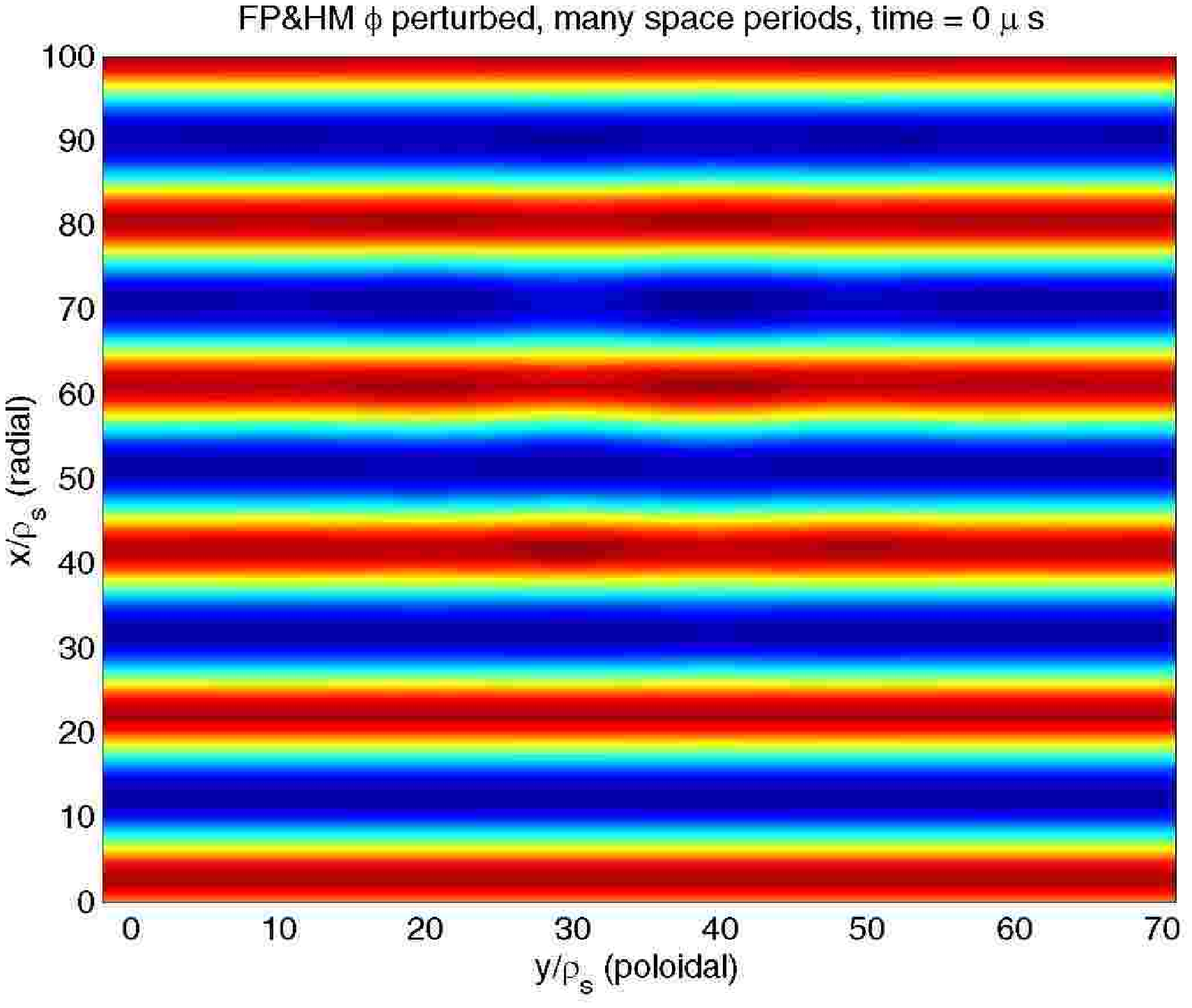}\hfill\includegraphics[width=7cm]{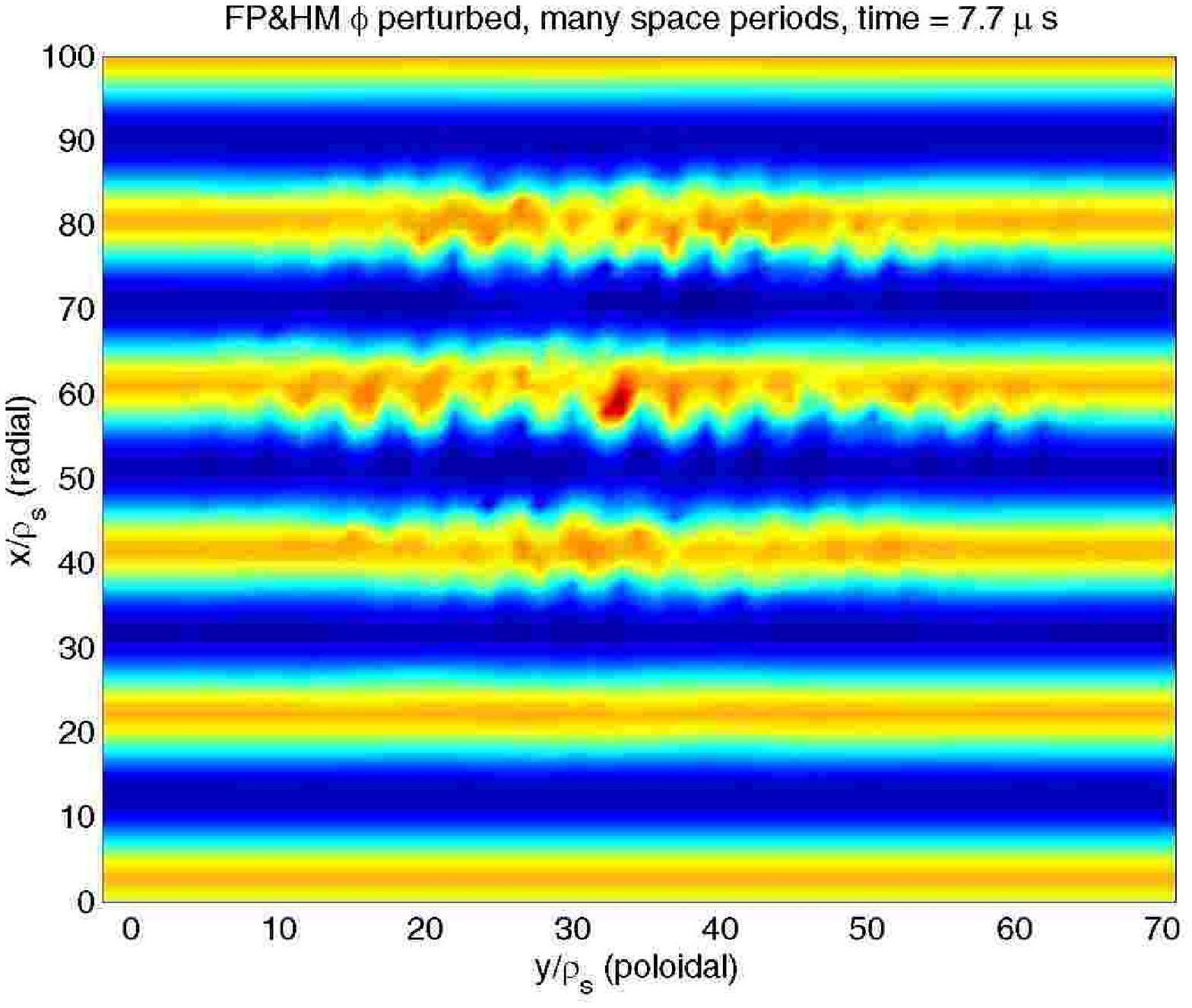}
}
\caption{Initial perturbation taken as a space oscillation of relatively small
wavelength.}
\end{figure}

\subsection{Conclusions from the numerical studies}

Extensive numerical simulations (which however must be considered
preliminary) have been done for studying the stability of $\phi _{s}$. In a
series of runs the time dependent equations with pure scalar nonlinearity
has been used with $\phi _{s}$ perturbed by $\phi _{p}$, taken first as a
monopolar vortex . For small amplitudes (\emph{e.g}. $\phi _{s}\sim 0.03$
and $\phi _{p}\sim 0.005$) the total flow is stable for more than $12\times
10^{3}\Omega _{i}$. The monopole is reshaped at early stages for equality of
the tangential flows and then is stably advected ($v_{\ast }/u\lesssim 1$).
This result may be relevant for the stability of the Red Spot, the
long-lived vortex embedded in the zonal flows of the atmosphere of Jupiter,
a geometry very similar to ours. A small amplitude dipolar initial
perturbation is also stable for similar durations.

The structural stability of $\phi _{s}$ is investigated using the time
dependent equation with both scalar and polarisation drift nonlinearities
retaind. The flow is stable on only shorter duration and a new effect
arises: at late stages, the flow tends to break up into a set of monopolar
vortices. Taking $\phi _{p}$ of high amplitude ($\phi _{s}\sim 0.06$ , $\phi
_{p}\sim 0.12$) monopolar and strongly elongated on $r$ there is a clear
tendency to rotate it such as to align with the background flow. At $6\,\mu
s $ the process of generation of vortices is pronounced. Similar conclusions
can be drawn from runs with $\phi _{p}$ taken as dipolar and as spatial
periodic structures. Examination of many runs leads to the conclusion that
the spatial extension of the generated vortices depends only weakly on the
initial conditions and of the time duration of quasi-stability and is $%
\lambda \sim 6\rho _{s}$ . Higher numerical precision is required to study
their evolution.

\newpage 

\section{Appendix}

\subsection{Derivation of the equation with time dependence}

\renewcommand{\theequation}{A.\arabic{equation}} \setcounter{equation}{0}

The aim of this Appendix is to review the derivation of the various forms of
the equations used in this work. It will be reviewed the stationary and the
time dependent equations, with scalar and polarisation drift nonlinearities,
etc. The main practical purpose is to compare the units used and to
facilitate the present numerical computation and the future possible
changes. For this reason some calculations are trivially simple and may be
skiped.

Consider the equations for the ITG model in two-dimensions with adiabatic
electrons: 
\begin{eqnarray*}
\frac{\partial n_{i}}{\partial t}+\mathbf{\nabla \cdot }\left( \mathbf{v}%
_{i}n_{i}\right) &=&0 \\
\frac{\partial \mathbf{v}_{i}}{\partial t}+\left( \mathbf{v}_{i}\cdot 
\mathbf{\nabla }\right) \mathbf{v}_{i} &=&\frac{e}{m_{i}}\left( -\mathbf{%
\nabla }\phi \right) +\frac{e}{m_{i}}\mathbf{v}_{i}\times \mathbf{B}
\end{eqnarray*}
We assume the quasineutrality 
\begin{equation*}
n_{i}\approx n_{e}
\end{equation*}
and the Boltzmann distribution of the electrons along the magnetic field
line 
\begin{equation*}
n_{e}=n_{0}\exp \left( -\frac{\left| e\right| \phi }{T_{e}}\right)
\end{equation*}
In general the electron temperature can be a function of the radial variable 
\begin{equation*}
T_{e}\equiv T_{e}\left( x\right)
\end{equation*}

The velocity of the ion fluid is perpendicular on the magnetic field and is
composed of the diamagnetic, electric and polarization drift terms 
\begin{eqnarray*}
\mathbf{v}_{i} &=&\mathbf{v}_{\perp i} \\
&=&\mathbf{v}_{dia,i}+\mathbf{v}_{E}+\mathbf{v}_{pol,i} \\
&=&\frac{T_{i}}{\left| e\right| B}\frac{1}{n_{i}}\frac{dn_{i}}{dr}\widehat{%
\mathbf{e}}_{y} \\
&&+\frac{-\mathbf{\nabla }\phi \times \widehat{\mathbf{n}}}{B} \\
&&-\frac{1}{B\Omega _{i}}\left( \frac{\partial }{\partial t}+\left( \mathbf{v%
}_{E}\cdot \mathbf{\nabla }_{\perp }\right) \right) \mathbf{\nabla }_{\perp
}\phi
\end{eqnarray*}
The diamagnetic velocity will be neglected. Introducing this velocity into
the continuity equation, one obtains an equation for the electrostatic
potential $\phi $.

Before writting this equation we introduce new dimensional units for the
variables. 
\begin{equation}
\phi ^{phys}\rightarrow \phi ^{\prime }=\frac{\left| e\right| \phi ^{phys}}{%
T_{e}}  \label{sc1}
\end{equation}
\begin{equation}
\left( x^{phys},y^{phys}\right) \rightarrow \left( x^{\prime },y^{\prime
}\right) =\left( \frac{x^{phys}}{\rho _{s}},\frac{y^{phys}}{\rho _{s}}\right)
\label{sc2}
\end{equation}
\begin{equation}
t^{phys}\rightarrow t^{\prime }=t^{phys}\Omega _{i}  \label{sc3}
\end{equation}
The new variables $\left( t,x,y\right) $ and the function $\phi $ are
non-dimensional. In the following the \emph{primes} are not written. With
these variables the equation obtained is 
\begin{eqnarray*}
&&\frac{\partial }{\partial t}\left( 1-\mathbf{\nabla }_{\perp }^{2}\right)
\phi \\
&&-\left( -\mathbf{\nabla }_{\perp }\phi \times \widehat{\mathbf{n}}\right)
\cdot \mathbf{v}_{\ast } \\
&&+\left( -\mathbf{\nabla }_{\perp }\phi \times \widehat{\mathbf{n}}\right)
\cdot \mathbf{v}_{T}\phi \\
&&+\left[ \left( -\mathbf{\nabla }_{\perp }\phi \times \widehat{\mathbf{n}}%
\right) \cdot \mathbf{\nabla }_{\perp }\right] \left( -\mathbf{\nabla }%
_{\perp }^{2}\phi \right) \\
&=&0
\end{eqnarray*}
where 
\begin{eqnarray}
\mathbf{v}_{\ast } &\equiv &-\mathbf{\nabla }_{\perp }\ln n_{0}-\mathbf{%
\nabla }_{\perp }\ln T_{e}  \label{vstarvt} \\
\mathbf{v}_{T} &\equiv &-\mathbf{\nabla }_{\perp }\ln T_{e}  \notag
\end{eqnarray}
(This is Eq.(8) from the paper \cite{LaedkeSpatschek1}).

\bigskip

\textbf{NOTE}

Before continuing we compare this equation with the equation of paper \cite
{LaedkeSpatschek2}, Eq.(16). Here taking still the units to be physical, the
form of the latter equation is (Eq.(12) from that paper) 
\begin{eqnarray}
&&\frac{\partial }{\partial t}\frac{\left| e\right| \phi }{T_{e}}-\frac{%
\partial }{\partial t}\frac{1}{B\Omega _{i}}\mathbf{\nabla }_{\perp }^{2}\phi
\label{Eq12} \\
&&+\frac{-\mathbf{\nabla }_{\perp }\phi \times \widehat{\mathbf{n}}}{B}\cdot 
\mathbf{\nabla }_{\perp }\ln n_{0}  \notag \\
&&+\frac{-\mathbf{\nabla }_{\perp }\phi \times \widehat{\mathbf{n}}}{B}\cdot 
\mathbf{\nabla }_{\perp }\frac{\left| e\right| \phi }{T_{e}}  \notag \\
&&+\frac{1}{B^{2}\Omega _{i}}\left[ \left( -\mathbf{\nabla }_{\perp }\phi
\times \widehat{\mathbf{n}}\right) \cdot \mathbf{\nabla }_{\perp }\right]
\left( -\mathbf{\nabla }_{\perp }^{2}\phi \right)  \notag \\
&=&0  \notag
\end{eqnarray}
The term containing the gradient of the equilibrium density comes from the
continuity equation, as convection of the equilibrium density by the
fluctuating $E\times B$ velocity. The term containing the gradient of the
equilibrium temperature comes from the continuity equation by the divergence
of the fluctuating flow of particles, where the adiabaticity has been
assumed, 
\begin{equation*}
\frac{\widetilde{n}}{n_{0}}=\frac{\left| e\right| \phi }{T_{e}\left(
x\right) }
\end{equation*}
The equation of \cite{LaedkeSpatschek2} remains expressed in these terms.
However, this is strange in relation with the derivation made by Lakhin et
al, where the temperature term is canceled. See below a discussion of the
Eq.(8) from Laedke and Spatschek 1986.

For the second term we have 
\begin{eqnarray*}
\frac{1}{B\Omega _{i}}\mathbf{\nabla }_{\perp }^{2}\phi &=&\frac{1}{B\Omega
_{i}}\frac{T_{e}}{\left| e\right| }\mathbf{\nabla }_{\perp }^{2}\frac{\left|
e\right| \phi }{T_{e}}=\frac{1}{\Omega _{i}}\frac{1}{\frac{\left| e\right| B%
}{m_{i}}}\frac{T_{e}}{m_{i}}\mathbf{\nabla }_{\perp }^{2}\frac{\left|
e\right| \phi }{T_{e}} \\
&=&\frac{1}{\Omega _{i}^{2}}c_{s}^{2}\mathbf{\nabla }_{\perp }^{2}\frac{%
\left| e\right| \phi }{T_{e}}=\rho _{s}^{2}\mathbf{\nabla }_{\perp }^{2}%
\frac{\left| e\right| \phi }{T_{e}}
\end{eqnarray*}
This will become (with its sign) 
\begin{equation}
-\frac{\partial }{\partial t}\mathbf{\nabla }_{\perp }^{\prime 2}\phi
^{\prime }  \label{second}
\end{equation}
in the new variables Eqs.(\ref{sc1})-(\ref{sc3}).

The third term is 
\begin{eqnarray*}
\frac{-\mathbf{\nabla }_{\perp }\phi \times \widehat{\mathbf{n}}}{B}\cdot 
\mathbf{\nabla }_{\perp }\ln n_{0} &=&\frac{1}{B}\frac{T_{e}}{\left|
e\right| }\left( -\mathbf{\nabla }_{\perp }\frac{\left| e\right| \phi }{T_{e}%
}\times \widehat{\mathbf{n}}\right) \cdot \mathbf{\nabla }_{\perp }\ln n_{0}
\\
&=&\frac{1}{\frac{\left| e\right| B}{m_{i}}}\frac{T_{e}}{m_{i}}\left( -%
\mathbf{\nabla }_{\perp }\frac{\left| e\right| \phi }{T_{e}}\times \widehat{%
\mathbf{n}}\right) \cdot \mathbf{\nabla }_{\perp }\ln n_{0} \\
&=&\Omega _{i}\frac{c_{s}^{2}}{\Omega _{i}^{2}}\left( -\mathbf{\nabla }%
_{\perp }\frac{\left| e\right| \phi }{T_{e}}\times \widehat{\mathbf{n}}%
\right) \cdot \mathbf{\nabla }_{\perp }\ln n_{0} \\
&=&\Omega _{i}\rho _{s}^{2}\left( \mathbf{\nabla }_{\perp }\frac{\left|
e\right| \phi }{T_{e}}\times \widehat{\mathbf{n}}\right) \cdot \left( -%
\mathbf{\nabla }_{\perp }\ln n_{0}\right)
\end{eqnarray*}
This will become 
\begin{equation*}
-\Omega _{i}\left( -\rho _{s}\mathbf{\nabla }_{\perp }\phi \times \widehat{%
\mathbf{n}}\right) \cdot \left( -\rho _{s}\mathbf{\nabla }_{\perp }\ln
n_{0}\right)
\end{equation*}
in the new variables 
\begin{equation}
-\Omega _{i}\left( -\mathbf{\nabla }_{\perp }^{\prime }\phi ^{\prime }\times 
\widehat{\mathbf{n}}\right) \cdot \left( -\mathbf{\nabla }_{\perp }^{\prime
}\ln n_{0}\right)  \label{third}
\end{equation}

The fourth term is 
\begin{eqnarray*}
\frac{-\mathbf{\nabla }_{\perp }\phi \times \widehat{\mathbf{n}}}{B}\cdot 
\mathbf{\nabla }_{\perp }\frac{\left| e\right| \phi }{T_{e}} &=&\frac{1}{%
\frac{\left| e\right| B}{m_{i}}}\frac{T_{e}}{m_{i}}\left( -\mathbf{\nabla }%
_{\perp }\frac{\left| e\right| \phi }{T_{e}}\times \widehat{\mathbf{n}}%
\right) \cdot \mathbf{\nabla }_{\perp }\frac{\left| e\right| \phi }{T_{e}} \\
&=&\Omega _{i}\frac{c_{s}^{2}}{\Omega _{i}^{2}}\left( -\mathbf{\nabla }%
_{\perp }\frac{\left| e\right| \phi }{T_{e}}\times \widehat{\mathbf{n}}%
\right) \cdot \mathbf{\nabla }_{\perp }\frac{\left| e\right| \phi }{T_{e}} \\
&=&\Omega _{i}\rho _{s}^{2}\left( -\mathbf{\nabla }_{\perp }\frac{\left|
e\right| \phi }{T_{e}}\times \widehat{\mathbf{n}}\right) \cdot \mathbf{%
\nabla }_{\perp }\frac{\left| e\right| \phi }{T_{e}} \\
&=&\Omega _{i}\rho _{s}^{2}\left( -\mathbf{\nabla }_{\perp }\frac{\left|
e\right| \phi }{T_{e}}\times \widehat{\mathbf{n}}\right) \cdot \left[ \frac{%
\left| e\right| }{T_{e}}\mathbf{\nabla }_{\perp }\phi -\frac{\left| e\right|
\phi }{T_{e}}\frac{\mathbf{\nabla }_{\perp }T_{e}}{T_{e}}\right] \\
&=&\Omega _{i}\rho _{s}^{2}\left( -\mathbf{\nabla }_{\perp }\frac{\left|
e\right| \phi }{T_{e}}\times \widehat{\mathbf{n}}\right) \cdot \frac{\left|
e\right| \phi }{T_{e}}\left( -\frac{\mathbf{\nabla }_{\perp }T_{e}}{T_{e}}%
\right)
\end{eqnarray*}
This term will become 
\begin{equation}
\Omega _{i}\left( -\mathbf{\nabla }_{\perp }^{\prime }\phi ^{\prime }\times 
\widehat{\mathbf{n}}\right) \phi ^{\prime }\cdot \left( -\mathbf{\nabla }%
_{\perp }^{\prime }\ln T_{e}\right)  \label{fourth}
\end{equation}
after introducing the new units.

The last term (with the polarization nonlinearity) is in physical units 
\begin{equation*}
\frac{1}{B^{2}\Omega _{i}}\left[ \left( -\mathbf{\nabla }_{\perp }\phi
\times \widehat{\mathbf{n}}\right) \cdot \mathbf{\nabla }_{\perp }\right]
\left( -\mathbf{\nabla }_{\perp }^{2}\phi \right)
\end{equation*}
This is converted to non-dimensional variables 
\begin{equation*}
\frac{1}{B^{2}\Omega _{i}}\left[ \left( -\frac{1}{\rho _{s}}\frac{T_{e}}{%
\left| e\right| }\rho _{s}\mathbf{\nabla }_{\perp }\frac{\left| e\right|
\phi }{T_{e}}\times \widehat{\mathbf{n}}\right) \cdot \frac{1}{\rho _{s}}%
\rho _{s}\mathbf{\nabla }_{\perp }\right] \frac{1}{\rho _{s}^{2}}\frac{T_{e}%
}{\left| e\right| }\left( -\rho _{s}^{2}\mathbf{\nabla }_{\perp }^{2}\frac{%
\left| e\right| \phi }{T_{e}}\right)
\end{equation*}
Collecting the physical coefficient we have 
\begin{eqnarray*}
\frac{1}{B^{2}\Omega _{i}}\left( \frac{T_{e}}{\left| e\right| }\right) ^{2}%
\frac{1}{\rho _{s}^{4}} &=&\frac{1}{\left( \frac{\left| e\right| B}{m_{i}}%
\right) ^{2}}\frac{1}{\Omega _{i}}\left( \frac{T_{e}}{m_{i}}\right) ^{2}%
\frac{1}{\rho _{s}^{4}} \\
&=&\Omega _{i}\frac{c_{s}^{4}}{\Omega _{i}^{4}}\frac{1}{\rho _{s}^{4}} \\
&=&\Omega _{i}
\end{eqnarray*}
Then, in the new variables, this term becomes 
\begin{equation}
\Omega _{i}\left[ \left( -\mathbf{\nabla }_{\perp }^{\prime }\phi ^{\prime
}\times \widehat{\mathbf{n}}\right) \cdot \mathbf{\nabla }_{\perp }^{\prime }%
\right] \left( -\mathbf{\nabla }_{\perp }^{\prime 2}\phi ^{\prime }\right)
\label{last}
\end{equation}

Then the Eqs.(\ref{Eq12}) with the new form of its terms (\ref{second}), (%
\ref{third}), (\ref{fourth}) and (\ref{last}) becomes 
\begin{eqnarray}
&&\frac{\partial }{\partial t}\phi ^{\prime }-\frac{\partial }{\partial t}%
\mathbf{\nabla }_{\perp }^{\prime 2}\phi ^{\prime }  \label{Eq12prim} \\
&&-\Omega _{i}\left( -\mathbf{\nabla }_{\perp }^{\prime }\phi ^{\prime
}\times \widehat{\mathbf{n}}\right) \cdot \left( -\mathbf{\nabla }_{\perp
}^{\prime }\ln n_{0}\right)  \notag \\
&&+\Omega _{i}\left( -\mathbf{\nabla }_{\perp }^{\prime }\phi ^{\prime
}\times \widehat{\mathbf{n}}\right) \phi ^{\prime }\cdot \left( -\mathbf{%
\nabla }_{\perp }^{\prime }\ln T_{e}\right)  \notag \\
&&+\Omega _{i}\left[ \left( -\mathbf{\nabla }_{\perp }^{\prime }\phi
^{\prime }\times \widehat{\mathbf{n}}\right) \cdot \mathbf{\nabla }_{\perp
}^{\prime }\right] \left( -\mathbf{\nabla }_{\perp }^{\prime 2}\phi ^{\prime
}\right)  \notag \\
&=&0  \notag
\end{eqnarray}
Introducing the time unit $\Omega _{i}^{-1}$, and eliminating the \emph{%
primes} 
\begin{eqnarray}
&&\frac{\partial }{\partial t}\left( 1-\mathbf{\nabla }_{\perp }^{2}\right)
\phi  \label{Eq12norm} \\
&&-\left( -\mathbf{\nabla }_{\perp }\phi \times \widehat{\mathbf{n}}\right)
\cdot \left( -\mathbf{\nabla }_{\perp }\ln n_{0}\right) +\left( -\mathbf{%
\nabla }_{\perp }\phi \times \widehat{\mathbf{n}}\right) \cdot \left( -%
\mathbf{\nabla }_{\perp }\ln T_{e}\right) \phi  \notag \\
&&+\left[ \left( -\mathbf{\nabla }_{\perp }\phi \times \widehat{\mathbf{n}}%
\right) \cdot \mathbf{\nabla }_{\perp }\right] \left( -\mathbf{\nabla }%
_{\perp }^{2}\phi \right)  \notag \\
&=&0  \notag
\end{eqnarray}

However this is not the equation (16) of Laedke and Spatschek 1988. This is
discussed in the following.

\textbf{END OF THE NOTE}

\bigskip

We return to the \emph{physical} equation, Eq.(\ref{Eq12}) or Eq.(12) from 
\cite{LaedkeSpatschek2} (1988) 
\begin{eqnarray}
&&\frac{\partial }{\partial t}\frac{\left| e\right| \phi }{T_{e}}-\frac{%
\partial }{\partial t}\frac{1}{B\Omega _{i}}\mathbf{\nabla }_{\perp }^{2}\phi
\label{Toteq12} \\
&&+\frac{-\mathbf{\nabla }_{\perp }\phi \times \widehat{\mathbf{n}}}{B}\cdot 
\mathbf{\nabla }_{\perp }\ln n_{0}  \notag \\
&&+\frac{-\mathbf{\nabla }_{\perp }\phi \times \widehat{\mathbf{n}}}{B}\cdot 
\mathbf{\nabla }_{\perp }\frac{\left| e\right| \phi }{T_{e}}  \notag \\
&&+\frac{1}{B^{2}\Omega _{i}}\left[ \left( -\mathbf{\nabla }_{\perp }\phi
\times \widehat{\mathbf{n}}\right) \cdot \mathbf{\nabla }_{\perp }\right]
\left( -\mathbf{\nabla }_{\perp }^{2}\phi \right)  \notag \\
&=&0  \notag
\end{eqnarray}
We make a change of variables 
\begin{equation*}
\eta =y-ut
\end{equation*}
which means 
\begin{equation*}
\frac{\partial }{\partial t}=\frac{\partial }{\partial t^{\prime }}-u\frac{%
\partial }{\partial \eta }
\end{equation*}
The equation becomes, after removing the \emph{primes} from $t$ 
\begin{eqnarray*}
&&\frac{\partial }{\partial t^{\prime }}\frac{\left| e\right| \phi }{T_{e}}-%
\frac{\partial }{\partial t^{\prime }}\frac{1}{B\Omega _{i}}\mathbf{\nabla }%
_{\perp }^{2}\phi -u\frac{\partial }{\partial \eta }\frac{\left| e\right|
\phi }{T_{e}}+u\frac{\partial }{\partial \eta }\frac{1}{B\Omega _{i}}\mathbf{%
\nabla }_{\perp }^{2}\phi \\
&&+\frac{-\mathbf{\nabla }_{\perp }\phi \times \widehat{\mathbf{n}}}{B}\cdot 
\mathbf{\nabla }_{\perp }\ln n_{0} \\
&&+\frac{-\mathbf{\nabla }_{\perp }\phi \times \widehat{\mathbf{n}}}{B}\cdot 
\mathbf{\nabla }_{\perp }\frac{\left| e\right| \phi }{T_{e}} \\
&&+\frac{1}{B^{2}\Omega _{i}}\left[ \left( -\mathbf{\nabla }_{\perp }\phi
\times \widehat{\mathbf{n}}\right) \cdot \mathbf{\nabla }_{\perp }\right]
\left( -\mathbf{\nabla }_{\perp }^{2}\phi \right) \\
&=&0
\end{eqnarray*}
The first term is 
\begin{eqnarray}
\frac{\partial }{\partial t}\frac{\left| e\right| \phi }{T_{e}} &=&\frac{%
\partial }{\partial t}\frac{1}{B}\frac{\left| e\right| B}{m_{i}}\frac{m_{i}}{%
T_{e}}\phi  \label{first1} \\
&=&\frac{\partial }{\partial t}\frac{1}{B}\frac{\Omega _{i}}{\Omega _{i}}%
\frac{\Omega _{i}}{c_{s}^{2}}\phi  \notag \\
&=&\frac{1}{B\Omega _{i}}\frac{\partial }{\partial t}\frac{1}{\rho _{s}^{2}}%
\phi  \notag
\end{eqnarray}
The third term is 
\begin{equation*}
-u\frac{\partial }{\partial \eta }\frac{\left| e\right| \phi }{T_{e}}=-\frac{%
1}{B\Omega _{i}}u\frac{\partial }{\partial \eta }\frac{1}{\rho _{s}^{2}}\phi
\end{equation*}
the second and fourth terms remains unchanged 
\begin{equation*}
-\frac{\partial }{\partial t}\frac{1}{B\Omega _{i}}\mathbf{\nabla }_{\perp
}^{2}\phi +u\frac{\partial }{\partial \eta }\frac{1}{B\Omega _{i}}\mathbf{%
\nabla }_{\perp }^{2}\phi
\end{equation*}
Since the product $1/\left( \Omega _{i}B\right) $ arises systematically, we
will extract it from the third term 
\begin{equation*}
\frac{-\mathbf{\nabla }_{\perp }\phi \times \widehat{\mathbf{n}}}{B}\cdot 
\mathbf{\nabla }_{\perp }\ln n_{0}=-\frac{1}{B\Omega _{i}}u\frac{\Omega _{i}%
}{u}\kappa _{n}\frac{\partial \phi }{\partial \eta }
\end{equation*}
where we define 
\begin{equation*}
\kappa _{n}\widehat{\mathbf{e}}_{x}\equiv \mathbf{\nabla }_{\perp }\ln n_{0}
\end{equation*}
with $\kappa _{n}$ measured in $m^{-1}$.

The term with the gradient of temperature 
\begin{equation*}
\frac{-\mathbf{\nabla }_{\perp }\phi \times \widehat{\mathbf{n}}}{B}\cdot 
\mathbf{\nabla }_{\perp }\frac{\left| e\right| \phi }{T_{e}}=\frac{-\mathbf{%
\nabla }_{\perp }\phi \times \widehat{\mathbf{n}}}{B}\left| e\right| \phi
\cdot \left( -\frac{1}{T_{e}^{2}}\mathbf{\nabla }_{\perp }T_{e}\right)
\end{equation*}
after introducing the definition 
\begin{equation*}
\kappa _{T}\widehat{\mathbf{e}}_{x}\equiv \mathbf{\nabla }_{\perp }\ln T_{e}
\end{equation*}
we obtain 
\begin{eqnarray*}
\frac{-\mathbf{\nabla }_{\perp }\phi \times \widehat{\mathbf{n}}}{B}\cdot 
\mathbf{\nabla }_{\perp }\frac{\left| e\right| \phi }{T_{e}} &=&\frac{\kappa
_{T}}{B\Omega _{i}}\frac{1}{B}\frac{\Omega _{i}\left| e\right| B}{\frac{T_{e}%
}{m_{i}}m_{i}}\phi \frac{\partial \phi }{\partial \eta } \\
&=&\frac{1}{B\Omega _{i}}\frac{1}{\rho _{s}^{2}}\frac{1}{B}\kappa _{T}\phi 
\frac{\partial \phi }{\partial \eta }
\end{eqnarray*}
The term with the polarisation nonlinearity is 
\begin{equation*}
\frac{1}{B^{2}\Omega _{i}}\left[ \left( -\mathbf{\nabla }_{\perp }\phi
\times \widehat{\mathbf{n}}\right) \cdot \mathbf{\nabla }_{\perp }\right]
\left( -\mathbf{\nabla }_{\perp }^{2}\phi \right)
\end{equation*}
The form of the equation is 
\begin{eqnarray*}
&&\frac{1}{B\Omega _{i}}\frac{\partial }{\partial t}\frac{1}{\rho _{s}^{2}}%
\phi -\frac{\partial }{\partial t}\frac{1}{B\Omega _{i}}\mathbf{\nabla }%
_{\perp }^{2}\phi \\
&&-\frac{1}{B\Omega _{i}}u\frac{\partial }{\partial \eta }\frac{1}{\rho
_{s}^{2}}\phi -\frac{1}{B\Omega _{i}}u\frac{\Omega _{i}}{u}\kappa _{n}\frac{%
\partial \phi }{\partial \eta } \\
&&+u\frac{\partial }{\partial \eta }\frac{1}{B\Omega _{i}}\mathbf{\nabla }%
_{\perp }^{2}\phi \\
&&+\frac{1}{B\Omega _{i}}\frac{1}{\rho _{s}^{2}}\frac{1}{B}\kappa _{T}\phi 
\frac{\partial \phi }{\partial \eta } \\
&&+\frac{1}{B^{2}\Omega _{i}}\left[ \left( -\mathbf{\nabla }_{\perp }\phi
\times \widehat{\mathbf{n}}\right) \cdot \mathbf{\nabla }_{\perp }\right]
\left( -\mathbf{\nabla }_{\perp }^{2}\phi \right) \\
&=&0
\end{eqnarray*}
Multiplying by $\Omega _{i}$ 
\begin{eqnarray}
&&\frac{\partial }{\partial t}\left[ \left( \frac{1}{\rho _{s}^{2}}-\mathbf{%
\nabla }_{\perp }^{2}\right) \frac{\phi }{B}\right]  \label{eqphys} \\
&&-u\frac{\partial }{\partial \eta }\left[ \left( \frac{1}{\rho _{s}^{2}}+%
\frac{\Omega _{i}\kappa _{n}}{u}\right) \frac{\phi }{B}\right]  \notag \\
&&+u\frac{\partial }{\partial \eta }\mathbf{\nabla }_{\perp }^{2}\frac{\phi 
}{B}  \notag \\
&&+\frac{1}{\rho _{s}^{2}}\frac{\kappa _{T}}{B^{2}}\phi \frac{\partial \phi 
}{\partial \eta }  \notag \\
&&+\frac{1}{B^{2}}\left[ \left( -\mathbf{\nabla }_{\perp }\phi \times 
\widehat{\mathbf{n}}\right) \cdot \mathbf{\nabla }_{\perp }\right] \left( -%
\mathbf{\nabla }_{\perp }^{2}\phi \right)  \notag \\
&=&0  \notag
\end{eqnarray}

\bigskip

\textbf{NOTE ON THE\ COMPARISON\ WITH}\ Su and Horton

We can compare this equation with the corresponding one, Eq. (45), from Ref. 
\cite{SuHorton}. This equation is 
\begin{eqnarray*}
&&\left( \frac{1}{T\left( x\right) }-\mathbf{\nabla }_{\perp }^{2}\right) 
\frac{\partial \varphi }{\partial t}+ \\
&&+\left( v_{d0}+v_{d0}^{\prime }x-\kappa _{T}\varphi \right) \frac{\partial
\varphi }{\partial y} \\
&&-\left[ \left( -\mathbf{\nabla }_{\perp }\varphi \times \widehat{\mathbf{n}%
}\right) \cdot \mathbf{\nabla }_{\perp }\right] \left( -\mathbf{\nabla }%
_{\perp }^{2}\varphi \right) \\
&=&0
\end{eqnarray*}
where 
\begin{equation*}
T\left( x\right) =\frac{T_{e}\left( x\right) }{T_{0}}
\end{equation*}
\begin{equation*}
\varphi =\frac{L_{n}}{\rho _{s0}}\frac{\left| e\right| \Phi }{T_{0}}
\end{equation*}
\begin{equation*}
\varepsilon _{n}=\frac{\rho _{s0}}{L_{n}}
\end{equation*}
This is the small parameter of the expansion. 
\begin{equation*}
v_{d}=1+v_{d0}^{\prime }x
\end{equation*}
\begin{equation*}
v_{d0}^{\prime }=\rho _{s0}\frac{dv_{d}}{dx}\sim \varepsilon _{n}v_{d0}
\end{equation*}

\textbf{END OF\ THE\ NOTE}

\bigskip

In Ref.\cite{LaedkeSpatschek2} the potential is redefined as 
\begin{equation*}
\psi \equiv \frac{\phi }{B}
\end{equation*}
and the equation becomes 
\begin{eqnarray*}
&&\frac{\partial }{\partial t}\left[ \left( \frac{1}{\rho _{s}^{2}}-\mathbf{%
\nabla }_{\perp }^{2}\right) \psi \right] -u\frac{\partial }{\partial \eta }%
\left[ \left( \frac{1}{\rho _{s}^{2}}+\frac{\Omega _{i}\kappa _{n}}{u}%
\right) \psi \right] \\
&&+u\frac{\partial }{\partial \eta }\mathbf{\nabla }_{\perp }^{2}\psi +\frac{%
1}{\rho _{s}^{2}}\frac{\kappa _{T}}{B^{2}}\psi \frac{\partial \psi }{%
\partial \eta } \\
&&+\left[ \left( -\mathbf{\nabla }_{\perp }\psi \times \widehat{\mathbf{n}}%
\right) \cdot \mathbf{\nabla }_{\perp }\right] \left( -\mathbf{\nabla }%
_{\perp }^{2}\psi \right) \\
&=&0
\end{eqnarray*}

In Ref.\cite{LaedkeSpatschek2} the following operator is introduced 
\begin{equation*}
P\equiv \frac{1}{\rho _{s}^{2}}-\mathbf{\nabla }_{\perp }^{2}
\end{equation*}
and the equation is expressed as 
\begin{equation*}
P\frac{\partial }{\partial t}\psi =\left[ \mathbf{\nabla }_{\perp }\left(
\psi -ux\right) \times \widehat{\mathbf{n}}\right] \cdot \mathbf{\nabla }%
_{\perp }\left( P\psi +\Omega _{i}\ln n_{0}\right)
\end{equation*}
In this way the term containing the temperature gradient $\kappa _{T}$ is
hidden in the application of the operator $\mathbf{\nabla }_{\perp }$ on $%
1/\rho _{s}^{2}$ contained in $P$.

\subsection{The stationary form of the equation}

The most general form of the solution at stationarity is 
\begin{eqnarray*}
\frac{\partial \psi }{\partial t} &=&0 \\
P\psi +\Omega _{i}\ln n_{0} &=&\Omega _{i}\ln \left[ n_{0}\left( x-\frac{%
\psi }{u}\right) \right]
\end{eqnarray*}

In the natural logarithm we expand the density at $x$ and then parform an
expansion 
\begin{equation*}
\ln \left( 1+\varepsilon \right) \approx \varepsilon -\frac{\varepsilon ^{2}%
}{2}
\end{equation*}
\begin{eqnarray*}
\Omega _{i}\ln \left[ n_{0}\left( x-\frac{\psi }{u}\right) \right] &=&\Omega
_{i}\ln \left[ n_{0}\left( x\right) -\frac{\psi }{u}\frac{dn_{0}\left(
x\right) }{dx}+\frac{1}{2}\frac{\psi ^{2}}{u^{2}}\frac{d^{2}n_{0}\left(
x\right) }{dx^{2}}\right] \\
&=&\Omega _{i}\ln \left\{ n_{0}\left( x\right) \left[ 1-\frac{\psi }{u}\frac{%
1}{n_{0}}\frac{dn_{0}\left( x\right) }{dx}+\frac{1}{2}\frac{\psi ^{2}}{u^{2}}%
\frac{1}{n_{0}}\frac{d^{2}n_{0}\left( x\right) }{dx^{2}}\right] \right\} \\
&=&\Omega _{i}\ln \left[ n_{0}\left( x\right) \right] \\
&&+\Omega _{i}\ln \left[ 1-\frac{\psi }{u}\frac{1}{n_{0}}\frac{dn_{0}\left(
x\right) }{dx}+\frac{1}{2}\frac{\psi ^{2}}{u^{2}}\frac{1}{n_{0}}\frac{%
d^{2}n_{0}\left( x\right) }{dx^{2}}\right]
\end{eqnarray*}
\begin{eqnarray*}
&&\Omega _{i}\ln \left[ n_{0}\left( x-\frac{\psi }{u}\right) \right] \\
&=&\Omega _{i}\ln \left[ n_{0}\left( x\right) \right] \\
&&+\Omega _{i}\left[ -\frac{\psi }{u}\frac{1}{n_{0}}\frac{dn_{0}\left(
x\right) }{dx}+\frac{1}{2}\frac{\psi ^{2}}{u^{2}}\frac{1}{n_{0}}\frac{%
d^{2}n_{0}\left( x\right) }{dx^{2}}-\frac{1}{2}\frac{\psi ^{2}}{u^{2}}\frac{1%
}{n_{0}^{2}}\left( \frac{dn_{0}}{dx}\right) ^{2}\right]
\end{eqnarray*}
The last two terms are 
\begin{eqnarray*}
\frac{1}{2}\frac{1}{n_{0}}\frac{d^{2}n_{0}\left( x\right) }{dx^{2}}-\frac{1}{%
2}\frac{1}{n_{0}^{2}}\left( \frac{dn_{0}}{dx}\right) ^{2} &=&\frac{1}{2}%
\frac{d}{dx}\kappa _{n} \\
&\equiv &\frac{1}{2}\kappa _{n}^{\prime }
\end{eqnarray*}
The final form is 
\begin{eqnarray*}
&&\Omega _{i}\ln \left[ n_{0}\left( x-\frac{\psi }{u}\right) \right] \\
&=&\Omega _{i}\ln \left[ n_{0}\left( x\right) \right] \\
&&+\Omega _{i}\left( -\frac{\psi }{u}\kappa _{n}+\frac{1}{2}\frac{\psi ^{2}}{%
u^{2}}\kappa _{n}^{\prime }\right)
\end{eqnarray*}
In this expression the units are physical: 
\begin{eqnarray*}
&&u\;\left( m/s\right) \\
&&\kappa _{n}\;\left( m^{-1}\right) \\
&&\kappa _{n}^{\prime }\;\left( m^{-2}\right) \\
\psi &\equiv &\frac{\phi }{B}\;\left( m^{2}/s\right)
\end{eqnarray*}

The equation is then 
\begin{eqnarray*}
P\psi +\Omega _{i}\ln n_{0} &=&\Omega _{i}\ln \left[ n_{0}\left( x-\frac{%
\psi }{u}\right) \right] \\
&=&\Omega _{i}\ln \left[ n_{0}\left( x\right) \right] +\Omega _{i}\left( -%
\frac{\psi }{u}\kappa _{n}+\frac{1}{2}\frac{\psi ^{2}}{u^{2}}\kappa
_{n}^{\prime }\right)
\end{eqnarray*}
or 
\begin{equation}
\left( \frac{1}{\rho _{s}^{2}}-\mathbf{\nabla }_{\perp }^{2}\right) \psi =-%
\frac{\psi }{u}\Omega _{i}\kappa _{n}+\frac{1}{2}\frac{\psi ^{2}}{u^{2}}%
\Omega _{i}\kappa _{n}^{\prime }  \label{eqsta}
\end{equation}

\subsection{The equation with time dependence and polarisation nonlinearity
retained}

In order to write an equation for the potential with time dependence, we
return to Eq.(\ref{eqphys}). We multiply by $B$%
\begin{eqnarray}
&&\frac{\partial }{\partial t}\left[ \left( \frac{1}{\rho _{s}^{2}}-\mathbf{%
\nabla }_{\perp }^{2}\right) \phi \right]  \label{eqphysB} \\
&&-u\frac{\partial }{\partial \eta }\left[ \left( \frac{1}{\rho _{s}^{2}}+%
\frac{\Omega _{i}\kappa _{n}}{u}\right) \phi \right]  \notag \\
&&+u\frac{\partial }{\partial \eta }\mathbf{\nabla }_{\perp }^{2}\phi  \notag
\\
&&+\frac{1}{\rho _{s}^{2}}\frac{\kappa _{T}}{B}\phi \frac{\partial \phi }{%
\partial \eta }  \notag \\
&&+\frac{1}{B}\left[ \left( -\mathbf{\nabla }_{\perp }\phi \times \widehat{%
\mathbf{n}}\right) \cdot \mathbf{\nabla }_{\perp }\right] \left( -\mathbf{%
\nabla }_{\perp }^{2}\phi \right)  \notag \\
&=&0  \notag
\end{eqnarray}
Then we derivate Eq.(\ref{eqsta}) with respect to $\eta $, multiply by $u$
and return from $\psi $ to $\phi \equiv \psi B$ which will be noted $\phi
_{s}$ (stationary solution) 
\begin{equation}
u\frac{1}{\rho _{s}^{2}}\frac{\partial }{\partial \eta }\phi _{s}-u\mathbf{%
\nabla }_{\perp }^{2}\frac{\partial }{\partial \eta }\phi _{s}+\Omega
_{i}\kappa _{n}\frac{\partial }{\partial \eta }\phi _{s}-\frac{\Omega
_{i}\kappa _{n}^{\prime }}{uB}\frac{\partial }{\partial \eta }\frac{\phi
_{s}^{2}}{2}=0  \label{eqphissu}
\end{equation}
Now we add this equation from Eq.(\ref{eqphysB}) and combine the two
functions $\phi _{s}$ and $\phi $ in order to obtain an equation for their
difference.

We first examine the last term 
\begin{eqnarray*}
&&\left\{ \left[ -\mathbf{\nabla }_{\perp }\left( \phi -\phi _{s}\right)
\times \widehat{\mathbf{n}}\right] \cdot \mathbf{\nabla }_{\perp }\right\} %
\left[ -\mathbf{\nabla }_{\perp }^{2}\left( \phi -\phi _{s}\right) \right] \\
&=&\left[ \left( -\mathbf{\nabla }_{\perp }\phi \times \widehat{\mathbf{n}}%
\right) \cdot \mathbf{\nabla }_{\perp }\right] \left( -\mathbf{\nabla }%
_{\perp }^{2}\phi \right) - \\
&&-\left[ \left( -\mathbf{\nabla }_{\perp }\phi _{s}\times \widehat{\mathbf{n%
}}\right) \cdot \mathbf{\nabla }_{\perp }\right] \left( -\mathbf{\nabla }%
_{\perp }^{2}\phi \right) \\
&&-\left[ \left( -\mathbf{\nabla }_{\perp }\phi \times \widehat{\mathbf{n}}%
\right) \cdot \mathbf{\nabla }_{\perp }\right] \left( -\mathbf{\nabla }%
_{\perp }^{2}\phi _{s}\right) \\
&&+\left[ \left( -\mathbf{\nabla }_{\perp }\phi _{s}\times \widehat{\mathbf{n%
}}\right) \cdot \mathbf{\nabla }_{\perp }\right] \left( -\mathbf{\nabla }%
_{\perp }^{2}\phi _{s}\right)
\end{eqnarray*}
The last term is zero. Expressing the term we need (the first one) 
\begin{eqnarray*}
&&\left[ \left( -\mathbf{\nabla }_{\perp }\phi \times \widehat{\mathbf{n}}%
\right) \cdot \mathbf{\nabla }_{\perp }\right] \left( -\mathbf{\nabla }%
_{\perp }^{2}\phi \right) \\
&=&\left\{ \left[ -\mathbf{\nabla }_{\perp }\left( \phi -\phi _{s}\right)
\times \widehat{\mathbf{n}}\right] \cdot \mathbf{\nabla }_{\perp }\right\} %
\left[ -\mathbf{\nabla }_{\perp }^{2}\left( \phi -\phi _{s}\right) \right] \\
&&+\left[ \left( -\mathbf{\nabla }_{\perp }\phi _{s}\times \widehat{\mathbf{n%
}}\right) \cdot \mathbf{\nabla }_{\perp }\right] \left( -\mathbf{\nabla }%
_{\perp }^{2}\phi \right) \\
&&+\left[ \left( -\mathbf{\nabla }_{\perp }\phi \times \widehat{\mathbf{n}}%
\right) \cdot \mathbf{\nabla }_{\perp }\right] \left( -\mathbf{\nabla }%
_{\perp }^{2}\phi _{s}\right)
\end{eqnarray*}
After adding the equations we obtain 
\begin{eqnarray*}
&&\frac{\partial }{\partial t}\left[ \left( \frac{1}{\rho _{s}^{2}}-\mathbf{%
\nabla }_{\perp }^{2}\right) \left( \phi -\phi _{s}\right) \right] \\
&&+u\frac{\partial }{\partial \eta }\left[ \left( -\frac{1}{\rho _{s}^{2}}+%
\mathbf{\nabla }_{\perp }^{2}\right) \left( \phi -\phi _{s}\right) \right] \\
&&+\frac{\partial }{\partial \eta }\left[ -\Omega _{i}\kappa _{n}\left( \phi
-\phi _{s}\right) \right] \\
&&+\frac{\partial }{\partial \eta }\left[ \frac{1}{\rho _{s}^{2}}\frac{%
\kappa _{T}}{B}\frac{\phi ^{2}}{2}-\frac{\Omega _{i}\kappa _{n}^{\prime }}{uB%
}\frac{\phi _{s}^{2}}{2}\right] \\
&&+\frac{1}{B}\left\{ \left[ -\mathbf{\nabla }_{\perp }\left( \phi -\phi
_{s}\right) \times \widehat{\mathbf{n}}\right] \cdot \mathbf{\nabla }_{\perp
}\right\} \left[ -\mathbf{\nabla }_{\perp }^{2}\left( \phi -\phi _{s}\right) %
\right] \\
&&+\frac{1}{B}\left[ \left( -\mathbf{\nabla }_{\perp }\phi _{s}\times 
\widehat{\mathbf{n}}\right) \cdot \mathbf{\nabla }_{\perp }\right] \left( -%
\mathbf{\nabla }_{\perp }^{2}\phi \right) +\frac{1}{B}\left[ \left( -\mathbf{%
\nabla }_{\perp }\phi \times \widehat{\mathbf{n}}\right) \cdot \mathbf{%
\nabla }_{\perp }\right] \left( -\mathbf{\nabla }_{\perp }^{2}\phi
_{s}\right) \\
&=&0
\end{eqnarray*}

We define the new potential representing the difference between the time
dependent solution and the stationary solution 
\begin{equation*}
\varphi \equiv \phi -\phi _{s}
\end{equation*}
Then the last two terms of the above equation will become 
\begin{eqnarray*}
&&\left[ \left( -\mathbf{\nabla }_{\perp }\phi _{s}\times \widehat{\mathbf{n}%
}\right) \cdot \mathbf{\nabla }_{\perp }\right] \left( -\mathbf{\nabla }%
_{\perp }^{2}\phi _{s}-\mathbf{\nabla }_{\perp }^{2}\varphi \right) \\
&=&\left[ \left( -\mathbf{\nabla }_{\perp }\phi _{s}\times \widehat{\mathbf{n%
}}\right) \cdot \mathbf{\nabla }_{\perp }\right] \left( -\mathbf{\nabla }%
_{\perp }^{2}\varphi \right)
\end{eqnarray*}
and 
\begin{eqnarray*}
&&\left[ \left( -\mathbf{\nabla }_{\perp }\phi _{s}\times \widehat{\mathbf{n}%
}\right) \cdot \mathbf{\nabla }_{\perp }+\left( -\mathbf{\nabla }_{\perp
}\varphi \times \widehat{\mathbf{n}}\right) \cdot \mathbf{\nabla }_{\perp }%
\right] \left( -\mathbf{\nabla }_{\perp }^{2}\phi _{s}\right) \\
&=&\left[ \left( -\mathbf{\nabla }_{\perp }\varphi \times \widehat{\mathbf{n}%
}\right) \cdot \mathbf{\nabla }_{\perp }\right] \left( -\mathbf{\nabla }%
_{\perp }^{2}\phi _{s}\right)
\end{eqnarray*}
We can use in the last term the equation defining the stationary solution, $%
\phi _{s}$. 
\begin{eqnarray*}
\mathbf{\nabla }_{\perp }^{2}\phi _{s} &=&\left( \frac{1}{\rho _{s}^{2}}+%
\frac{\Omega _{i}\kappa _{n}}{u}\right) \phi _{s}-\frac{\Omega _{i}\kappa
_{n}^{\prime }}{u^{2}B}\frac{\phi _{s}^{2}}{2} \\
&\equiv &\alpha \phi _{s}-\beta \phi _{s}^{2}
\end{eqnarray*}
with notations 
\begin{eqnarray*}
\alpha &\equiv &\frac{1}{\rho _{s}^{2}}+\frac{\Omega _{i}\kappa _{n}}{u} \\
\beta &\equiv &\frac{\Omega _{i}\kappa _{n}^{\prime }}{2u^{2}B}
\end{eqnarray*}
Then the equation is 
\begin{eqnarray}
&&\frac{\partial }{\partial t}\left( \frac{1}{\rho _{s}^{2}}-\mathbf{\nabla }%
_{\perp }^{2}\right) \varphi -  \label{timeqphys} \\
&&-u\frac{\partial }{\partial \eta }\left( \frac{1}{\rho _{s}^{2}}-\mathbf{%
\nabla }_{\perp }^{2}\right) \varphi  \notag \\
&&-\Omega _{i}\kappa _{n}\frac{\partial }{\partial \eta }\varphi  \notag \\
&&+\frac{\partial }{\partial \eta }\left[ \frac{\kappa _{T}}{2\rho _{s}^{2}B}%
\left( \phi _{s}+\varphi \right) -\frac{\Omega _{i}\kappa _{n}^{\prime }}{2uB%
}\phi _{s}^{2}\right]  \notag \\
&&+\frac{1}{B}\left[ \left( -\mathbf{\nabla }_{\perp }\varphi \times 
\widehat{\mathbf{n}}\right) \cdot \mathbf{\nabla }_{\perp }\right] \left( -%
\mathbf{\nabla }_{\perp }^{2}\varphi \right)  \notag \\
&&+\frac{1}{B}\left[ \left( -\mathbf{\nabla }_{\perp }\phi _{s}\times 
\widehat{\mathbf{n}}\right) \cdot \mathbf{\nabla }_{\perp }\right] \left( -%
\mathbf{\nabla }_{\perp }^{2}\varphi \right)  \notag \\
&&+\frac{1}{B}\left[ \left( -\mathbf{\nabla }_{\perp }\varphi \times 
\widehat{\mathbf{n}}\right) \cdot \mathbf{\nabla }_{\perp }\right] \left( -%
\mathbf{\nabla }_{\perp }^{2}\phi _{s}\right)  \notag \\
&=&0  \notag
\end{eqnarray}
In this equation, the quantitities are in \emph{physical units} 
\begin{equation*}
\begin{array}{cc}
\left( x,\eta \right) & m \\ 
t & s \\ 
\varphi & V \\ 
u & m/s
\end{array}
\end{equation*}
We change to the units 
\begin{equation*}
\begin{array}{cc}
\left( \frac{x^{phys}}{\rho _{s}},\frac{\eta ^{phys}}{\rho _{s}}\right)
\rightarrow & \left( x^{\prime },\eta ^{\prime }\right) \\ 
\Omega _{i}t^{phys}\rightarrow & t^{\prime } \\ 
\frac{\left| e\right| \varphi ^{phys}}{T_{e}}\rightarrow & \varphi ^{\prime }
\end{array}
\end{equation*}
Also 
\begin{equation*}
\frac{\left| e\right| \phi _{s}^{phys}}{T_{e}}\rightarrow \phi _{s}^{\prime }
\end{equation*}
After some arrangement, we have 
\begin{eqnarray}
&&\frac{\partial }{\partial t^{\prime }}\left( 1-\mathbf{\nabla }_{\perp
}^{^{\prime }2}\right) \varphi ^{\prime }  \label{primeeq} \\
&&-\frac{u^{phys}}{\Omega _{i}\rho _{s}}\frac{\partial }{\partial \eta
^{\prime }}\left( 1-\mathbf{\nabla }_{\perp }^{^{\prime }2}\right) \varphi
^{\prime }  \notag \\
&&-\rho _{s}\kappa _{n}^{phys}\frac{\partial }{\partial \eta ^{\prime }}%
\varphi ^{\prime }  \notag \\
&&+\left( \frac{\kappa _{T}^{phys}T_{e}}{B\left| e\right| \rho _{s}\Omega
_{i}}\right) \frac{1}{2}\frac{\partial }{\partial \eta ^{\prime }}\left(
\phi _{s}^{\prime }+\varphi ^{\prime }\right) ^{2}  \notag \\
&&-\left( \frac{\rho _{s}T_{e}}{B\left| e\right| u^{phys}}\kappa
_{n}^{\prime phys}\right) \frac{1}{2}\frac{\partial }{\partial \eta ^{\prime
}}\left( \phi _{s}^{\prime }\right) ^{2}  \notag \\
&&+\left( \frac{T_{e}}{B\rho _{s}^{2}\left| e\right| \Omega _{i}}\right) 
\left[ \left( -\mathbf{\nabla }_{\perp }^{\prime }\varphi ^{\prime }\times 
\widehat{\mathbf{n}}\right) \cdot \mathbf{\nabla }_{\perp }^{\prime }\right]
\left( -\mathbf{\nabla }_{\perp }^{\prime 2}\varphi ^{\prime }\right)  \notag
\\
&&+\left( \frac{T_{e}}{B\rho _{s}^{2}\left| e\right| \Omega _{i}}\right) 
\left[ \left( -\mathbf{\nabla }_{\perp }^{\prime }\phi _{s}^{\prime }\times 
\widehat{\mathbf{n}}\right) \cdot \mathbf{\nabla }_{\perp }^{\prime }\right]
\left( -\mathbf{\nabla }_{\perp }^{\prime 2}\varphi ^{\prime }\right)  \notag
\\
&&+\left( \frac{T_{e}}{B\rho _{s}^{2}\left| e\right| \Omega _{i}}\right) 
\left[ \left( -\mathbf{\nabla }_{\perp }^{\prime }\varphi ^{\prime }\times 
\widehat{\mathbf{n}}\right) \cdot \mathbf{\nabla }_{\perp }^{\prime }\right]
\left( -\mathbf{\nabla }_{\perp }^{\prime 2}\phi _{s}^{\prime }\right) 
\notag \\
&=&0  \notag
\end{eqnarray}
In this form, the \emph{prime} in $\phi _{s}^{\prime }$ means that it will
be calculated for $\left( x^{\prime },\eta ^{\prime }\right) $, \emph{i.e.}
with normalised variables.

We introduce new units for the speed of the reference system, $u$%
\begin{equation}
u\equiv \frac{u^{phys}}{\Omega _{i}\rho _{s}}  \label{newu}
\end{equation}
Also 
\begin{equation*}
\kappa _{n}\equiv \rho _{s}\kappa _{n}^{phys}
\end{equation*}
\begin{eqnarray*}
\frac{\kappa _{T}^{phys}T_{e}}{B\left| e\right| \rho _{s}\Omega _{i}} &=&%
\frac{\kappa _{T}^{phys}c_{s}^{2}}{\Omega _{i}^{2}\rho _{s}} \\
&=&\rho _{s}\kappa _{T}^{phys}
\end{eqnarray*}
Then we introduce 
\begin{equation*}
\kappa _{T}\equiv \rho _{s}\kappa _{T}^{phys}
\end{equation*}
\begin{equation*}
\frac{\rho _{s}T_{e}}{\left| e\right| B}\frac{1}{u^{phys}}\kappa
_{n}^{\prime phys}=\frac{1}{u}\rho _{s}^{2}\kappa _{n}^{\prime phys}
\end{equation*}
where $u$ is the adimensional speed, Eq.(\ref{newu}). It is then useful to
introduce 
\begin{equation*}
\kappa _{n}^{\prime }\equiv \rho _{s}^{2}\kappa _{n}^{\prime phys}
\end{equation*}
and the coefficient in the fifth line of Eq.(\ref{primeeq}) is, in terms of
adimensional quantities 
\begin{equation*}
\frac{\rho _{s}T_{e}}{\left| e\right| B}\frac{1}{u^{phys}}\kappa
_{n}^{\prime phys}=\frac{\kappa _{n}^{\prime }}{u}
\end{equation*}
Finally we note that 
\begin{equation*}
\frac{T_{e}}{B\rho _{s}^{2}\left| e\right| \Omega _{i}}\equiv 1
\end{equation*}
The same adimensionalization is done for $\alpha $ and $\beta $. The
equation is multiplied by $\rho _{s}^{2}$ and $\phi _{s}$ is normalized to $%
\phi _{s}^{\prime }$. 
\begin{eqnarray*}
\rho _{s}^{2}\alpha ^{phys} &=&\rho _{s}^{2}\left( \frac{1}{\rho _{s}^{2}}+%
\frac{\Omega _{i}\kappa _{n}^{phys}}{u^{phys}}\right) \rightarrow \alpha
^{\prime }=1+\frac{\rho _{s}\kappa _{n}^{phys}}{u^{phys}/\left( \Omega
_{i}\rho _{s}\right) } \\
&=&1+\frac{\kappa _{n}}{u}
\end{eqnarray*}
\begin{eqnarray*}
\beta ^{phys}\frac{T_{e}}{\left| e\right| }\rho _{s}^{2} &=&\frac{\Omega
_{i}\kappa _{n}^{\prime phys}}{2\left( u^{phys}\right) ^{2}B}\frac{T_{e}}{%
\left| e\right| }\rho _{s}^{2}\rightarrow \beta ^{\prime } \\
\beta ^{\prime } &=&\frac{1}{2}\left( \rho _{s}^{2}\kappa _{n}^{\prime
phys}\right) c_{s}^{2}\frac{1}{\Omega _{i}^{2}}\frac{1}{\left(
u^{phys}\right) ^{2}/\left( \Omega _{i}^{2}\right) } \\
&=&\frac{1}{2}\left( \rho _{s}^{2}\kappa _{n}^{\prime phys}\right) \frac{1}{%
\left[ u^{phys}/\left( \Omega _{i}\rho _{s}\right) \right] ^{2}} \\
&=&\frac{1}{2}\frac{\kappa _{n}^{\prime }}{u^{2}}
\end{eqnarray*}

The equation in adimensional variables (and removing the primes from the
variables and operators) becomes 
\begin{eqnarray}
&&\frac{\partial }{\partial t}\left( 1-\mathbf{\nabla }_{\perp }^{2}\right)
\varphi  \label{timeq} \\
&&-u\frac{\partial }{\partial \eta }\left( 1-\mathbf{\nabla }_{\perp
}^{2}\right) \varphi  \notag \\
&&-\kappa _{n}\frac{\partial }{\partial \eta }\varphi  \notag \\
&&+\kappa _{T}\frac{1}{2}\frac{\partial }{\partial \eta }\left( \phi
_{s}+\varphi \right) ^{2}  \notag \\
&&-\frac{\kappa _{n}^{\prime }}{u}\frac{1}{2}\frac{\partial }{\partial \eta }%
\phi _{s}^{2}  \notag \\
&&+\left[ \left( -\mathbf{\nabla }_{\perp }\varphi \times \widehat{\mathbf{n}%
}\right) \cdot \mathbf{\nabla }_{\perp }\right] \left( -\mathbf{\nabla }%
_{\perp }^{2}\varphi \right)  \notag \\
&&+\left[ \left( -\mathbf{\nabla }_{\perp }\phi _{s}\times \widehat{\mathbf{n%
}}\right) \cdot \mathbf{\nabla }_{\perp }\right] \left( -\mathbf{\nabla }%
_{\perp }^{2}\varphi \right)  \notag \\
&&+\left[ \left( -\mathbf{\nabla }_{\perp }\varphi \times \widehat{\mathbf{n}%
}\right) \cdot \mathbf{\nabla }_{\perp }\right] \left( -\mathbf{\nabla }%
_{\perp }^{2}\phi _{s}\right)  \notag \\
&=&0  \notag
\end{eqnarray}
In contrast to Eq.(\ref{timeqphys}) this equation is only expressed in terms
of adimensional variables.

\bigskip

Let us introduce a new function 
\begin{equation}
\psi \equiv \left( 1-\mathbf{\nabla }_{\perp }^{2}\right) \varphi
\label{defpsi}
\end{equation}
Then 
\begin{eqnarray}
&&\frac{\partial \psi }{\partial t}-u\frac{\partial \psi }{\partial \eta }%
-\kappa _{n}\frac{\partial \varphi }{\partial \eta }  \label{psiphi3} \\
&&+\kappa _{T}\frac{1}{2}\frac{\partial }{\partial \eta }\left( \phi
_{s}+\varphi \right) ^{2}-\frac{\kappa _{n}^{\prime }}{u}\frac{1}{2}\frac{%
\partial }{\partial \eta }\phi _{s}^{2}  \notag \\
&&+\left[ \left( -\mathbf{\nabla }_{\perp }\varphi \times \widehat{\mathbf{n}%
}\right) \cdot \mathbf{\nabla }_{\perp }\right] \left( \psi -\varphi \right)
\notag \\
&&+\left[ \left( -\mathbf{\nabla }_{\perp }\phi _{s}\times \widehat{\mathbf{n%
}}\right) \cdot \mathbf{\nabla }_{\perp }\right] \left( \psi -\varphi \right)
\notag \\
&&+\left[ \left( -\mathbf{\nabla }_{\perp }\varphi \times \widehat{\mathbf{n}%
}\right) \cdot \mathbf{\nabla }_{\perp }\right] \left( -\mathbf{\nabla }%
_{\perp }^{2}\phi _{s}\right)  \notag \\
&=&0  \notag
\end{eqnarray}
The last three nonlinear terms are 
\begin{equation}
\left( -\mathbf{\nabla }_{\perp }\varphi \times \widehat{\mathbf{n}}\right)
\cdot \left( \mathbf{\nabla }_{\perp }\psi -\mathbf{\nabla }_{\perp }\varphi
\right) =-\frac{\partial \varphi }{\partial \eta }\frac{\partial \psi }{%
\partial x}+\frac{\partial \varphi }{\partial x}\frac{\partial \psi }{%
\partial \eta }  \label{t31}
\end{equation}
\begin{eqnarray}
\left( -\mathbf{\nabla }_{\perp }\phi _{s}\times \widehat{\mathbf{n}}\right)
\cdot \left( \mathbf{\nabla }_{\perp }\psi -\mathbf{\nabla }_{\perp }\varphi
\right) &=&-\frac{\partial \phi _{s}}{\partial \eta }\frac{\partial \psi }{%
\partial x}+\frac{\partial \phi _{s}}{\partial x}\frac{\partial \psi }{%
\partial \eta }  \label{t32} \\
&&+\frac{\partial \phi _{s}}{\partial \eta }\frac{\partial \varphi }{%
\partial x}-\frac{\partial \phi _{s}}{\partial x}\frac{\partial \varphi }{%
\partial \eta }  \notag
\end{eqnarray}
\begin{eqnarray}
&&\left[ \left( -\mathbf{\nabla }_{\perp }\varphi \times \widehat{\mathbf{n}}%
\right) \cdot \mathbf{\nabla }_{\perp }\right] \left( -\alpha \phi
_{s}+\beta \phi _{s}^{2}\right)  \label{t33} \\
&=&\left( -\alpha +2\beta \phi _{s}\right) \left( -\frac{\partial \varphi }{%
\partial \eta }\frac{\partial \phi _{s}}{\partial x}+\frac{\partial \varphi 
}{\partial x}\frac{\partial \phi _{s}}{\partial \eta }\right)  \notag
\end{eqnarray}

The final form of the equation for $\psi $ is 
\begin{eqnarray}
&&\frac{\partial \psi }{\partial t}-u\frac{\partial \psi }{\partial \eta }%
-\kappa _{n}\frac{\partial \varphi }{\partial \eta }  \label{psiphi4} \\
&&-\frac{\partial \varphi }{\partial \eta }\frac{\partial \psi }{\partial x}+%
\frac{\partial \varphi }{\partial x}\frac{\partial \psi }{\partial \eta }-%
\frac{\partial \phi _{s}}{\partial \eta }\frac{\partial \psi }{\partial x}+%
\frac{\partial \phi _{s}}{\partial x}\frac{\partial \psi }{\partial \eta } 
\notag \\
&&+\kappa _{T}\frac{1}{2}\frac{\partial }{\partial \eta }\left( \phi
_{s}+\varphi \right) ^{2}-\frac{\kappa _{n}^{\prime }}{u}\frac{1}{2}\frac{%
\partial }{\partial \eta }\phi _{s}^{2}+\frac{\partial \phi _{s}}{\partial
\eta }\frac{\partial \varphi }{\partial x}-\frac{\partial \phi _{s}}{%
\partial x}\frac{\partial \varphi }{\partial \eta }  \notag \\
&&+\left( -\alpha +2\beta \phi _{s}\right) \left( -\frac{\partial \varphi }{%
\partial \eta }\frac{\partial \phi _{s}}{\partial x}+\frac{\partial \varphi 
}{\partial x}\frac{\partial \phi _{s}}{\partial \eta }\right)  \notag \\
&=&0  \notag
\end{eqnarray}

\subsection{The time evolution of a perturbation around the stationary
poloidal flow}

In this case, $\phi _{s}$ is the periodic solution expressed in terms of the
Weierstrass function, and with parameters chosen such that the pattern of
the flow is parallel with the poloidal direction. Then $\phi _{s}$ only
depends on $x$ and 
\begin{equation}
\frac{\partial \phi _{s}}{\partial \eta }\equiv 0  \label{phisnoy}
\end{equation}
which considerably simplifies the equation.

In addition we will take again the fixed (laboratory frame, for this case 
\begin{eqnarray*}
t &\rightarrow &t^{\prime }=t \\
\eta &\rightarrow &y=\eta +ut
\end{eqnarray*}
which gives 
\begin{eqnarray*}
\frac{\partial }{\partial \eta } &=&\frac{\partial }{\partial y} \\
\frac{\partial }{\partial t^{\prime }} &=&\frac{\partial }{\partial t}-u%
\frac{\partial }{\partial y}
\end{eqnarray*}
and the equation becomes 
\begin{eqnarray*}
&&\frac{\partial \psi }{\partial t}-\frac{\partial \varphi }{\partial y}%
\frac{\partial \psi }{\partial x}+\frac{\partial \varphi }{\partial x}\frac{%
\partial \psi }{\partial y}+\frac{\partial \phi _{s}}{\partial x}\frac{%
\partial \psi }{\partial y} \\
&&-\frac{\partial \phi _{s}}{\partial x}\frac{\partial \varphi }{\partial y}%
-\kappa _{n}\frac{\partial \varphi }{\partial y}+\kappa _{T}\frac{1}{2}\frac{%
\partial }{\partial y}\left( \phi _{s}+\varphi \right) ^{2} \\
&&+\left( -\alpha +2\beta \phi _{s}\right) \left( -\frac{\partial \varphi }{%
\partial y}\frac{\partial \phi _{s}}{\partial x}\right) \\
&=&0
\end{eqnarray*}
or 
\begin{eqnarray}
&&\frac{\partial \psi }{\partial t}+\frac{\partial \psi }{\partial x}\left( -%
\frac{\partial \varphi }{\partial y}\right) +\frac{\partial \psi }{\partial y%
}\left( \frac{\partial \varphi }{\partial x}+\frac{\partial \phi _{s}}{%
\partial x}\right)  \label{eqpsi} \\
&&-\left[ \left( 1-\alpha +2\beta \phi _{s}\right) \frac{\partial \phi _{s}}{%
\partial x}+\kappa _{n}\right] \frac{\partial \varphi }{\partial y}+\kappa
_{T}\frac{1}{2}\frac{\partial }{\partial y}\left( \phi _{s}+\varphi \right)
^{2}  \notag \\
&=&0  \notag
\end{eqnarray}
This can be written 
\begin{equation}
\frac{D\psi }{Dt}+C\left[ \varphi \right] =0  \label{eqlag}
\end{equation}
where 
\begin{equation*}
\frac{D\psi }{Dt}\equiv \frac{\partial \psi }{\partial t}+v_{x}\frac{%
\partial \psi }{\partial x}+v_{y}\frac{\partial \psi }{\partial y}
\end{equation*}
\begin{eqnarray}
v_{x} &\equiv &-\frac{\partial \varphi }{\partial y}  \label{vxvy} \\
v_{y} &\equiv &\frac{\partial \varphi }{\partial x}+\frac{\partial \phi _{s}%
}{\partial x}  \notag
\end{eqnarray}
\begin{eqnarray}
C\left[ \varphi \right] &\equiv &-\left[ \left( 1-\alpha +2\beta \phi
_{s}\right) \frac{\partial \phi _{s}}{\partial x}+\kappa _{n}\right] \frac{%
\partial \varphi }{\partial y}+\kappa _{T}\frac{1}{2}\frac{\partial }{%
\partial y}\left( \phi _{s}+\varphi \right) ^{2}  \label{cmaredephi} \\
&=&\left[ -\kappa _{n}-\left( 1-\alpha +2\beta \phi _{s}\right) \frac{%
\partial \phi _{s}}{\partial x}+\kappa _{T}\left( \phi _{s}+\varphi \right) %
\right] \frac{\partial \varphi }{\partial y}  \notag
\end{eqnarray}
The solution of Eq.(\ref{eqlag}) is 
\begin{equation}
\psi \left( x,y\right) =\psi \left( x_{0},y_{0}\right)
-\int_{t_{0}}^{t}d\tau C\left[ \varphi \left( x_{0}+\tau v_{x},y_{0}+\tau
v_{y}\right) \right]  \label{solpsi}
\end{equation}

The Eq.(\ref{solpsi}) gives a possible way to solve numerically the
equations. The final system is 
\begin{eqnarray}
-\mathbf{\nabla }_{\perp }^{2}\varphi +\varphi &=&\psi  \label{sys1} \\
\psi \left( x,y\right) &=&\psi \left( x_{0},y_{0}\right)  \notag \\
&&-\int_{t_{0}}^{t}d\tau \left. \left[ -\kappa _{n}-\left( 1-\alpha +2\beta
\phi _{s}\right) \frac{\partial \phi _{s}}{\partial x}+\kappa _{T}\left(
\phi _{s}+\varphi \right) \right] \frac{\partial \varphi }{\partial y}%
\right| _{\substack{ x=x_{0}+\tau v_{x}  \\ y=y_{0}+\tau v_{y}}}
\end{eqnarray}

We note that the vector field of Eq.(\ref{vxvy}) does not have zero
divergence 
\begin{equation}
\mathbf{\nabla }_{\perp }\cdot \mathbf{v=-}\frac{\partial ^{2}\phi _{s}}{%
\partial x^{2}}  \label{divv}
\end{equation}

\subsection{The equation of Petviashvili with time dependence due to the
temperature gradient}

\subsubsection{Physical model of the ion drift instability}

This equation has been derived in several papers and used in numerical
calculations (Laedke Spatschek 1986). From the dynamical equations of ion
density (as above) it is obtained 
\begin{eqnarray}
&&\frac{\partial }{\partial t}\left( 1-\mathbf{\nabla }_{\perp }^{2}\right)
\phi  \label{dynion2} \\
&&-\left( -\mathbf{\nabla }_{\perp }\phi \times \widehat{\mathbf{n}}\right)
\cdot \mathbf{v}_{d}^{\ast }+\left( -\mathbf{\nabla }_{\perp }\phi \times 
\widehat{\mathbf{n}}\right) \cdot \mathbf{v}_{dT}\phi  \notag \\
&&+\left[ \left( -\mathbf{\nabla }_{\perp }\phi \times \widehat{\mathbf{n}}%
\right) \cdot \mathbf{\nabla }_{\perp }\right] \left( -\mathbf{\nabla }%
_{\perp }^{2}\phi \right)  \notag \\
&=&0  \notag
\end{eqnarray}
where 
\begin{equation*}
\mathbf{v}_{d}^{\ast }=-\mathbf{\nabla }_{\perp }\ln n_{0}-\mathbf{\nabla }%
_{\perp }\ln T_{e}
\end{equation*}
\begin{equation*}
\mathbf{v}_{dT}=-\mathbf{\nabla }_{\perp }\ln T_{e}
\end{equation*}
This is Eq.(8) from \cite{LaedkeSpatechek1}. It differs of what we have
obtained before.

Later the equation is expressed as, explained in the article as the limit 
\begin{equation*}
L_{T}\rightarrow \infty
\end{equation*}
\begin{equation}
\frac{\partial }{\partial t}\left( 1-\mathbf{\nabla }_{\perp }^{2}\right)
\phi +v_{\ast }\frac{\partial \phi }{\partial y}-v_{dT}\frac{1}{2}\frac{%
\partial }{\partial y}\phi ^{2}=0  \label{eqvt}
\end{equation}
This is the equation (12) from paper of 1986 by Laedke and Spatschek \cite
{LaedkeSpatschek1}. Here the units are: time is measured in $\Omega
_{i}^{-1} $; space in $\rho _{s}$; the potential $\phi $ is $\left| e\right|
\phi ^{phys}/T_{e}$.

\subsubsection{The transformation of the equation}

A referential moving with the velocity $u$ is introduced 
\begin{eqnarray*}
y &\rightarrow &y^{\prime }=y-ut \\
t &\rightarrow &t^{\prime }=t
\end{eqnarray*}
Then 
\begin{equation*}
\left( \frac{\partial }{\partial t^{\prime }}-u\frac{\partial }{\partial
y^{\prime }}\right) \left( 1-\mathbf{\nabla }_{\perp }^{2}\right) \phi
+v_{\ast }\frac{\partial \phi }{\partial y^{\prime }}-v_{dT}\frac{1}{2}\frac{%
\partial }{\partial y^{\prime }}\phi ^{2}=0
\end{equation*}
or 
\begin{equation*}
\frac{\partial }{\partial t^{\prime }}\left( 1-\mathbf{\nabla }_{\perp
}^{2}\right) \phi -u\frac{\partial }{\partial y^{\prime }}\left( 1-\mathbf{%
\nabla }_{\perp }^{2}\right) \phi +v_{\ast }\frac{\partial \phi }{\partial
y^{\prime }}-v_{dT}\frac{1}{2}\frac{\partial }{\partial y^{\prime }}\phi
^{2}=0
\end{equation*}
We will drop the prime 
\begin{equation}
\frac{\partial }{\partial t}\left( 1-\mathbf{\nabla }_{\perp }^{2}\right)
\phi =\frac{\partial }{\partial y}\left[ -u\mathbf{\nabla }_{\perp
}^{2}+\left( u-v_{\ast }\right) +\frac{v_{dT}}{2}\phi ^{2}\right]
\label{equv}
\end{equation}
We divide by $u$ 
\begin{equation*}
\frac{1}{u}\frac{\partial }{\partial t}\left( 1-\mathbf{\nabla }_{\perp
}^{2}\right) \phi =\frac{\partial }{\partial y}\left[ -\mathbf{\nabla }%
_{\perp }^{2}+\frac{\left( u-v_{\ast }\right) }{u}+\frac{v_{dT}}{2u}\phi ^{2}%
\right]
\end{equation*}
and replace the time variable by 
\begin{equation*}
t\rightarrow t^{\prime }\equiv ut
\end{equation*}
Here the time in Eq.(\ref{equv}) had been already adimensionalized by
expressing it in units of $\Omega _{i}^{-1}$. Since $y$ in the same equation
is measured in $\rho _{s}$, the velocity is measured in units of 
\begin{equation*}
\frac{\rho _{s}}{\Omega _{i}^{-1}}
\end{equation*}
This is the case for both $u$ and $v_{\ast }$. The equation becomes 
\begin{equation*}
\frac{\partial }{\partial t}\left( 1-\mathbf{\nabla }_{\perp }^{2}\right)
\phi =\frac{\partial }{\partial y}\left[ -\mathbf{\nabla }_{\perp
}^{2}+4\eta ^{2}+\frac{v_{dT}}{2u}\phi ^{2}\right]
\end{equation*}
\begin{equation}
\frac{\partial }{\partial t}\left( 1-\mathbf{\nabla }_{\perp }^{2}\right)
\phi =\frac{\partial }{\partial y}\left( -\mathbf{\nabla }_{\perp
}^{2}+4\eta ^{2}\right) \phi +\frac{v_{dT}}{u}\phi \frac{\partial \phi }{%
\partial y}  \label{eqvt2}
\end{equation}
with 
\begin{equation}
4\eta ^{2}\equiv \frac{u-v_{\ast }}{u}>0  \label{patrueta2}
\end{equation}
We now renormalise the potential $\phi $ by including in it the factor $%
-v_{dT}/u$. 
\begin{equation}
\phi \rightarrow \phi ^{\prime }\equiv -\frac{v_{dT}}{u}\phi
\label{scalephi}
\end{equation}
For this both sides of the equation are multiplied by $-v_{dT}/u$ and the
equation is 
\begin{equation}
\frac{\partial }{\partial t}\left( 1-\mathbf{\nabla }_{\perp }^{2}\right)
\phi ^{\prime }=\frac{\partial }{\partial y}\left( -\mathbf{\nabla }_{\perp
}^{2}+4\eta ^{2}\right) \phi ^{\prime }-\phi ^{\prime }\frac{\partial \phi
^{\prime }}{\partial y}  \label{FPeqtime}
\end{equation}
The units are: 
\begin{eqnarray}
\phi |_{phys} &\rightarrow &\frac{e\phi |_{phys}}{T_{e}}\rightarrow -\frac{%
v_{dT}}{u}\frac{e\phi |_{phys}}{T_{e}}\;\;\text{or}  \label{phitrans} \\
\phi ^{\prime } &=&-\frac{v_{dT}}{u}\frac{e\phi |_{phys}}{T_{e}}  \notag
\end{eqnarray}
The velocities are measured as 
\begin{eqnarray}
u|_{phys} &\rightarrow &\frac{1}{\rho _{s}\Omega _{i}^{-1}}u|_{phys}\;\;%
\text{or}  \label{utrans} \\
u^{\prime } &=&\frac{1}{\rho _{s}\Omega _{i}^{-1}}u|_{phys}  \notag
\end{eqnarray}
This quantity disappears from the equation, being absorbed into the new time
variable. The time is measured as 
\begin{eqnarray}
t|_{phys} &\rightarrow &\Omega _{i}^{-1}t|_{phys}\rightarrow u^{\prime
}\Omega _{i}^{-1}t|_{phys}\;\;\text{or}  \label{ttrans} \\
t^{\prime } &=&\frac{1}{\rho _{s}\Omega _{i}^{-1}}u|_{phys}\Omega
_{i}^{-1}t|_{phys}  \notag \\
&=&\frac{1}{\rho _{s}}u|_{phys}t|_{phys}  \notag
\end{eqnarray}
The distances are measured in $\rho _{s}$%
\begin{equation}
y^{\prime }=\frac{1}{\rho _{s}}y|_{phys}  \label{ytrans}
\end{equation}
Naturally, the physical variables are measured in SI 
\begin{equation*}
\begin{array}{cc}
y|_{phys} & m \\ 
t|_{phys} & s \\ 
u,v_{\ast }\,|_{phys} & m/s \\ 
\phi |_{phys} & V
\end{array}
\end{equation*}
and at this point the \emph{primes} are suppressed.

If we have a result about some time duration $\Delta t^{\prime }$ (\emph{i.e.%
} expressed in terms of the non-dimensional variables) then in order to
recuperate the \emph{physical} duration, we have to do 
\begin{equation*}
\Delta t|_{phys}=\frac{\rho _{s}}{u|_{phys}}\Delta t^{\prime }
\end{equation*}

\bigskip

Only for 
\begin{equation*}
u>v_{\ast }
\end{equation*}
the localised stationary solutions are posible.

The stationary form of the Petviashvili equation is 
\begin{equation}
\frac{\partial }{\partial y}\left( -\mathbf{\nabla }_{\perp }^{2}+4\eta
^{2}\right) \phi -\phi \frac{\partial \phi }{\partial y}=0  \label{eqvtstat}
\end{equation}
\begin{equation}
\Delta \phi =4\eta ^{2}\phi -\frac{1}{2}\phi ^{2}  \label{eqvtstat2}
\end{equation}
which is the same as the general form for

These can be found in \cite{LaedkeSpatschek1}.

\subsection{The equation derived from Ertel 's theorem}

The derivation of the equation is done on the same basis as the derivation
of Laedke and Spatschek 1988. However, in a series of papers, Su and Horton 
\cite{Su}, \cite{HH}, develop a more physical justification of the scalar
nonlinearity model, although their primary aim was to investigate the
stability of the dipolar (Larichev-Reznik) solution of the Hasegawa-Mima
equation.

The equations for the ion instabilities are 
\begin{eqnarray*}
\left[ \frac{\partial }{\partial t}+\left( \mathbf{v}\cdot \mathbf{\nabla }%
\right) \right] \mathbf{v} &=&-\frac{\left| e\right| }{m_{i}}\mathbf{\nabla }%
_{\perp }\Phi +\Omega _{i}\mathbf{v\times }\widehat{\mathbf{n}} \\
\frac{\partial n}{\partial t}+\mathbf{\nabla }_{\perp }\left( n\mathbf{v}%
\right) &=&0
\end{eqnarray*}
It is derived the Ertel's theorem 
\begin{equation}
\frac{d}{dt}\left[ \frac{\Omega _{i}+\widehat{\mathbf{n}}\cdot \left( 
\mathbf{\nabla }_{\perp }\times \mathbf{v}\right) }{n\left( x\right) }\right]
=0  \label{Ertel}
\end{equation}
This essentially means that: the sum of the cyclotron frequency and the
vertical component of the vorticity, divided by the density, is constant
along the line of Lagrangean evolution with the velocity $E\times B$.

If one neglects the cyclotron frequency and takes the density constant, this
theorem is equivalent with the Euler theorem for ideal fluid 
\begin{equation*}
\frac{d\mathbf{\omega }}{dt}=0
\end{equation*}

The ordering is 
\begin{equation*}
\varepsilon _{t}\equiv \frac{1}{\Omega _{i}}\frac{\partial }{\partial t}\sim 
\frac{\mathbf{v\cdot \nabla }_{\perp }}{\Omega _{i}}\ll 1
\end{equation*}
The equation becomes 
\begin{equation}
\frac{\partial }{\partial t}\left( \frac{1+\varepsilon _{n}\mathbf{\nabla }%
_{\perp }^{2}\varphi }{n}\right) +\left[ \left( -\mathbf{\nabla }_{\perp
}\varphi \times \widehat{\mathbf{n}}\right) \cdot \mathbf{\nabla }_{\perp }%
\right] \left( \frac{1+\varepsilon _{n}\mathbf{\nabla }_{\perp }^{2}\varphi 
}{n}\right) =0  \label{eqionfromertel}
\end{equation}

It is assumed the Boltzmann density distribution for the ions 
\begin{eqnarray}
n &=&n_{0}\left( x\right) \exp \left( -\frac{\left| e\right| \Phi }{T_{e}}%
\right)  \label{nisexp} \\
&=&n_{0}\left( x\right) \exp \left[ -\varepsilon _{n}\frac{\varphi }{T\left(
x\right) }\right]  \notag
\end{eqnarray}

\subsection{Stationary traveling solutions}

This means solutions of the type 
\begin{equation*}
\varphi =\varphi \left( x,y-ut\right)
\end{equation*}
where $u$ is the speed of the system of reference. The \emph{stationary }%
equation in the moving frame is 
\begin{equation*}
-u\frac{\partial }{\partial y}\left( \frac{1+\varepsilon _{n}\mathbf{\nabla }%
_{\perp }^{2}\varphi }{n_{0}\left( x\right) \exp \left[ \frac{\varepsilon
_{n}\varphi }{T\left( x\right) }\right] }\right) +\left[ \left( -\mathbf{%
\nabla }_{\perp }\varphi \times \widehat{\mathbf{n}}\right) \cdot \mathbf{%
\nabla }_{\perp }\right] \left( \frac{1+\varepsilon _{n}\mathbf{\nabla }%
_{\perp }^{2}\varphi }{n_{0}\left( x\right) \exp \left[ \frac{\varepsilon
_{n}\varphi }{T\left( x\right) }\right] }\right) =0
\end{equation*}

This equation is identically verified if the expression in the brackets is a
general function of 
\begin{equation*}
\varphi -ux
\end{equation*}
\emph{i.e.} 
\begin{equation*}
\frac{1+\varepsilon _{n}\mathbf{\nabla }_{\perp }^{2}\varphi }{n_{0}\left(
x\right) \exp \left[ \frac{\varepsilon _{n}\varphi }{T\left( x\right) }%
\right] }=F\left( \varphi -ux\right)
\end{equation*}
One can take this function to be 
\begin{equation*}
F\left( \varphi -ux\right) =n_{0}\left( x-\frac{\varphi }{u}\right)
\end{equation*}
which gives 
\begin{equation}
\varepsilon _{n}\mathbf{\nabla }_{\perp }^{2}\varphi =\frac{n_{0}\left(
x\right) }{n_{0}\left( x-\frac{\varphi }{u}\right) }\exp \left[ \frac{%
\varepsilon _{n}\varphi }{T\left( x\right) }\right] -1  \label{eqexp}
\end{equation}
In the case where the density profile is exponential 
\begin{equation*}
n_{0}\left( x\right) =\exp \left( -\varepsilon _{n}x\right)
\end{equation*}
the equation becomes 
\begin{equation}
\varepsilon _{n}\mathbf{\nabla }_{\perp }^{2}\varphi =\exp \left[
\varepsilon _{n}\left( \frac{1}{T\left( x\right) }-\frac{1}{u}\right)
\varphi \right] -1  \label{AHeq}
\end{equation}
The dimensional parameters are 
\begin{equation}
\kappa _{T}\equiv \rho _{s0}\frac{d}{dx}\left( \frac{1}{T}\right)  \label{kt}
\end{equation}
\begin{equation}
v_{d0}^{\prime }\equiv \rho _{s0}\frac{dv_{d}}{dx}\sim \varepsilon _{n}v_{d0}
\label{vdprim}
\end{equation}
Thaking the logarithm of the Eq.(\ref{eqexp}) 
\begin{equation*}
\ln \left( 1+\varepsilon _{n}\mathbf{\nabla }_{\perp }^{2}\varphi \right)
=\ln n_{0}\left( x\right) -\ln n_{0}\left( x-\frac{\varphi }{u}\right)
+\varepsilon _{n}\frac{\varphi }{T\left( x\right) }
\end{equation*}
The density profile is expanded 
\begin{equation*}
\ln n_{0}\left( x\right) \approx -\varepsilon _{n}\left( v_{d0}x+\frac{%
v_{d0}^{\prime }}{2}x^{2}+\cdots \right)
\end{equation*}
gives 
\begin{equation*}
\ln \left( 1+\varepsilon _{n}\mathbf{\nabla }_{\perp }^{2}\varphi \right)
\approx \varepsilon _{n}k^{2}\left( u,x\right) \varphi +\varepsilon _{n}%
\frac{v_{d0}^{\prime }}{2u^{2}}\varphi ^{2}+\cdots
\end{equation*}
\begin{equation}
k^{2}\left( u,x\right) \equiv \frac{1}{T\left( x\right) }-\frac{v_{d}\left(
x\right) }{u}  \label{k2su}
\end{equation}
\begin{eqnarray*}
v_{d}\left( x\right) &=&v_{d0}+v_{d0}^{\prime }x \\
&=&1+v_{d0}^{\prime }x
\end{eqnarray*}

The ordering is 
\begin{equation}
\rho _{s0}^{2}\mathbf{\nabla }^{2}\sim \frac{\left| e\right| \Phi }{T_{e}}%
\sim \kappa _{T}\sim v_{d0}^{\prime }\sim \frac{\rho _{s0}}{L_{n}}\equiv
\varepsilon _{n}\sim \varepsilon \ll 1  \label{order1}
\end{equation}
The velocity of the moving referential $u$ and of $v_{d0}$ are of the same
amplitude, normalized at $1$. 
\begin{equation*}
u\sim v_{d0}=1
\end{equation*}
Keeping only terms of order $\varepsilon ^{2}$ 
\begin{equation}
\mathbf{\nabla }_{\perp }^{2}\varphi =k_{0}^{2}\varphi +\frac{v_{d0}^{\prime
}}{2u^{2}}\varphi ^{2}  \label{FPaccSu}
\end{equation}
where 
\begin{equation*}
k^{2}\left( u,x\right) =k_{0}^{2}+\alpha x+\cdots
\end{equation*}
with the notations 
\begin{equation*}
k_{0}^{2}=1-\frac{v_{d0}}{u}\sim \varepsilon
\end{equation*}
\begin{equation*}
\alpha =\kappa _{T}-\frac{v_{d0}}{u}\sim \varepsilon ^{2}
\end{equation*}
The gradient of the drift velocity gives rise to a linear damping 
\begin{equation*}
-\frac{v_{d0}}{u}x\varphi
\end{equation*}
The temperature gradient generates also a linear term that compensates the
linear term of the gradient of the drift velocity.

\subsection{The energy of the solutions}

The density of energy has the expression 
\begin{equation*}
E\left( x,y,t\right) =\frac{1}{2}\left[ \frac{\varphi ^{2}\left( x\right) }{%
T\left( x\right) }+\left( \mathbf{\nabla }_{\perp }\varphi \right) ^{2}%
\right]
\end{equation*}
The units are 
\begin{equation*}
\varphi =\frac{L_{n}}{\rho _{s0}}\frac{\left| e\right| \Phi }{T_{0}}
\end{equation*}
and $T\left( x\right) $ is adimansional. The distances are measured in $\rho
_{s0}$. This means that 
\begin{equation*}
E\left( x,y,t\right) =\left( \frac{L_{n}}{\rho _{s0}}\right) ^{2}\frac{1}{2}%
\left[ \frac{1}{T\left( x\right) }\left( \frac{\left| e\right| \Phi }{T_{0}}%
\right) ^{2}+\left( \mathbf{\nabla }_{\perp }\frac{\left| e\right| \Phi }{%
T_{0}}\right) ^{2}\right]
\end{equation*}

\end{document}